\documentclass[aps,prb,nobalancelastpage,superscriptaddress, 10pt, twocolumn]{revtex4-2}

\usepackage{color,ifthen,amsthm,amsmath,amsxtra,amsfonts,dsfont,graphicx,bm,tikz,scalerel,wasysym,bbm,graphicx,amsthm,braket,physics,empheq,CircuitTikz,comment,soul}

\usepackage[inline]{enumitem}
\usepackage[colorlinks=true,linkcolor=blue, citecolor=blue, urlcolor=blue, bookmarks]{hyperref}
\def\rhoAQ{\rho_{A,Q}}

\newcommand{\be}{\begin{equation}}
\newcommand{\ee}{\end{equation}}
\newcommand{\bea}{\begin{eqnarray}}
\newcommand{\eea}{\end{eqnarray}}
\newcommand{\bse}{\begin{subequations}}
\newcommand{\ese}{\end{subequations}}
\newcommand{\conL}{\star}
\newcommand{\conR}{\ast}

\theoremstyle{plain}
\newcommand{\1}{\mathbbm{1}}

\newtheorem*{lemma}{Lemma}
\theoremstyle{plain}

\theoremstyle{plain}
\newtheorem*{observation}{Observation}
\theoremstyle{plain}

\begin{document}

\title{Dynamics of charge fluctuations from asymmetric initial states}

\author{Bruno Bertini}
\affiliation{School of Physics and Astronomy, University of Nottingham, Nottingham, NG7 2RD, UK}
\affiliation{Centre for the Mathematics and Theoretical Physics of Quantum Non-Equilibrium Systems, University of Nottingham, Nottingham, NG7 2RD, UK}

\author{Katja Klobas}
\affiliation{School of Physics and Astronomy, University of Nottingham, Nottingham, NG7 2RD, UK}
\affiliation{Centre for the Mathematics and Theoretical Physics of Quantum Non-Equilibrium Systems, University of Nottingham, Nottingham, NG7 2RD, UK}

\author{Mario Collura}
\affiliation{SISSA and INFN Sezione di Trieste, via Bonomea 265, 34136 Trieste, Italy}

\author{Pasquale Calabrese}
\affiliation{SISSA and INFN Sezione di Trieste, via Bonomea 265, 34136 Trieste, Italy}
\affiliation{International Centre for Theoretical Physics (ICTP), Strada Costiera 11, 34151 Trieste, Italy}

\author{Colin Rylands}
\affiliation{SISSA and INFN Sezione di Trieste, via Bonomea 265, 34136 Trieste, Italy}

\begin{abstract}

Conserved-charge densities are very special observables in quantum many-body systems as, by construction, they encode information about the dynamics. Therefore, their evolution is expected to be of much simpler interpretation than that of generic observables and to return universal information on the state of the system at any given time. Here we study the dynamics of the fluctuations of conserved $U(1)$ charges in systems that are prepared in charge-asymmetric initial states. We characterise the charge fluctuations in a given subsystem using the full-counting statistics of the truncated charge and the quantum entanglement between the subsystem and the rest resolved to the symmetry sectors of the charge. We show that, even though the initial states considered are homogeneous in space, the charge fluctuations generate an effective inhomogeneity due to the charge-asymmetric nature of the initial states. We use this observation to map the problem into that of charge fluctuations on inhomogeneous, charge-symmetric states and treat it using a recently developed space-time duality approach. Specialising the treatment to interacting integrable systems we combine the space-time duality approach with generalised hydrodynamics to find explicit predictions. 

\end{abstract}

\maketitle


\section{Introduction}
\label{sec:intro}

Finding an efficient description for the non-equilibrium dynamics of interacting quantum matter is one of the main challenges of modern theoretical physics~\cite{PolkovnikovReview, calabrese2016introduction, essler2016quench,gogolin2016equilibration, rigol2008thermalization, serbyn2021quantum}. Even though this problem has been at the centre of attention since the inception of quantum mechanics~\cite{vonneumann2010proof}, and with the turn of the millennium it has become amenable to experimental investigations~\cite{bloch2008many}, it remains to date largely open. Indeed, apart from a few remarkable special cases~\cite{bertini2019entanglement, bertini2019exact, piroli2020exact, klobas2021entanglement,klobas2021exact, klobas2021exactrelaxation, klobas2021entanglement}, an efficient description for the finite-time dynamics of quantum matter prepared in an out-of-equilibrium state has not yet been found. The only regime that can be efficiently described in generic systems is the quasi-stationary regime emerging at late times, where quantum matter behaves as a classical fluid~\cite{VidmarRigol, doyon2020lecture, bertini2021finitetemperature, bastianello2022introduction, alba2021generalized}. 
 
Recently, a remarkable breakthrough came from exploiting a duality between space and time~\cite{bertini2023nonequilibrium, bertini2022growth} (see also Refs.~\cite{banuls2009matrix, bertini2018exact,bertini2019entanglement,ippoliti2021postselectionfree, lu2021spacetime, garratt2021manybody, lerose2021influence} for related approaches). In essence, the idea is to describe the finite-time dynamics of a system in terms of the ``space-dynamics" of the ``dual system" obtained exchanging the roles of space and time in its path integral. In this way the far from equilibrium regime of a large system is mapped into the quasi-stationary regime of its dual counterpart. This approach works naturally for one-dimensional systems (although one can imagine to extend it to higher dimensions by exchanging the roles of time and one particular spatial dimension) and leads to predictions for the time evolution of ``universal" properties of the system such as entanglement among subsystems~\cite{bertini2022growth} and fluctuations of conserved U(1) charges~\cite{bertini2023nonequilibrium}. This approach is particularly powerful for interacting-integrable systems treatable via thermodynamic Bethe ansatz (TBA)~\cite{korepin1997quantum, takahashi1999thermodynamics}, where the predictions can be efficiently evaluated by solving few suitable integral equations.

Up to now, however, the space-time duality approach has been able to capture the dynamics of charge fluctuations only when {the initial state has no charge fluctuations within a subsystem.} This constraint poses serious limitations on the observable physics: the fluctuation of the charge in a certain region can only originate at the region's boundary rather than throughout its bulk as it happens for asymmetric initial states. The physics of charge fluctuations emerging from asymmetric initial states is consequently much richer and, in a sense, much more ``out-of-equilibrium". For instance, a natural question that one can study in this setting is to what extent the symmetry is broken by the initial state and whether or not it gets restored at large times~\cite{ares2022entanglement,ares2023lack}.

{Here we propose a significant extension of the space-time duality approach that is able to solve this problem. Our key observation is that measuring the evolution of charge fluctuations from states that are spatially homogeneous but not charge-symmetric makes the problem effectively \emph{spatially inhomogeneous}}. This suggests it can be treated combining the space-time duality approach with {generalised hydrodynamics}~\cite{bertini2016transport,castroalvaredo2016emergent}, the nowadays standard theory for inhomogeneous quenches~\cite{alba2021generalized}. In the following we show that this intuition can be made precise and find closed-form predictions for {the dynamics of the full counting statistics of the conserved charge and the growth of entanglement resolved to each symmetry sector. In particular, this allows us to provide a closed-form prediction for the evolution of the so-called ``entanglement asymmetry''~\cite{ares2022entanglement}, which characterizes the restoration of the symmetry at late times.} To the best of our knowledge, the one presented here is the first analytical characterisation of the dynamics of  charge fluctuations in the presence of interactions and for generic initial states.  

An interesting highlight of our approach is to connect full counting statistics of conserved charges after quantum quenches to current fluctuations on non-equilibrium steady states. The latter are attracting increasing attention~\cite{doyon2020fluctuations,myers2020transport,doyon2023ballistic, gopalakrishnan2024distinct,krajnik2022exact,krajnik2024universal, samajdar2023quantum, krajnik2024dynamical, mcculloch2023full,rylands2023transport} as they give a characterisation of non-equilibrium steady states that is directly accessible in current experimental setups, see, e.g., Ref.~\cite{wei2022quantum}. In particular, we show that the space-time duality approach recovers the results found via the so-called ballistic fluctuation formalism~\cite{doyon2020fluctuations, myers2020transport}. Differently from the latter, however, it is immediately applicable to multi-replica quantities and can be used to characterise the interplay between charge fluctuations and quantum entanglement.  

{In the following two subsections we define more precisely the observables of interest and present our main results.}

\subsection{Observables of Interest }
\label{subsec:observables}

The main objective of this paper is to characterise the fluctuations of a $U(1)$ charge, $Q$, within a finite spatial region, $A$, of a quantum many-body system out of equilibrium. A natural quantity to consider is then the full counting statistics {(FCS)}~\cite{klich2009quantum,eisler2013full,eisler2013universality,lovas2017full,najafi2017full,collura2017full,bastianello2018from,calabrese2020full,oshima2023disordered}
\begin{equation} \label{eq:FCS}
  Z_{\beta}(A,t)= \tr[ e^{\beta Q_A} \rho_A(t) ],\qquad \beta\in \mathbb{R},
\end{equation}
where $Q_A$ is the charge truncated to the region $A$ and $\rho_A(t)=\tr_{\bar{A}}[\rho(t)]$ is the density matrix at time $t$ reduced to $A$ at time $t$ ($\bar{A}$ denotes the complement of $A$). 
 
We focus on the standard case where a system is brought out of equilibrium by means of a quantum quench protocol, i.e., the state at time $t>0$ is taken to be 
\be
\rho(t)= \mathbb U^{t} \ketbra{\Psi_0} \mathbb U^{-t},
\label{eq:quench}
\ee
where $\mathbb U$ is the time-evolution operator and the initial state $\ket{\Psi_0}$ is not one of its eigenstates~\footnote{In fact, it is the superposition of a macroscopic number of eigenstates of $\mathbb U$.}. In addition, here we mainly consider the generic situation in which $\ket{\Psi_0}$ is not an eigenstate of the charge $Q$, i.e., we admit 
\be
[\rho_A(t), Q_A]\neq 0.
\ee 
The FCS encodes all the charge fluctuations in a \emph{single replica} of the system. Indeed, upon taking derivatives of Eq.~\eqref{eq:FCS} with respect to $\beta$ we find all the moments of the charge
\be
\partial_\beta^n  Z_{\beta}(A,t)|_{\beta=0} = \tr[ Q_A^n \rho_A(t)]\,.
\ee
To characterise more general ``multi-replica fluctuations" one can introduce a
richer family of observables dubbed \emph{charged
moments}~\cite{ares2022entanglement,ares2023lack,goldstein2018symmetry,xavier2018equipartition,parez2021quasiparticle}
\begin{equation} \label{eq:chargedmoments}
  \begin{gathered}
  Z_{\bm{\beta}}(A,t)= \tr[\prod_{j=1}^n 
 \left( e^{\beta_j  Q_A} \rho_A(t)\right) ],\\
    \bm{\beta}=\begin{bmatrix}\beta_1&\beta_2&\cdots&\beta_n\end{bmatrix},\qquad
      n\in\mathbb{N},\qquad \beta_j\in\mathbb{R},
  \end{gathered}
\end{equation}
where the product of non-commuting operators should be interpreted from left to right, i.e., 
\begin{equation}
  \prod_{j=1}^n  \left( e^{\beta_j  Q_A} \rho_A(t)\right)
  =  e^{\beta_1 Q_A} \rho_A(t)\cdots  e^{\beta_n  Q_A} \rho_A(t)\,.  
\end{equation}
Charged moments return information about the interplay between charge
fluctuations within the region $A$ and the entanglement between $A$ and the
rest of the system. They characterise regular R\'enyi entropies
\begin{equation}\label{eq:Renyi}
  S^{(n)}_A(t) = \frac{\log \tr[\rho_A^n(t)]}{1-n}=
  \frac{\log Z_{\bm{0}}(A,t)}{1-n}, 
\end{equation}
as well as R\'enyi entropies of the reduced density matrix projected to a given
charge sector~\cite{goldstein2018symmetry,xavier2018equipartition,parez2021quasiparticle}.
Indeed, defining  
\be
\rho_{A,q}(t) = \Pi_q \rho_A(t) \Pi_q,  
\ee
where $\Pi_q=\int_{-\pi}^\pi\frac{{\rm d}\beta}{2\pi}e^{i\beta (Q_A-q)}$ is the projector to the sector of charge $q\in\mathbb
Z$~\footnote{Here and in the following we assume that the $U(1)$ charge of interest has an integer spectrum, i.e., it is like a particle number.},
we find 
\begin{align}
S^{(n)}_{A,q}(t) &= \frac{\log \tr[\rho_{A,q}^n(t)]}{1-n} \notag\\ 
&=  \frac{1}{1-n}\log \prod_{j=1}^n \int_{-\pi}^{\pi} \frac{{\rm d} \beta_j}{2\pi} Z_{i \bm{\beta}}(A,t) e^{-i q \sum_j \beta_j }.
\label{eq:Renyiq}
\end{align}
Here $i \bm{\beta}$ denotes the vector $\bm{\beta}$ multiplied by the
imaginary unit. To lighten notation $n$ is not explicitly reported in the r.h.s.\ of Eq.~\eqref{eq:Renyi}: the subscript in $Z_{\mathbf{0}}$ refers to the $n$-dimensional vector $\mathbf{0}=[0,0,\ldots,0]$. 

Importantly, the R\'enyi entropies \eqref{eq:Renyiq} characterise the symmetry-resolved entanglement only when the reduced density matrix is block diagonal with respect to the charge in the subsystem, i.e., when $\ket{\Psi_0}$ is an eigenstate of the charge. Whenever this is not the case, the charged moments \eqref{eq:chargedmoments} can be used to investigate the interplay between the breaking of the symmetry in the initial state and how this evolves in time. Indeed, they specify the so-called \emph{entanglement asymmetry}~\cite{ares2022entanglement, ares2023lack}. The latter, typically denoted by $\Delta S_A(t)$, is defined as  a relative entropy between two related reduced density matrices, i.e., 
\begin{align}
\Delta S_A(t)&= \tr[\rho_A(t)(\log \rho_A(t)-\log\rhoAQ(t))] \notag\\
& =S(\rhoAQ(t))-S(\rho_A(t)). 
\label{eq:asymm}
\end{align}
Here ${S(\rho)=-\tr[\rho \log \rho]}$ is the von-Neumann entropy, $\rho_A(t)$ is the reduced density matrix of the subsystem $A$, while 
\begin{eqnarray}
\rhoAQ(t)=\sum_q\Pi_q\rho_A(t)\Pi_q
\end{eqnarray}
is this density matrix projected to a block diagonal form. From its definition as a relative entropy one can immediately observe that $\Delta S_A(t)\geq 0$ (see e.g., Ref.~\cite{nielsen2010quantum}) with equality attained only when $\rho_A(t)$ commutes with ${Q}_A$. Thus, the asymmetry quantifies how much the symmetry is broken by the initial state and allows one to determine if and on what time scales it is restored at large times. There are other ways in which one could quantify such information, for example one might use the distance between reduced density matrices. The entanglement asymmetry, however, packages this information in a very convenient and accessible fashion. Indeed, $\Delta S_A(t)$ has already been used to illuminate a number of exotic phenomena including counter intuitive relaxation dynamics known as the Mpemba effect~\cite{ares2022entanglement} and the lack of symmetry restoration in spin chains~\cite{ares2023lack}.

Eq.~\eqref{eq:asymm} can be expressed in terms of the charged moments via a replica trick
\be
\label{eq:asymmetryrep}
  \!\!\!\Delta S_A(t)=\lim_{n\to1}\frac{1}{1-n}\left[\log{\text{tr}[\rhoAQ^n(t)]}\!-\!\log{\text{tr}[\rho^n_{A}(t)]}\right]\!.
\ee
For $\mathbb{N} \ni n \geq 2$ the second term on the r.h.s.\ is written in terms of the charged moments using Eq.~\eqref{eq:Renyi}, while a simple calculation reveals that the first is given by 
\be
\label{eq:trrhobarn}
 \tr[\rhoAQ^n(t)]= \int_{-\pi}^\pi\frac{{\rm d}\boldsymbol{\beta}}{(2\pi)^{n-1}}Z_{i\boldsymbol{\beta}}(A,t)\delta_p(\sum_{j=1}^n\beta_j),
\ee
where $\delta_p(x)\equiv\sum_{q\in\mathbb{Z}}e^{-i q x}/2\pi$ is the $2\pi$-periodic delta function.  Computing these functions, continuing the replica index $n$ to real values, and
taking ${n\to 1}$ then gives the entanglement asymmetry.

\subsection{Summary of main results}
\label{subsec:mainresults}

Throughout this paper we shall derive several new results on the fluctuations of $U(1)$ conserved charges in far from equilibrium quantum systems. These range from universal properties for generic systems to specific predictions for integrable models as well as applications of these to the quantities of interest discussed above. In this section we briefly present and discuss some of the most significant of these results with the details of their derivation presented in the succeeding sections and appendices.

Our results are obtained building on the space-time duality approach of Refs.~\cite{bertini2023nonequilibrium, bertini2022growth}. The latter is based on the observation, put forward in Ref.~\cite{bertini2022growth}, that during the evolution of a quantum state there exists a regime --- referred to as the ``nonequilibrium regime'' --- where generalised purities (i.e., traces of integer powers $\rho_A(t)$) can be mapped onto generalised purities of a dual system that is instead at equilibrium. Specifically, Ref.~\cite{bertini2022growth} showed that for ${1\ll t\ll |A|}$ we have 
\begin{equation}\label{eq:PuritiesPreview}
  \tr\mkern-0mu[ {\rho}_{A}^n(t)]\!\simeq \! 
  \tr\mkern-0mu[\tilde{\rho}^n_{{\rm st},t}]^2,\qquad n\in \mathbb N\,,
  \mkern-6mu
\end{equation}
where $\tilde{\rho}_{{\rm st},t}$ is loosely speaking the stationary state of the dual system and $\simeq$ means that the equality holds at the leading order. More precisely, the dual system arises from exchanging the roles of space and time in the original system and its dynamics are not generically unitary. Instead, they are determined by a quantum channel whose boundary action is set by the initial state of the time evolution. This means that generically its evolution has two different stationary states, or fixed points, a left and a right one: $\tilde{\rho}_{{\rm st},t}$ is the (normalised) product of them. The second power on the r.h.s.\ of Eq.~\eqref{eq:PuritiesPreview} comes from the fact that one gets two equivalent contributions from each of the boundaries between $A$ and $\bar A$ (two in our setting). If one considers open boundary conditions and $A$ starting from the edge, there is only one boundary between $A$ and $\bar A$ and the power of 2 does not appear. 

Ref.~\cite{bertini2023nonequilibrium} widened the scope of this observation by showing that the same conclusion applies for the FCS in systems with conserved $U(1)$ charges evolving from a state with no charge fluctuations inside $A$. Namely, for ${1\ll t\ll |A|}$ one has 
\begin{equation}\label{eq:FCSsymPreview}
 \!\!\!\!{Z_{\beta}(A,t)}\!=\! 
  \tr\mkern-0mu[e^{\beta Q_A} \rho_A(t)]\!\simeq\! 
  \tr\mkern-0mu[e^{\beta \tilde{Q}_t}\tilde{\rho}_{{\rm st},t}]
  \tr\mkern-0mu[e^{-\beta \tilde{Q}_t}\tilde{\rho}_{{\rm st},t}],
  \mkern-6mu
\end{equation}
where $\tilde{Q}_t$ is the conserved $U(1)$ charge of the dual system (it is always present if the original system is $U(1)$ invariant). Once again, each of the terms on the r.h.s.\ is the contribution of one boundary between $A$ and $\bar A$. The result continues to hold for $\rho_A(t)\mapsto \rho_A^n(t)$.

Here we consider the more challenging case of systems with a $U(1)$ charge evolving from a state that is \emph{not} an eigenstate of the charge. Remarkably, we find that, with appropriate modifications, an equation similar to Eq.~\eqref{eq:FCSsymPreview} applies also in this case. Specifically, for $1\ll t\ll |A|$ we obtain 
\begin{equation}\label{eq:FCSasymptPreview}
 \frac{Z_{\beta}(A,t)}{Z_{\beta}(A,0)}\!\simeq\! 
  \tr\mkern-0mu[e^{\beta \tilde{Q}_t}\tilde{\rho}_{{\rm st},t}(\beta,0)]
  \tr\mkern-0mu[e^{-\beta \tilde{Q}_t}\tilde{\rho}_{{\rm st},t}(0,\beta)],
  \mkern-6mu
\end{equation}
together with the appropriate generalisation for higher charged moments. Here $\tilde{\rho}_{{\rm st},t}(\beta_1,\beta_2)$ is again the product of left and right fixed points of the space-evolving quantum channel, however, in this case the channels have twisted boundary conditions parameterised by $\beta_1$ and $
\beta_2$. For instance $\tilde{\rho}_{{\rm st},t}(0,\beta)$ is obtained by multiplying the left fixed point of the space-evolution with no twist and the right fixed point of the space evolution with twist $\beta$. Interestingly, and this is our second main result, we find that $\tilde{\rho}_{{\rm st},t}(\beta_1,\beta_2)$ can be characterised by solving a standard \emph{bipartitioning protocol}, i.e., the quantum quench problem where the two halves of the system are prepared in different homogeneous states and, from $t=0$, the whole system is let to evolve with a homogeneous Hamiltonian~\cite{vasseur2016nonequilibrium, bernard2016conformal, bertini2021finitetemperature}. This special kind of quench provides a controlled model for inhomogeneous settings and has been studied intensely over the last few years~\cite{bertini2016transport,castroalvaredo2016emergent, bastianello2022introduction, alba2021generalized, bertini2021finitetemperature}. 

More precisely, we argue  
\be
  \tr\mkern-0mu[e^{\beta \tilde{Q}_t}\tilde{\rho}_{{\rm st},t}(\beta_1,\beta_2)] \overset{ t \leftrightarrow x}{\longleftrightarrow}   \tr\mkern-0mu[e^{\beta {Q}_A} {\rho}_{{\rm st},A}(\beta_1,\beta_2)].
\label{eq:connectionPreview}
\ee
Here $\overset{ t \leftrightarrow x}{\longleftrightarrow}$ denotes a space-time swap, i.e., an exchange of space and time, and ${\rho}_{{\rm st}}(\beta_1,\beta_2)$ is the stationary state reached by the region around $x=0$ after a bipartitioning protocol from the initial state 
\be
{\rho}_{0}(\beta_1,\beta_2)=e^{\beta_1 Q_L} \rho_{\rm st, L} \otimes e^{\beta_2 Q_R} \rho_{\rm st, R},
\ee
where $R/L$ denote quantities reduced to the left/right half of the system and $\rho_{\rm st}$ is the stationary state describing local observables after a quench from $\ket{\Psi_0}$. Combined together, our results establish an interesting connection between charge fluctuations from charge-asymmetric but homogeneous initial states and bipartitioning protocols. 

Similarly to what shown in Refs.~\cite{bertini2023nonequilibrium, bertini2022growth} for the cases of Eqs.~\eqref{eq:PuritiesPreview} and \eqref{eq:FCSsymPreview}, Eq.~\eqref{eq:FCSasymptPreview} can be used to determine some general properties of charge fluctuations by making physical assumptions on $\tilde{\rho}_{{\rm st},t}(\beta_1,\beta_2)$. For example, if this state has an extensive cumulant generating function (the logarithm of the FCS) then Eq.~\eqref{eq:FCSasymptPreview} implies that the FCS of the original system decays exponentially in time in the non-equilibrium regime. In cases where the r.h.s.\ of  Eq.~\eqref{eq:connectionPreview} can actually be computed, however, our results lead to explicit predictions for charge fluctuations in the nonequilibrium regime. We emphasise that this is a very powerful statement as the r.h.s.\ of  Eq.~\eqref{eq:connectionPreview} is entirely written in terms of equilibrium quantities. 

As a non-trivial example where the latter strategy can be successfully applied we consider \emph{interacting integrable models}, wherein the charged moments in the stationary states of the time evolution can be be calculated using the method of thermodynamic Bethe ansatz (TBA), see Sec. (\ref{sec:TBA} for a full review). In these systems the space-time swap can be conveniently performed in Fourier space, i.e., by exchanging the roles of energy and momentum of their quasiparticles. Moreover, ${\rho}_{{\rm st}}(\beta_1,\beta_2)$ can be characterised using the techniques of generalised hydrodynamics (GHD)~\cite{bertini2016transport,castroalvaredo2016emergent}. As a result, for a generic integrable model, with $M$ quasiparticle species labelled by a species index $m$ and rapidity $\lambda$ we find
\begin{equation} \label{eq:swapdeq2Preview}
    \mkern-3mu
    \lim_{t\to\infty}\mkern-2mu\frac{1}{t}\mkern-2mu\log\mkern-2mu\tr[{\tilde{ \rho}}_{{\rm st},t}(\beta,0) e^{\beta {\tilde Q}_t}]\mkern-4mu =\mkern-8mu \int\limits _0^\beta \mkern-6mu \mathrm{d}u
    \mkern-6mu
    \smashoperator{\sum_{m}}
    \mkern-6mu \int\mkern-8mu \mathrm{d}\lambda\, q_m \tilde{\rho}^{(u)}_m(\lambda),
    \mkern-3mu
\end{equation}
where $q_m$ is the bare charge associated to each quasiparticle and $\tilde{\rho}^{(u)}_m(\lambda)$ is the distribution of occupied quasiparticles of species $m$ in the stationary state of the space evolution corresponding to the value of $\beta=u$.  This latter function can be determined exactly in terms of a set of TBA equations and combining this with~\eqref{eq:FCSasymptPreview} gives an exact expression for the FCS in the nonequilibrium regime. In Sec.~\ref{sec:tests} this prediction is tested against independent analytical derivations in the case of non-interacting systems and the quantum cellular automaton Rule 54, and against tensor-network based numerical simulations in the XXZ model. 
 
Eq.~\eqref{eq:swapdeq2Preview} makes quantitative the aforementioned connection between charge fluctuations from asymmetric but homogeneous initial states and bipartitioning protocols. Once again, similar expressions can be written for the higher R\'enyi charged moments. Although they are more complex the key ingredient in their derivation is the use of the space-time swapped stationary state obtained from the GHD solution of an inhomogenous quench.

These observations on the form of the charged moments and the explicit formulae in the case of TBA integrable models can then be used to understand the behavior of the physical observables of interest.  In particular,  the probability distribution for measuring a charge $q$, different from the expectation value, $q=\left<Q_A\right>+\Delta q$, inside $A$ at time $t\ll|A|$  is given by the Fourier transform of the FCS,  $Z_{i\beta}(A,t)$.  This can be computed using a saddle point approximation provided $\Delta q\ll |A|$ with the result
\begin{eqnarray}\label{eq:chargeprobpreview}
P(\Delta q,t)\simeq\frac{1}{\sqrt{2\pi\mathcal{D}(t) }}e^{-\frac{\Delta q^2}{2\mathcal{D}(t)}},
\end{eqnarray}
where 
\begin{eqnarray}
\!\!\!\mathcal{D}(t)\!=\!2\sum_m\!\int\!\!{\rm d}\lambda\, \!\left(|A|-t|v_{m}(\lambda)|\right)\mathcal{X}_m(\lambda)\!,
\end{eqnarray}
and $\mathcal{X}_m(\lambda)$, $v_m(\lambda)$ are  the charge susceptibility  and velocity of a quasiparticle of species $m$ with rapidity $\lambda$ each of which can be explicitly determined.  This expression provides a transparent physical interpretation of the evolution of charge probability distribution: it is determined by the ballistic propagation of pairs of quasiparticles throughout the system which transport and disperse charge fluctuations, encoded in $\mathcal{X}_m(\lambda)$ as they propagate. 

Similarly,  we obtain an explicit formula for the entanglement asymmetry in the nonequilibrium regime 
\begin{equation}
  \begin{aligned}
    \Delta S_A(t)&=\frac{1}{2}+\frac{1}{2}\log \pi \chi(t),\\
    \chi(t)&=\sum_m\!\int\!\!{\rm d}\lambda\, \!\left(|A|-2t|v_{m}(\lambda)|\right)\mathcal{X}_m(\lambda).
  \end{aligned}
\end{equation}
The similarity between the expressions for $\mathcal{D}(t)$ and $\chi(t)$ thus allows one assign some physical intuition to the highly complicated $\Delta S_A(t)$ based upon the more readily understandable charge probability distribution. Accordingly, the interplay between the restoration of the broken symmetry and the spreading of entanglement can be studied in detail.


Having presented our main results and some of their applications we now turn to their derivation in the remainder of the paper which is laid out as follows.  In Sec.~\ref{sec:duality}, we explain the main ideas of the space-time duality approach: First, in Sec.~\ref{sec:circuits}, we introduce a class of many-body systems in discrete space-time --- brickwork quantum circuits~\cite{nahum2017quantum,chan2018solution,chan2018spectral,fisher2023random} ---
where the space-time duality is most easily implemented. Then, in Sec.~\ref{sec:discreteduality1}, we illustrate the space-time duality approach in brickwork quantum circuits. Finally, in Sec.~\ref{sec:trotter}, we argue that our results can directly be extended to locally interacting systems in continuous time. In Sec.~\ref{sec:integrablecircuits} we specialise the treatment to integrable systems and derive closed-form expressions for the asymptotic dynamics of full counting statistics (FCS) in the language of TBA, which are extensively tested in Sec.~\ref{sec:tests}. In Sec.~\ref{sec:higherrenyi} we use our results to produce explicit predictions for the entanglement asymmetry and charge probability distribution in interacting integrable systems and discuss their key physical features. Finally, Sec.~\ref{sec:conclusions} contains our conclusions.

\section{Space-time duality approach to charge fluctuations}
\label{sec:duality}

Having introduced the necessary concepts in the previous section, we are now in a position to explain how to characterise the charge fluctuations by means of the space-time duality. 

In essence, the approach is based on two observations 
\begin{enumerate}[label=(\roman*)]
  \item \label{it:twoObs1} For large $t$, the charge moments take two different
    asymptotic forms depending on whether or not $t$ is larger than the size of
    $A$ (both are taken to be large compared to microscopic scales). 
  \item \label{it:twoObs2} These two forms are mapped into each other upon
    performing a formal swap of space and time.  
\end{enumerate}
This means that if one can access one of the two regimes analytically, then
they can use~\ref{it:twoObs2} to access the other. As we discuss in
Sec.~\ref{sec:integrablecircuits}, this is the case for \emph{integrable}
systems. 

For the sake of clarity we proceed to illustrate these two observations focussing on the FCS~\eqref{eq:FCS} in a class of systems where space and time are treated on equal footings. Namely, we consider the so-called~\emph{brickwork
quantum} circuits~\cite{nahum2017quantum,chan2018solution,chan2018spectral,fisher2023random}, where
interactions are instantaneous in time and local in space, and where space and
time are both discrete. We then generalise the treatment to generic charged moments and argue that the same ideas continue to apply for systems in continuous time. Before proceeding, however, we provide a brief self contained introduction to brickwork quantum circuits.

\subsection{Brickwork Quantum Circuits}
\label{sec:circuits}

Brickwork quantum circuits are systems of $2L$ qudits, i.e., quantum systems with $d\geq 2$ internal states, where the time evolution is generated by the unitary operator  
\be
\label{eq:U}
{\mathbb U} = {\mathbb U}_{\rm e} {\mathbb U}_{\rm o},\quad {\mathbb U}_{\rm e}= {U}^{\otimes L}, \quad  {\mathbb U}_{\rm o}=  \Pi_{2L} U^{\otimes L} \Pi^{\dag}_{2L}\,. 
\ee
Here the ``local gate'' $U$ acts on two (neighbouring) qudits and $ \Pi_{x}$ is a periodic one-site shift in a lattice of $x$ sites, and, for simplicity, we have assumed $\mathbb U$ to be time-independent and invariant under two-site shifts. We emphasise that \eqref{eq:U} generates \emph{strictly causal} dynamics: there is a strict maximal speed for the propagation of information.

The physical properties of the time evolution are entirely determined by the local gate and, by varying it, one can observe a very rich spectrum of dynamical behaviours~\cite{nahum2017quantum, vonKeyserlingk2018operator, bertini2019entanglement, gopalakrishnan2019unitary, piroli2020exact, bertini2020operator, bertini2020operator2, claeys2020maximum, bertini2020scrambling, jonay2021triunitary, chan2018solution, khemani2018operator, rakovszky2018diffusive, zhou2020entanglement, foligno2023temporal, claeys2020maximum, reid2021entanglement, wang2019barrier, piroli2020exact, claeys2021ergodic} and spectral correlations~\cite{friedman2019spectral, bertini2018exact,chan2018spectral, flack2020statistics, bertini2021random, fritzsch2021eigenstate, kos2021thermalization, bertini2022exact, garratt2021manybody, garratt2021local}. In particular, for our purposes it is important to stress that there exist choices of $U$ making the quantum circuit Yang-Baxter integrable~\cite{faddeev1996how, vanicat2018integrable, ljubotina2022ballistic, gombor2021integrable, gombor2022superintegrable, aleiner2021bethe, claeys2022correlations, miao2024floquet} and treatable via thermodynamic Bethe ansatz~\cite{vernier2023integrable}. In fact, one can define an integrable brickwork quantum circuit corresponding to each fundamental spin model with Hamiltonian of range 2 (see, e.g., Sec. 11 in Ref.~\cite{faddeev1996how}). 

Besides their inherent importance, brickwork quantum circuits are also used as
computationally efficient approximations of locally interacting systems in
continuous time~\cite{trotter1959product,suzuki1991general} --- both in the
context of classical~\cite{schollwock2011density} and
quantum~\cite{feynman1982simulating,georgescu2014quantum,arute2019quantum}
simulation. Indeed, considering local gates of the form
\begin{equation} \label{eq:FTapprox}
  U = e^{-i \tau h},
\end{equation}
where $h=h^\dag$ is some $d^2\times d^2$ Hermitian matrix  acting on two sites. Performing the so-called Trotter limit 
\begin{equation}\label{eq:trotterlimit}
  \lim_{\rm Tr}:\quad\tau\to0, \quad t\to\infty, \quad \tau t = \mathfrak t = {\rm fixed}, 
\end{equation}
one has 
\begin{equation}
  \lim_{\rm Tr} {\mathbb U}^t =
  \mathrm{exp}\Big[{-i \mathfrak t\smashoperator{\sum_{x \in \mathbb Z_{L}/2}}  h_x}\Big]\,,
\end{equation}
where we labelled sites by half-integer numbers from $-L$ to $L$ and we
introduced the operator $h_x$ acting as the matrix $h$ at sites $x$ and
$x+1/2$ and as the identity elsewhere. In this limit the evolution of a quantum
circuit reproduces the one generated by the Hamiltonian 
\begin{equation}
  H = \smashoperator{\sum_{x \in \mathbb Z_{L}/2}}  h_x,
\end{equation}
up to time $\mathfrak t$. In Sec.~\ref{sec:trotter} we argue that considering
this limit one can apply our results to continuous time. 

\begin{figure}
  \includegraphics[width=\columnwidth]{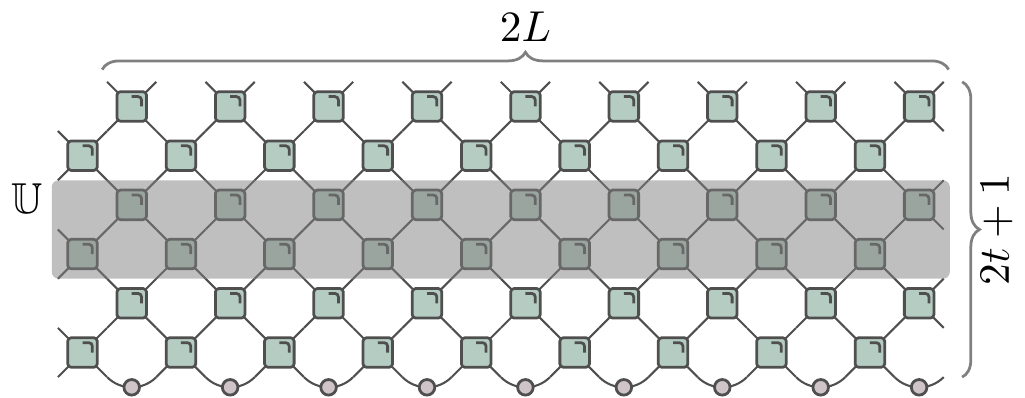}
  \caption{\label{fig:DiagTE} Diagrammatic representation of the time-evolved state $\ket{\Psi_t}$, with $t=3$, and $L=9$. The grey box denotes the time-evolution operator $\mathbb{U}$ defined in~\eqref{eq:U}, which is repeatedly applied on $\ket{\Psi_0}$ to give $\ket{\Psi_t}$. The initial state $\ket{\Psi_0}$ is assumed to be expressed as a product of pairs (cf.\ \eqref{eq:initialState}). We assume periodic boundary conditions in space, which we will for simplicity not explicitly represent graphically.}
\end{figure}

After being prepared in the state $\ket{\Psi_0}$ at time $t=0$ the state of the system at a (discrete) time $t$ is given by 
\begin{equation} \label{eq:statetimet}
\ket{\Psi_t} = {\mathbb U}^t \ket{\Psi_0}.
\end{equation}
We introduce the diagrammatic representation
\begin{equation} \label{eq:localU}
 U = \begin{tikzpicture}[baseline={([yshift=-0.6ex]current bounding box.center)},scale=0.65]
    \prop{0}{0}{colU}
  \end{tikzpicture},\qquad
  U^{\ast} =
  \begin{tikzpicture}[baseline={([yshift=-0.6ex]current bounding box.center)},scale=0.65]
    \prop{0}{0}{colUc}
  \end{tikzpicture},
\end{equation}
where different legs act on different spatial sites, and the matrix elements of $U$ are given as
\begin{equation} \label{eq:localUME}
  \mel{c,d}{U}{a,b} 
  = \begin{tikzpicture}[baseline={([yshift=-0.6ex]current bounding box.center)},scale=0.65]
    \prop{0}{0}{colU}
   \node at (-0.65,-0.65) {$a$};
   \node at (0.65,-0.65) {$b$};
   \node at (0.65,0.65) {$d$};
   \node at (-0.65,0.65) {$c$};
  \end{tikzpicture}.
\end{equation}
Transposition is given by flipping the gate upside down, i.e.,
\begin{equation} \label{eq:localUtranspose}
  U^{T} =
  \begin{tikzpicture}[baseline={([yshift=-0.6ex]current bounding box.center)},scale=0.65]
    \propT{0}{0}{colU}
  \end{tikzpicture},\qquad
  U^{\dagger} =
  \begin{tikzpicture}[baseline={([yshift=-0.6ex]current bounding box.center)},scale=0.65]
    \propT{0}{0}{colUc}
  \end{tikzpicture}.
\end{equation}
The matrix multiplication is represented by joining legs and goes from bottom to top, which, for example, gives the following diagrammatic representation for the unitarity condition
\begin{equation}
\label{eq:unitarity}
  U U^{\dagger}=U^{\dagger} U = \1 \otimes \1,\qquad
  \begin{tikzpicture}[baseline={([yshift=-0.6ex]current bounding box.center)},scale=0.65]
    \draw[semithick,colLines] (-0.5,0.25) arc (-45:45:-0.5/\sqrtTwo);
    \draw[semithick,colLines] (0.5,0.25) arc (45:-45:0.5/\sqrtTwo);
    \propT{0}{0.75}{colUc}
    \prop{0}{-0.75}{colU}
  \end{tikzpicture}
  =
  \begin{tikzpicture}[baseline={([yshift=-0.6ex]current bounding box.center)},scale=0.65]
    \draw[semithick,colLines] (-0.5,0.25) arc (-45:45:-0.5/\sqrtTwo);
    \draw[semithick,colLines] (0.5,0.25) arc (45:-45:0.5/\sqrtTwo);
    \prop{0}{0.75}{colU}
    \propT{0}{-0.75}{colUc}
  \end{tikzpicture}
  =
  \begin{tikzpicture}[baseline={([yshift=-0.6ex]current bounding box.center)},scale=0.65]
    \tgridLine{0.5}{-1.25}{0.5}{1.25}
    \tgridLine{0}{-1.25}{0}{1.25}
  \end{tikzpicture}\,,
\end{equation}
where the free horizontal legs represent the identity operator $\1\in\mathbb{C}^d$. Using these conventions, Eq.~\eqref{eq:statetimet} can be depicted as in Fig.~\ref{fig:DiagTE}. In the figure we conveniently considered dimer-product initial states of the form  
\begin{equation}\label{eq:initialState}
  \ket{\Psi_0}= \ket{\psi_0}^{\otimes L},\qquad \ket{\psi_0}=\sum_{i,j=1}^d m_{ij} \ket{i,j},
\end{equation}
where $\{\ket{i}\}$ is a basis of the Hilbert space of a single qudit and $m$ is an arbitrary $d\times d$ matrix fulfilling ${\rm tr}[mm^\dag]=1$ to ensure normalisation. The single two-site state $\ket{\psi_0}$ can be represented graphically as  
\be
\ket{\psi_0} =    
\begin{tikzpicture}[baseline={([yshift=-0.6ex]current bounding box.center)},scale=0.5]
  \tgridLine{0}{0}{-0.25}{0.25}
  \tgridLine{1}{0}{1.25}{0.25}
  \istate{0}{0}{colIst}
\end{tikzpicture}.
\ee
These states are particularly convenient for our purposes, as they have low entanglement and their physical properties are controlled by a single small matrix $m$, therefore we will from now on consider initial states of this form. In particular, in our treatment of Sec.~\ref{sec:integrablecircuits} we will eventually restrict ourselves to a subset of possible $m$, which generates the so-called \emph{integrable} initial states~\cite{piroli2017what,piroli2019integrableI,piroli2019integrableII,pozsgay2019integrable,rylands2022integrable,rylands2022solution} (see Sec.~\ref{sec:integrablecircuits}).

To study the dynamics of the charged moments, we focus on circuits with $U(1)$
charges of the form 
\begin{equation}
  { Q}= \smashoperator{\sum_{x\in \mathbb Z_L/2}}  q_{x},
\end{equation}
where the operator $ q_x$ acts as the $d\times d$ matrix $q$ at site $x$ and as
the identity elsewhere. Without loss of generality we can take $q$ to be
traceless. 

Because of the ultralocal nature of the charge and the strict causal structure
of the time evolution, in a quantum circuit the conservation of ${Q}$ is
implemented locally. Namely, as we show in Appendix~\ref{sec:proofOfCC}, the conservation of charge implies that
$q$ and $U$ satisfy the relation 
\begin{equation} \label{eq:chargecons}
  (e^{\beta  q}\otimes e^{\beta  q})  U =  U (e^{\beta  q}\otimes e^{\beta  q})\,, \qquad \forall \beta\in\mathbb R\,,
\end{equation}
which can be represented diagrammatically as 
\begin{equation} \label{eq:localrelation}
\begin{tikzpicture}[baseline={([yshift=-0.6ex]current bounding box.center)},scale=0.5]
  \tgridLine{0}{0+6}{0+0.75}{0+6-0.75}
  \tgridLine{0}{0+6}{0-0.75}{0+6-0.75}
  \tgridLine{0}{0+6}{0+0.75}{0+6+0.75}
  \tgridLine{0}{0+6}{0-0.75}{0+6+0.75}
  \prop{0}{0+6}{colU}
  \obs{-0.5}{6.5}{myred}
  \obs{0.5}{6.5}{myred}
\end{tikzpicture}
=
\begin{tikzpicture}[baseline={([yshift=-0.6ex]current bounding box.center)},scale=0.5]
  \tgridLine{0+3}{0+6}{0+3+0.75}{0+6-0.75}
  \tgridLine{0+3}{0+6}{0+3-0.75}{0+6-0.75}
  \tgridLine{0+3}{0+6}{0+3+0.75}{0+6+0.75}
  \tgridLine{0+3}{0+6}{0+3-0.75}{0+6+0.75}
  \prop{0+3}{0+6}{colU}
  \obs{2.5}{5.5}{myred}
  \obs{3.5}{5.5}{myred}
\end{tikzpicture},
\qquad e^{\beta q} = 
\begin{tikzpicture}[baseline={([yshift=-0.6ex]current bounding box.center)},scale=0.5]
  \tgridLine{0}{-0.5}{0}{0.5}
  \obs{0}{0}{myred}
\end{tikzpicture}.
\end{equation}
As shown in Appendix~\ref{sec:current}, this relation can be used to find the
following explicit expression for the current associated to the charge~${Q}$
\begin{equation} \label{eq:appcurrent}
j_x(t)=\begin{cases}
  \phantom{-} q_x(t) & x+t\in\mathbb Z\\
  - q_x(t) & x+t\in\mathbb Z+\frac{1}{2}
\end{cases}\,,
\end{equation}
where we adopted the Heisenberg picture 
\begin{align}
  &  O(t) ={\mathbb U}^{-t}  O {\mathbb U}^t,\quad  O(t+\tfrac{1}{2}) = {\mathbb U}^{-t} {\mathbb U}^{-1}_{\rm o}  O {\mathbb U}_{\rm o} {\mathbb U}^t\,. 
  \label{eq:Heisenberg}
\end{align}

\begin{figure*}
  \includegraphics[width=1.75\columnwidth]{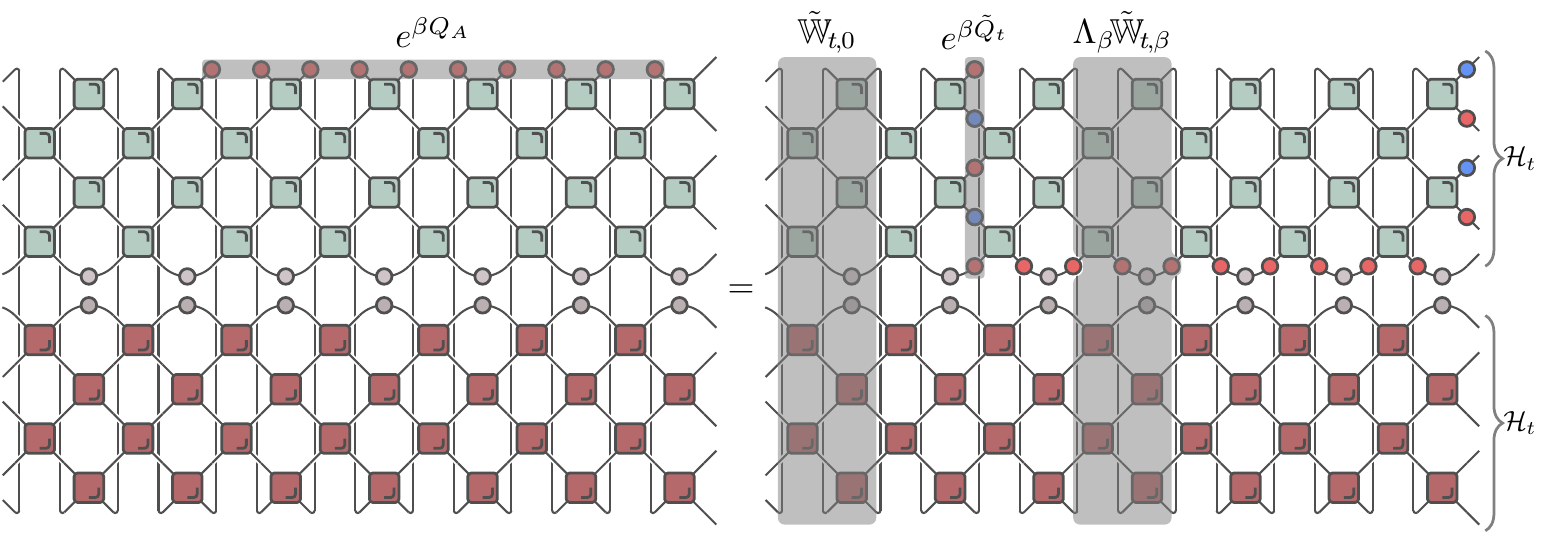}
  \caption{\label{fig:singleReplicaTN} Diagrammatic representation of $Z_{\beta}(A,t)$ for $t=2$. The diagram on the left follows directly from the definition~\eqref{eq:FCS}, by plugging in the state~\eqref{eq:statetimet} (see Fig.~\eqref{fig:DiagTE}). The r.h.s.\ is obtained by applying the conservation of $Q$ (cf.\ \eqref{eq:chargecons}, \eqref{eq:conservationQtilde}). The diagram on the right can be equivalently represented by contracting in space, as given in~\eqref{eq:WmatrixFormula}, with the transfer matrices $\tilde{\mathbb{W}}_{t,\beta}$, and $e^{\beta\tilde{Q}_t}$ highlighted in gray. The transfer matrices act on $\mathcal{H}_t\otimes\mathcal{H}_t$, with $\mathcal{H}_t$ being the Hilbert space of $2t+1$ qudits. Note that we are implicitly assuming periodic boundary conditions in space, which are not explicitly shown for clarity (i.e., the open legs on the left are connected to the open legs on the right).}
\end{figure*}

\subsection{Space-time duality in discrete time}
\label{sec:discreteduality1}
We begin to illustrate the space-time duality approach by noting that, because of the generic phenomenon of local
relaxation~\cite{PolkovnikovReview,calabrese2016introduction,VidmarRigol,
essler2016quench,doyon2020lecture,bastianello2022introduction,
alba2021generalized}, in the limit of infinite times and fixed $A$, the FCS in Eq.~\eqref{eq:FCS} becomes stationary. Namely
\begin{equation} \label{eq:FCSstat}
 \lim_{t\to\infty} Z_{\beta}(A,t)=
  \tr[ e^{\beta  Q_A} \rho_{{\rm st}, A} ],
\end{equation}
where $ \rho_{\rm st,A}$ is a stationary state of ${\mathbb U}$ that can
generically be expressed as a Generalised Gibbs Ensemble~\cite{VidmarRigol}.
This implies that FCS can be thought of as the ratio of two
partition functions and their logarithms are generically extensive in the size
of $A$. Therefore, we can capture their bulk features by considering the
following limit 
\begin{equation}
\label{eq:density1}
\begin{aligned}
d_{\beta}&:=\lim_{|A|\to\infty}\lim_{t\to\infty} 
  \frac{\log Z_{ \beta}(A,t)}{|A|}\\
  &\phantom{:}= \lim_{|A|\to\infty} \frac{1}{|A|}
  \log\tr[e^{\beta  Q_A} \rho_{{\rm st}, A} ],
\end{aligned}
\end{equation}
so that 
\begin{equation} \label{eq:Z1equilibrium}
  Z_{\beta}(A,t) \simeq e^{|A| d_{\beta}}, \qquad t\gg |A| \gg 1.
\end{equation}
Here we used that, by continuity, the limit \eqref{eq:density1} describes the leading order in the asymptotic regime. Here $|A|$ denotes the size of $A$, which we conveniently define as the number
of its sites divided by two. On the other hand, for $|A|\gg t \gg 1$, the FCS is
observed to decay exponentially in time, with a possible prefactor
$\Lambda_{\beta}^{|A|}$~\cite{parez2021quasiparticle,ares2022entanglement,bertini2022growth,bertini2023nonequilibrium} (to be specified later).
This behaviour can be captured by defining  
\begin{equation} \label{eq:slope1}
s_{\beta}:=\lim_{t\to\infty}\lim_{|A|\to\infty} 
  \frac{\log Z_{\beta}(A,t)- |A|\log \Lambda_{\beta}}{t},
\end{equation}
such that 
\begin{equation}\label{eq:Z1nonequilibrium}
  Z_{\beta}(A,t) \simeq \Lambda_{\beta}^{|A|} e^{t s_{\beta}}, \qquad 1 \ll t\ll |A|. 
\end{equation}
The asymptotic forms \eqref{eq:Z1equilibrium} and \eqref{eq:Z1nonequilibrium}
are those anticipated in~\ref{it:twoObs1} and, for obvious reasons, we refer to
the two regimes in which they hold as ``equilibrium'' and ``non-equilibrium'' respectively.

To explain~\ref{it:twoObs2} we now rewrite the rate \eqref{eq:slope1} in a form that is similar to the second line of~\eqref{eq:density1} but where the roles of space and time are swapped.  We begin by formulating the FCS in terms of the evolution in \emph{space}. This can be done by exploiting the fact that in quantum circuits the dynamics is discrete both in space and \emph{in time}, therefore we can straightforwardly express $Z_{\beta}(A,t)$ as a trace of powers of \emph{space}-transfer matrices acting column to column~\cite{banuls2009matrix,muellerhermes2012tensor,hastings2015connecting,bertini2018exact,bertini2019exact,bertini2022entanglement,bertini2022growth,ippoliti2021postselectionfree,ippoliti2021fractal}. The latter are given in terms of \emph{space-evolution} operator $\tilde{U}$, obtained from $U$ by a reshuffle of its indices as
\begin{equation} \label{eq:localUtilde}
  \mel{a,c}{\tilde{U}}{b,d}=
 \begin{tikzpicture}[baseline={([yshift=-0.6ex]current bounding box.center)},scale=0.65]
    \prop{0}{0}{colU}
    \node at (-0.65,-0.65) {$a$};
    \node at (0.65,-0.65) {$b$};
    \node at (-0.65,0.65) {$c$};
    \node at (0.65,0.65) {$d$};
  \end{tikzpicture}=
  \mel{c,d}{U}{a,b}.
\end{equation}
Note that a transpose of the gate $\tilde{U}$ is obtained by left-right reflection, i.e.,
\begin{equation} \label{eq:localUtildeTranspose}
  \mel{a,c}{\left.\tilde{U}\!\right.^T}{b,d}=
  \mel{b,d}{\tilde{U}}{a,c}=
 \begin{tikzpicture}[baseline={([yshift=-0.6ex]current bounding box.center)},scale=0.65]
    \propL{0}{0}{colU}
    \node at (-0.65,-0.65) {$a$};
    \node at (0.65,-0.65) {$b$};
    \node at (-0.65,0.65) {$c$};
    \node at (0.65,0.65) {$d$};
  \end{tikzpicture},
\end{equation}
and in general $\widetilde{U^T}$ does not coincide with $\left.\tilde{U}\!\right.^T$.

Transfer matrices $\tilde{\mathbb{W}}_{\beta}\in\mathrm{End}(\mathcal{H}_t\otimes \mathcal{H}_t)$ act on two copies of the ``temporal chain'' of $2t+1$ qudits, $\mathcal{H}_t=\left.\mathbb{C}^d\right.^{\otimes (2t+1)}$, and are given as
\begin{equation} \label{eq:WmatrixFormula}
  \begin{split}
    \tilde{\mathbb{W}}_{t,\beta}\!=\!
    \frac{1}{\Lambda_{\beta}}
    \!
    \left(
    \smashoperator[r]{\sum_{s_1,s_2=0}^{d-1}}
    \ketbra{s_1}{s_2}\otimes
    \tilde{U}^{\otimes t}\!\otimes
    \widetilde{U^{\dagger}}^{\otimes t}\!\otimes
    \ketbra{s_1}{s_2}
    \right)\\
    \cross
    \left(
    \tilde{U}^{\otimes t}\!\otimes
    (e^{\beta q^{T}} m e^{\beta q}) \otimes
    m^* \otimes
    \widetilde{U^{\dagger}}^{\otimes t}\!
    \right).
  \end{split}
\end{equation}
Here $m$ is the $d\times d$ matrix defining the initial state (cf.\ \eqref{eq:initialState}), and the normalization factor $\Lambda_{\beta}$ is the expectation value of $e^{\beta Q}$ in the initial state
\begin{equation}
  \Lambda_\beta = \expval{e^{\beta q} \otimes e^{\beta q}}{\psi_0}.
\end{equation}
See the r.h.s.\ of Fig.~\ref{fig:singleReplicaTN} for a diagrammatic representation. 

\begin{figure}
  \includegraphics[width=\columnwidth]{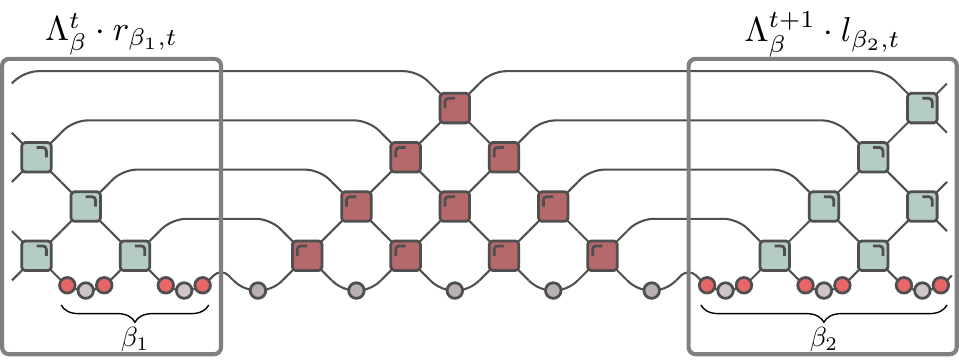}
  \caption{\label{fig:rhotilde}
  Diagrammatic representation of $\left.\rho_{\mathrm{st},t}(\beta_1,\beta_2)\right|_{t=2}$ (up to the scalar prefactor), as given by Eqs.~\eqref{eq:defRhot}, \eqref{eq:defRhot2}. The values of $\beta$ implied in the red circles (cf.\ \eqref{eq:localrelation}) are indicated by the underbraces. Note that $\beta=0$ gives the identity operator and therefore there are no red circles in the central part corresponding to $L_{0,t}$, $R_{0,t}$.
  }
\end{figure}

The conservation of $Q$, given by Eqs.~\eqref{eq:chargecons}, and~\eqref{eq:localrelation}, also implies
\begin{equation}\label{eq:conservationQtilde}
  (e^{\beta q}\otimes \1) U 
  (e^{-\beta q}\otimes \1) =
  (\1 \otimes e^{-\beta q}) U (\1 \otimes e^{\beta q}),
\end{equation}
or equivalently
\begin{equation}\label{eq:transposeMapping}
  (e^{-\beta q}\otimes e^{\beta q^{T}}) \tilde{U} =
  \tilde{U} (e^{\beta q^T}\otimes e^{-\beta q}),
\end{equation}
expressed diagrammatically as
\begin{equation} \label{eq:timelatticecons1}
  \begin{tikzpicture}[baseline={([yshift=-0.6ex]current bounding box.center)},scale=0.5]
    \tgridLine{0}{6}{0.75}{6-0.75}
    \tgridLine{0}{6}{-0.75}{6-0.75}
    \tgridLine{0}{6}{+0.75}{6+0.75}
    \tgridLine{0}{+6}{-0.75}{+6+0.75}
    \prop{0}{6}{colU}
    \draw [thick,rounded corners=1,colLines,fill=myred] (-.5,+6.5) circle (4.5pt);
    \draw [thick,rounded corners=1,colLines,fill=myblue] (-.5,+5.5) circle (4.5pt);
  \end{tikzpicture}
  =
  \begin{tikzpicture}[baseline={([yshift=-0.6ex]current bounding box.center)},scale=0.5]
    \tgridLine{3}{6}{3+0.75}{6-0.75}
    \tgridLine{3}{6}{3-0.75}{6-0.75}
    \tgridLine{3}{6}{3+0.75}{6+0.75}
    \tgridLine{3}{6}{3-0.75}{6+0.75}
    \prop{3}{6}{colU}
    \draw [thick,rounded corners=1,colLines,fill=myred] (3.5,5.5) circle (4.5pt);
    \draw [thick,rounded corners=1,colLines,fill=myblue] (3.5,6.5) circle (4.5pt);
  \end{tikzpicture},
  \qquad 
  e^{-\beta q} = 
  \begin{tikzpicture}[baseline={([yshift=-0.6ex]current bounding box.center)},scale=0.5]
    \tgridLine{0}{-0.5}{0}{0.5}
    \draw [thick,rounded corners=1,colLines,fill=myblue] (0,0) circle (4.5pt);
  \end{tikzpicture}.
\end{equation}
Repeatedly using this relation (together with~\eqref{eq:chargecons}), one can show that the FCS can be represented in terms of the transfer-matrix~\eqref{eq:WmatrixFormula} as
\begin{equation} \label{eq:defSpaceEvolution}
    Z_{\beta}(A,t) \!=\! \Lambda_{\beta}^{|A|}
    {\rm tr}[
      \tilde{\mathbb{W}}_{t,0}^{|\bar{A}|} (e^{\beta \tilde{Q}_t} \!\otimes\!\1)
    \tilde{\mathbb{W}}_{t,\beta}^{|A|} (e^{-\beta \tilde{Q}_t} \!\otimes\!\1)],
\end{equation}
which is depicted in the r.h.s.\ of Fig.~\ref{fig:singleReplicaTN}.
Here, $\tilde{Q}_t$ is the space-time swapped analogue of $Q$, i.e., the U(1) charge of the space evolution, and is given by
\begin{equation}
  \tilde{Q}_t=\sum_{\tau\in\mathbb{Z}_t}
  q_{\tau}-q^{T}_{\tau+\frac{1}{2}}.
  \label{eq:Qtilde}
\end{equation}
The form~\eqref{eq:defSpaceEvolution} is completely equivalent to the original expression and in general provides no immediate advantage. {However, as shown explicitly in Appendix~\ref{sec:proofEI}}, unitarity and locality imply that large powers of the transfer matrix factorise into a rank-one object, and therefore, whenever $|A|,|\bar{A}|>2t$, the FCS splits into the product of the contributions from the two edges between the subsystem $A$ and the rest
\begin{equation}\label{eq:FCSasymptFINAL}
  \mkern-6mu Z_{\beta}(A,t)\!=\!\Lambda_{\beta}^{|A|}\!
  \tr\mkern-0mu[e^{\beta \tilde{Q}_t}\tilde{\rho}_{{\rm st},t}(\beta,0)]
  \tr\mkern-0mu[e^{-\beta \tilde{Q}_t}\tilde{\rho}_{{\rm st},t}(0,\beta)].
  \mkern-6mu
\end{equation}
Here $\tilde{\rho}_{\mathrm{st},t}(\beta_1,\beta_2)$ is given as
(see Fig.~\ref{fig:rhotilde})
\begin{equation}\label{eq:defRhot}
  \tilde{\rho}_{\mathrm{st},t}(\beta_1,\beta_2)=  r^{\phantom{\dag}}_{t, \beta_1} r^{\dag}_{t, 0} l^\dag_{t, 0} l_{t, \beta_2}^{\phantom{\dagger}},
\end{equation}
with
\begin{equation}\label{eq:defRhot2}
  \mkern-12mu
  \begin{aligned}
    &r_{t, \beta}=
    \frac{1}{\Lambda_{\beta}^t}
  \prod_{\tau=0}^{t-1}
    \Big [\mkern-6mu
  \left(\1^{\otimes (2\tau+1)} \otimes \tilde{U}^{\otimes (t-\tau)}\right)\\
    &\times
  \left(\1^{\otimes (2\tau+2)} \otimes \tilde{U}^{\otimes (t-\tau-1)}
    \otimes (e^{\beta q^{T}}m e^{\beta q})
  \right)\mkern-6mu\Big ],\\
    &l_{t, \beta}=
    \frac{1}{\Lambda_{\beta}^{t+1}}
    \prod_{\tau=0}^{t-1}
    \Big [\mkern-6mu
    \left(\1^{\otimes(2t-2\tau)} \otimes \tilde U^{\otimes \tau}
    \otimes (e^{\beta q^{T}} m e^{\beta q})\right)\\
    &\times
    \left(\1^{\otimes(2t-2\tau-1)} \otimes \tilde U^{\otimes (\tau+1)}\right)
    \mkern-6mu\Big ]
  \left(\tilde U^{\otimes t} \otimes (e^{\beta q^{T}} m e^{\beta q})\right).
  \end{aligned}
  \mkern-12mu
\end{equation}
Note that here the terms in the product \emph{do not} commute, but rather they are assumed to be multiplied from left to right with increasing $\tau$.

Mathematically $\tilde{\rho}_{\mathrm{st},t}(\beta_1,\beta_2)$ is the product of the right fixed point of $ \tilde{\mathbb{W}}_{t,\beta_1}$ and the left fixed point of $\tilde{\mathbb{W}}_{t,\beta_2}$. From the physical point of view it can be understood as the space-evolution analogue of $\rho_{{\rm st}}$. Indeed, the latter is the fixed point (both left and right) of the time evolution. Moreover, by comparing with the expression for a stationary FCS given in Eq.~\eqref{eq:FCSstat} we see that each one of the traces in Eq.~\eqref{eq:FCSasymptFINAL} can be interpreted as the stationary FCS for the system of temporal lattice. Therefore Eq.~\eqref{eq:FCSasymptFINAL} can be interpreted as the statement that the FCS in the non-equilibrium regime can be written as the product of two stationary FCS for the system on the time lattice. The fact that we have the product of two of them is due to the fact that there are two boundaries between $A$ and $\bar A$. Consistently, in the case of a single boundary (e.g., for open boundary conditions and $A$ at the edge) one finds a single stationary FCS on the r.h.s.\ of Eq.~\eqref{eq:FCSasymptFINAL} (see, e.g., Sec.~5 in the Supplemental Material of Ref.~\cite{klobas2021exact}). Using this relation to connect non-equilibrium properties of the system with stationary properties of its space-time swapped counterpart is the main idea of the space-time duality approach. Due to its the instrumental role in this approach we refer to Eq.~\eqref{eq:FCSasymptFINAL} as the \emph{the fundamental identity of space time duality}.

Besides being conceptually intriguing, relating out-of-equilibrium properties to equilibrium ones is of great practical utility as the latter are much easier to study. In particular, this observation can be used in two different directions~\cite{bertini2022growth, bertini2023nonequilibrium}: (A) Invoke general properties of equilibrium states to infer qualitative features of the FCS in generic systems; (B) Find quantitative predictions whenever ${\tilde \rho}_{{\rm st},t}{(\beta_1,\beta_2)}$ can be accessed. 

As an example of (A), one can argue that the slope $s_{\beta}$ defined in Eq.~\eqref{eq:slope1} should generically strictly be smaller than zero, which follows from the extensivity and positivity of the free energy of equilibrium states. This implies that the FCS in non-equilibrium regime should in general decay exponentially
(even though there are known examples where the temporal free energy is sub-extensive and the slope vanishes~\cite{rakovszky2019sub, huang2020dynamics,znidaric2020entanglement}). As shown in Ref.~\cite{bertini2023nonequilibrium}, other examples are obtained by plugging the representation in Eq.~\eqref{eq:FCSasymptFINAL} back into Eq.~\eqref{eq:Renyiq} to infer general features of the symmetry resolved entanglement entropies such as the presence of a delay time for activation for symmetric initial states or the logarithmic growth of number entropy. 

Instead, to obtain the quantitative predictions (B) we proceed as follows. We identify a stationary state ${\rho}_{{\rm st},A}(\beta_1,\beta_2)$ of the system on the spatial lattice that corresponds to ${\tilde \rho}_{{\rm st},t}(\beta_1,\beta_2)$ upon swapping space and time, we compute its FCS analytically, and then exchange the roles of space and time to obtain an expression for the terms in the r.h.s.\ of Eq.~\eqref{eq:FCSasymptFINAL}. The agreement of the quantitative predictions obtained in this way with exact analytical and numerical results (cf.~Sec.~\ref{sec:tests}), constitutes the main justification for this approach. 

The procedure outlined requires two main ingredients: (1) an analytic expression for ${\rho}_{{\rm st},A}(\beta_1,\beta_2)$; (2) an analytic expression for its FCS. To secure (2) we consider interacting integrable models treatable by TBA. Indeed, as we review in Sec.~\ref{sec:integrablecircuits}, in these systems the FCS of any stationary state can be accessed analytically. Determining ${\rho}_{{\rm st},A}(\beta_1,\beta_2)$, instead, is a non-trivial task that so far has only been achieved in special cases~\cite{bertini2022growth, bertini2023nonequilibrium}. Here we solve this problem in the general case beginning from the following observation 
\begin{observation}
The expectation value on ${\tilde \rho}_{{\rm st},t}{(\beta_1,\beta_2)}$ of any product operator on the temporal lattice can be written as the expectation value of a time-ordered product over a deformed initial state. 
\end{observation}

More precisely, denoting a local operator at position $\tau$ of the time lattice as~\footnote{
  The transpose for half-integer times follows from the $\tilde{\cdot}$ mapping: the operators at integer and half-integer times act in opposite directions when seen in space. This is also the reason for $q^T$ in Eq.~\eqref{eq:transposeMapping}.}
\begin{equation}
  \tilde a_\tau  = \begin{cases}
    \1^{\otimes 2\tau} \otimes a \otimes \1^{\otimes 2t-2\tau} , &\tau \in\mathbb Z,\\
    \1^{\otimes 2\tau} \otimes a^T \otimes \1^{\otimes 2t-2\tau}, &\tau\in\mathbb Z+\tfrac{1}{2},
  \end{cases}
\end{equation}
we can rewrite an expectation value of a string of operators $\tilde{a}^{(j)}$ on the time lattice as
\begin{equation}
  \label{eq:coefficients}
  \begin{aligned}
    &\tr[{\tilde \rho}_{{\rm st},t}{(\beta_1,\beta_2)} \tilde a^{(0)}_0
    \cdots \tilde a^{({2t})}_t] \\
    &= \frac{
      \expval{a^{(0)}_0(0)  \cdots a^{({2t})}_0({t})e^{\beta_1 Q_{L}+ \beta_2 Q_{R}}}{\Psi_0}
    }{\expval{e^{\beta_1 Q_{L}+ \beta_2 Q_{R}}}{\Psi_0}},
  \end{aligned}
\end{equation}
where the two sides of the equation are the restatement of the same expectation value as viewed in terms of space- and time-evolution respectively.  Here $a^{({j})}_x(t)$ are operators at position $x$ of the space lattice evolved in time according to the Heisenberg picture (cf.~Eq.~\eqref{eq:Heisenberg}), superscripts $(j)$ label different operators, and we introduced U(1) charges on the left and right half chains
\begin{equation}
  \begin{gathered}
    Q_{L} =\smashoperator{\sum_{x\in\mathbb Z_{L/2}/2}} q_{-x-1/2}, \qquad
    Q_{R}=\smashoperator{\sum_{x\in\mathbb Z_{L/2}/2}}q_{x},\\
    Q = Q_L+Q_R.
  \end{gathered}
\end{equation}
Eq.~\eqref{eq:coefficients} follows from a direct application of the definition of ${\tilde \rho}_{{\rm st},t}{(\beta_1,\beta_2)}$ in Eq.~\eqref{eq:defRhot}. Using Eq.~\eqref{eq:coefficients} we obtain 
\begin{equation}
  \label{eq:rhostlink}
  \begin{aligned}
    &\lim_{t\to\infty}\tr[{\tilde \rho}_{{\rm st},t}{(\beta_1,\beta_2)}
    \tilde a^{(1)}_{2t-\tau} \cdots \tilde a^{({\tau})}_{2t}] \\
    &= \tr[\rho^* (\beta_1,\beta_2) {a}^{(1)}_0(0)\cdots {a}^{(\tau)}_\tau(0)]\,,\quad\forall a_j\,,
  \end{aligned}
\end{equation}
where we introduced the state $\rho^* (\beta_1,\beta_2)$ such that 
\begin{equation} \label{eq:rhostbeta}
  \begin{aligned} 
    &\lim_{t\to\infty} \lim_{L\to\infty}
    \frac{\expval{ {\mathbb U}^{-t}{\mathcal O} {\mathbb U}^te^{\beta_1 Q_{L}+ \beta_2 Q_{R}}}{\Psi_0}}{\expval{e^{\beta_1 Q_{L}+\beta_2 Q_{R}}}{\Psi_0}} \\
    & = \lim_{L\to\infty} \tr[\rho^*(\beta_1,\beta_2) \mathcal O]\,,
  \end{aligned}
\end{equation}
for every local observable ${\mathcal O}$. Reasoning as in the case of
bipartitioning quench protocols, see, e.g., Ref.~\cite{alba2021generalized}, we
conclude that $\rho^*(\beta_1,\beta_2)$ is a stationary state of the
time-evolution operator ${\mathbb U}$. In particular, for integrable models it
can be explicitly determined using Generalised
Hydrodynamics~\cite{castroalvaredo2016emergent, bertini2016transport}. Since
Eq.~\eqref{eq:rhostlink} holds for every local operator $a_j$, we conclude that
${\tilde \rho}_{{\rm st},t}{(\beta_1,\beta_2)}$ is the space-time swap
correspondent of $\rho^* (\beta_1,\beta_2)$. Therefore we set 
\begin{equation}
  {\rho}_{{\rm st},A}{(\beta_1,\beta_2)} = \tr_{\bar A}[\rho^* (\beta_1,\beta_2)]\,.
\end{equation}
This equation fully specifies ${\rho}_{{\rm st},A}{(\beta_1,\beta_2)}$ for any initial state and represents the first main result of this paper. Before using it to find explicit predictions, however, we employ it to make another general observation. As special case of Eq.~\eqref{eq:coefficients}, one has    
\begin{equation}\label{eq:currentid}
  \begin{aligned}
    &\tr[{\tilde \rho}_{{\rm st},t}{(\beta_1,\beta_2)} e^{\beta \tilde Q_t}]  \\
    =&\frac{\expval{ e^{\beta j_0(\frac{1}{2})}\cdots e^{\beta j_0(t)} e^{\beta_1 Q_{L}+ \beta_2 Q_{R}}}{\Psi_0}}{\expval{e^{\beta_1 Q_{L}+\beta_2 Q_{R}}}{\Psi_0}}\,,
  \end{aligned}
\end{equation}
where $j_x$ is the current associated to the U(1) charge $Q$ (cf.\
Eq.~\eqref{eq:appcurrent}). This expression corresponds to
the expectation value of the time ordered exponential of the current in $x=0$
integrated in time from $0$ to $t$ multiplied by $\exp({\beta_1 Q_{L}+\beta_2
Q_{R}})$. As shown in Appendix~\ref{sec:equiv}, assuming local relaxation
Eq.~\eqref{eq:currentid} gives 
\begin{equation}
  \label{eq:chargecurrentFCS}
  \begin{aligned}
    &\lim_{t\to\infty}\frac{1}{t} \log\tr[{\tilde{ \rho}}_{{\rm st},t}(\beta_1,\beta_2)
    e^{\beta {\tilde Q}_t}] \\
    &=  \lim_{t\to\infty}\frac{1}{t} \log\tr[{{ \rho}}_{{\rm st}}(\beta_1,\beta_2)
    e^{\beta j_0(\frac{1}{2})}\cdots e^{\beta j_0(t)}]. 
  \end{aligned}
\end{equation}
The quantity on the r.h.s. of this equation is precisely the scaled cumulant
generating function of the current in $x=0$ in the non-equilibrium steady state
${\rho}_{{\rm st},A}{(\beta_1,\beta_2)}$. Therefore, we can establish a direct
link between our work and the recent literature on current fluctuations on
non-equilibrium steady
states~\cite{doyon2020fluctuations,myers2020transport,doyon2023ballistic,gopalakrishnan2024distinct,krajnik2022exact,krajnik2024universal,wei2022quantum,samajdar2023quantum,krajnik2024dynamical,mcculloch2023full}:
The FCS \eqref{eq:FCS} in the non-equilibrium regime is given by the product of the 
FCS of the currents at the two boundaries of the subsystem $A$. The current FCS
are evaluated in the stationary states  ${\rho}_{{\rm st},A}{(\beta,0)}$ and
${\rho}_{{\rm st},A}{(0,\beta)}$, which are non-equilibrium steady states
whenever the initial state is not an eigenstate of the charge. 

\subsubsection{Duality for higher charged moments}
\label{sec:discreteduality2}

The above discussion can be straightforwardly generalised to the case of more general charged moments~\eqref{eq:chargedmoments}. We can again define the stationary density 
\begin{equation}
\label{eq:densityn}
\begin{aligned}
  d_{\bm{\beta}}&:=\lim_{|A|\to\infty}\lim_{t\to\infty} 
  \frac{\log Z_{\bm{\beta}}(A,t)}{|A|}\\
  &\phantom{:}= \lim_{|A|\to\infty} \frac{1}{|A|}
  \log\tr[\prod_{j=1}^{n}\left( e^{\beta_j  Q_A} \rho_{{\rm st}, A} \right)],
\end{aligned}
\end{equation}
and the asymptotic slope
\begin{equation}
\label{eq:slopen}
  s_{\bm{\beta}}:=\lim_{t\to\infty}\lim_{|A|\to\infty} 
  \frac{\log Z_{\bm{\beta}}(A,t)-|A|\log\Lambda_{\bm{\beta}}}{t}.
\end{equation}
Using these definitions we can write the leading-order form of the charged moments as
\begin{equation}\label{eq:Zbothlimits}
  Z_{\bm{\beta}}(A,t)\simeq
  \begin{cases}
    \Lambda^{|A|}_{\bm{\beta}}e^{t s_{\bm{\beta}}},&
    1\ll t \ll |A|,\\
    e^{|A| d_{\bm{\beta}}},&
    1 \ll |A| \ll t,
  \end{cases}
\end{equation}
where $\Lambda_{\bm{\beta}}$ is determined by
  $Z_{\bm{\beta}}(A,0)=\Lambda^{|A|}_{\bm{\beta}}$
\begin{equation}
\label{eq:lambdan}
  \Lambda_{\bm{\beta}}=\prod_{j=1}^n 
  \mel{\psi_0}{e^{\beta_j q}\otimes e^{\beta_j q}}{\psi_0}=
  \prod_{j=1}^{n}\Lambda_{\beta_j}.
\end{equation}
As in the case of FCS, the slope $s_{\bm{\beta}}$ 
can be put in a form dual to~\eqref{eq:densityn}. To see this, we 
start by expressing $Z_{\bm{\beta}}(A,t)$ in terms of the time-evolved state
$\ket{\Psi_t}$ as
\begin{equation}\label{eq:Zdecomposition}
  Z_{\bm{\beta}}(A,t)=
  \tr_{\!A}\!\Big[
    \prod_{j=1}^{n}
    \left(\tr_{\!\bar{A}}\!\left[
      \mathrm{e}^{\beta_j Q_{A}}\ketbra{\Psi_t}
      \right]\right)
    \Big],
\end{equation}
where we denoted by $\bar{A}$ the complement of $A$, and used that $e^{\beta Q_A}$
acts as the identity in $\bar{A}$.
\begin{figure*}
  \includegraphics[width=\textwidth]{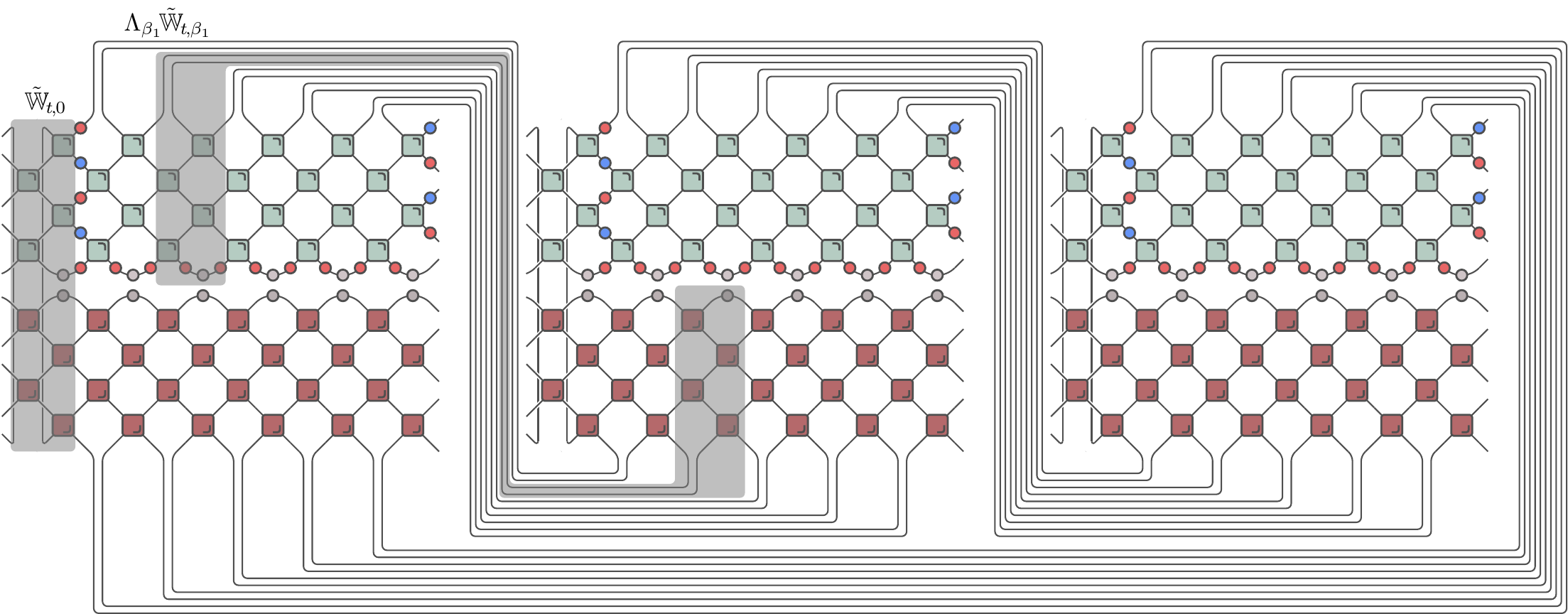}
  \caption{\label{fig:multiReplicaTN} Diagrammatic representation of
  $Z_{\bm{\beta}}(A,t)$, as defined in~\eqref{eq:Zdecomposition}. The same
  diagram can be alternatively generated by the repeated application of tensor
  products of space-transfer matrices $\tilde{\mathbb{W}}_{t,0}$ and
  $\tilde{\mathbb{W}}_{t,\beta_j}$, denoted by grey boxes in the diagram. The
  transfer matrices in $\bar{A}$ are coupling together the top and bottom part
  of each of the three copies, while the transfer matrices acting on $A$ are
  connecting the bottom part of each copy with the top half of the preceding
  one. This is accounted for by introducing the operator $\mathbb{P}_{\sigma}$
  that permutes the $2n$ replicas ($2$ for each of the $n$ copies above), which
  allows the expression in terms of powers of transfer matrices, as shown in
  Eq.~\eqref{eq:defSpaceEvolutionMulti}. Note that the boundary conditions in space
  are assumed to be periodic.
  }
\end{figure*}
Representing $Z_{\bm{\beta}}(A,t)$ in terms of space transfer matrix~\eqref{eq:WmatrixFormula} we obtain the following expression
\begin{equation}
  \label{eq:defSpaceEvolutionMulti}
\begin{split}
  Z_{\bm{\beta}}(A,t) &= \Lambda_{\bm{\beta}}^{|A|}
  {\rm tr}\Bigg[
    \mathbb{P}_{\sigma}^{\dagger}
    \bigg(\!\bigotimes_{j=1}^{n} \tilde {\mathbb W}_{t,0}^{|\bar A|}\bigg)
  \mathbb{P}_\sigma\\
  &\times\mkern-6mu\bigg(\mkern-6mu\bigotimes_{j=1}^{n}  
  (e^{\beta_j {\tilde Q}_t}\!\otimes\! \1) 
  \tilde{\mathbb W}_{t,\beta_j}^{|A|}
  (e^{-\beta_j {\tilde Q}_t}\!\otimes\! \1) \mkern-6mu
  \bigg)\mkern-6mu
  \Bigg],
\end{split}
\end{equation}
where the tensor product the operator $\mathbb P_\sigma$ implements the permutation
\be
\sigma = \begin{pmatrix}
  1 & 2 & 3 & 4 & \cdots & 2n-1 & 2 n \\
  2n-1 & 2 & 1 & 4  & \cdots & 2n-3 & 2 n\\
\end{pmatrix},
\ee
on the replicas. {Intuitively, this permutation arises because the indices pertaining respectively to the subsystem $A$ and the rest of the system $\bar A$ are contracted in a different way because of the partial trace in \eqref{eq:chargedmoments} inducing a modification of the space transfer matrix, see Fig.~\ref{fig:multiReplicaTN} for a diagrammatic illustration.}

Whenever $|A|, |\bar{A}|\ge 2t$ locality and unitarity of the interactions again imply that the expression factorizes into two contributions so that we obtain the analogue of Eq.~\eqref{eq:FCSasymptFINAL} as
\begin{equation} \label{eq:dualityCM}
  \begin{aligned}
    &Z_{\bm{\beta}}(A,t) = \Lambda_{\bm{\beta}}^{|A|} \tr[\prod_{j=1}^n 
    e^{\beta_j {\tilde Q}_t}
    {\tilde{ \rho}}_{{\rm st},t}{(\beta_j,0)} ]
    \\
    &\times
    \tr[\prod_{j=1}^n {\tilde \rho}_{{\rm st},t}{(0,\beta_j)} e^{- \beta_j {\tilde Q}_t}]\!.
  \end{aligned}
\end{equation}
We see that each of the two traces on the r.h.s.\ is the space-time swapped
version of that appearing in Eq.~\eqref{eq:densityn} with the only difference
that each replica is in an \emph{a priori} different stationary state.  

\subsection{Trotter limit}
\label{sec:trotter}

Let us conclude this general discussion by remarking on the generality
of Eq.~\eqref{eq:dualityCM}. To arrive at that expression, we assumed the dynamics
to be given in terms of a brickwork quantum circuit, but we expect it to hold also
for Hamiltonian dynamics which can be accessed via the Trotter limit in Eq.~\eqref{eq:trotterlimit}. When this limit is performed naively, however, the rescaling of space and time means that Eq.~\eqref{eq:dualityCM} only holds for $|A|=\infty$. Therefore, we need to refine our argument to show that it holds in a nontrivial regime.

To do this let us assume the dynamics is generated by a Hamiltonian $H$ with a
local $2$-site density, which we approximate with a quantum circuit obtained by
using two-site unitary gates $U_{\tau}$ (cf.\ \eqref{eq:FTapprox}) with
the label $\tau$ denoting the time-step. Then, as long as $\abs{A}>2t$, Eq.~\eqref{eq:dualityCM} holds. Here we conveniently express it as follows 
\begin{equation} \label{eq:dualityCMrewritten}
  \log\left[\frac{Z_{\bm{\beta},\tau}(A,t)}{\Lambda_{\bm{\beta}}^{|A|}}\right]=
  \log\left[Z^{(\mathrm{L})}_{\bm{\beta},\tau}(t)\right]+
  \log\left[Z^{(\mathrm{R})}_{\bm{\beta},\tau}(t)\right]. 
\end{equation}
For a fixed value of $\tau$, we can use the fact that both the contributions on the r.h.s.\ are independent of the size of
the subsystem $A$. Therefore, they are the same as the charged moments in a system of
size $2L$ (with sites labelled between $-L$ and $L$) with \emph{open} boundary conditions,
and when the subsystem $A$ is the right/left half of the chain
\begin{equation}
  \begin{aligned}
    Z^{(\mathrm{L})}_{\bm{\beta},\tau}(t)&=
  \lim_{L\to\infty}
  \frac{\tr\left[\prod_{j=1}^{n} e^{\beta_j Q_{[0,L]}}\rho_{[0,L]}\right]}
  {\Lambda_{\bm{\beta}}^L},\\
    Z^{(\mathrm{R})}_{\bm{\beta},\tau}(t)&=
  \lim_{L\to\infty}
  \frac{\tr\left[\prod_{j=1}^{n} e^{\beta_j Q_{[-L,0]}}\rho_{[-L,0]}\right]}
  {\Lambda_{\bm{\beta}}^L}.
  \end{aligned}
\end{equation}
These two contributions have a well defined Trotter limit,
\begin{equation}
  \log[Z^{(\mathrm{r})}_{\bm{\beta}}(\mathfrak{t})]=
  \lim_{\substack{\tau\to 0,\ t\to\infty \\ \mathfrak{t}=\tau t}}
  \tau \log\left[Z^{(\mathrm{r})}_{\bm{\beta},\tau}(t)\right].
\end{equation}
What remains to be argued is that for a non-zero value of $\mathfrak{t}/A$, the sum
of these contributions is equal to  the Trotter limit of the l.h.s.\ of Eq.~\eqref{eq:dualityCMrewritten}, i.e.,
\begin{equation}
  \log\left[\frac{Z_{\bm{\beta}}(A,\mathfrak{t})}{\Lambda_{\bm{\beta}}^{|A|}}\right]
  = \lim_{\substack{\tau\to 0,\ t\to\infty \\ \mathfrak{t}=\tau t}}
  \tau\log\left[\frac{Z_{\bm{\beta},\tau}(A,t)}{\Lambda_{\bm{\beta}}^{|A|}}\right].
\end{equation}
This is ensured by assuming that $H$ fulfils the Lieb-Robinson bound~\cite{lieb1972finite}: there
exists a velocity $v_{\mathrm{LB}}>0$ so that the local perturbations to the
initial state at the position $x+d$, $\abs{d}>v_{\mathrm{LB}} \mathfrak{t}$
will give corrections exponentially small in
$\abs{d}-v_{\mathrm{LB}}\mathfrak{t}$ to the local properties of the state at
time $\mathfrak{t}$ and position $x$. Intuitively, this implies that as long as
$\abs{A}>2 v_{\mathrm{LB}} \mathfrak{t}$, the information from one edge of 
the subsystem cannot propagate far enough to change the local properties of the
state at the other edge, and the charged moments (up to exponentially small corrections)
decouple into the two contributions
\begin{equation}
  \log\left[\frac{Z_{\bm{\beta}}(A,\mathfrak{t})}{\Lambda_{\bm{\beta}}^{|A|}}\right]
  =
  \log[Z^{(\mathrm{L})}_{\bm{\beta}}(\mathfrak{t})]+
  \log[Z^{(\mathrm{R})}_{\bm{\beta}}(\mathfrak{t})].
\end{equation}

\section{Integrable Systems}
\label{sec:integrablecircuits}

Let us now specialise the treatment of the previous section to the case of TBA-integrable systems. As we will briefly review, in these systems one can explicitly evaluate Eq.~\eqref{eq:density1}, as well as all other thermodynamic quantities, in terms of the solution of suitable integral equations. Here we want to argue that a similar treatment can also be performed for Eq.~\eqref{eq:slope1}, giving access to the charged moments in the non-equilibrium regime. 

Our discussion proceeds as follows. In Sec.~\ref{sec:TBA} we recall a number of basic facts concerning integrable systems and their TBA description. In Sec.~\ref{sec:TBArhotilde} we argue that the TBA description can also be applied to the system on the temporal lattice and derive the relevant equations. Finally, in Sec.~\ref{sec:TBApredictions} we report our closed-form predictions for the charged moments in the equilibrium and non-equilibrium regime. 

Note that, since the Bethe-ansatz solution has the same structure for both for integrable circuits and integrable Hamiltonians~\cite{faddeev1996how, aleiner2021bethe} we assume that the TBA treatment is the same. This assumption has been verified explicitly in Ref.~\cite{vernier2023integrable} for the case of the XXZ chain.  


\subsection{Thermodynamics via Bethe Ansatz}
\label{sec:TBA}

An integrable model possesses an extensive number of quasi-local conserved charges $\{Q^{(k)}\}_{k=0,1, \ldots}$. One can intuitively think of quasi-locality as the property of having an exponentially localised density, see, however, Ref.~\cite{ilievski2016quasilocal} for a more precise definition. From now on, we focus on the standard case where $\{Q^{(k)}\}$ commute and we specify the $\rm U(1)$ charge $Q$ to be the first one in the tower, i.e., $Q^{(0)}=Q$.

Because of the constraints on the scattering imposed by the conservation laws, integrable models admit stable quasiparticle excitations~\cite{korepin1997quantum}. More precisely, one can write a basis of scattering states  
\be
\ket{\bm{\lambda}} = \ket*{\lambda^{(1)}_1,\ldots,\lambda^{(1)}_{M_1}; \ldots; \lambda^{(N_s)}_1,\ldots,\lambda^{(N_s)}_{M_{N_s}}},
\label{eq:basis}
\ee
which are simultaneous eigenstates of all the conserved charges. Namely
\be
Q^{(k)} \ket{\bm{\lambda}} = \sum_{m=1}^{N_s}\sum_{j=1}^{M_m} q_m^{(k)}(\lambda^{(m)}_j) \ket{\bm{\lambda}}.
\ee 
Here $\lambda^{(m)}_{j}$ are real rapidities (fulfilling appropriate quantisation conditions when the system is confined to a finite volume), the superscript $m=1,\ldots,N_s$ labels different quasiparticle species, and $q_m^{(k)}(\lambda)$ are the quasiparticle charges. Particularly important for our purposes are the quasiparticle energy $\varepsilon_m(\lambda)$, momentum $p_m(\lambda)$ and  $U(1)$ charge $q_m$ (rapidity independent). Note that one can always parameterise the dispersion relation such that 
\be
p'_m(\lambda)>0.
\ee 
We emphasise that here we describe the states in terms of rapidities, rather than regular momenta, as this allows for a comprehensive treatment of all TBA-integrable models~\cite{korepin1997quantum, takahashi1999thermodynamics}.

The quasiparticles scatter non-trivially but elastically. We denote by $S_{ml}(\lambda,\mu)$ the $S$-matrix between quasiparticles of species $m$ and $l$ with rapidities $\lambda$ and $\mu$ and define the scattering phase shift as 
\be
T_{ml}(\lambda,\mu)=\frac{1}{2\pi i}\partial_{\lambda}\log S_{ml}(\lambda,\mu). 
\ee

A stationary state of the system is specified by a set of quasiparticles. In the thermodynamic limit the latter is characterised by a set of distribution functions in rapidity space $\rho_m(\lambda)$, where $m$ runs from one to $N_s$: the number of species of quasiparticles. It is also convenient to introduce  the distribution of unoccupied quasiparticles, $\rho_m^h(\lambda)$, as well as the distribution of available ``momentum slots"   
\begin{equation}
  \rho_m^t(\lambda)=\rho_m(\lambda)+\rho^h_m(\lambda), 
\end{equation}
and the filling function 
\begin{equation}
  \vartheta_m(\lambda)=\frac{\rho_m(\lambda)}{\rho_m(\lambda)+\rho^h_m(\lambda)} \in [0,1].  
\end{equation}
These distributions are not independent: they are related to each other through the thermodynamic Bethe-Takahashi equations, a set of coupled integral equations arising from the quantisation conditions for the system at finite size. In our notation they read as  
\begin{equation} \label{eq:BetheTakahashi}
\rho^t_m(\lambda)=\frac{p'_m(\lambda)}{2\pi}
  -(T\conR \rho)_{m}(\lambda),
\end{equation}
where we introduced the short-hand notation $\conR$ and $\conL$ to denote the generalised convolution over the second and first parameter respectively,
\begin{equation}
  \begin{aligned}
    (f\conR g)_m(\lambda)&=\sum_{l} \int {\rm d} \mu f_{m l}(\lambda,\mu) g_{l}(\mu),\\
    (f\conL g)_m(\lambda)&=\sum_{l} \int {\rm d} \mu f_{l m}(\mu,\lambda) g_{l}(\mu).
  \end{aligned}
\end{equation}
Eq.~\eqref{eq:BetheTakahashi} has to be combined with one specifying either
$\rho_m(\mu)$ or $\vartheta_m(\lambda)$  to fully characterise the state. For
instance, the filling functions describing the GGE
\begin{equation} \label{eq:GGE}
  \rho_{{\rm st},L} = \frac{e^{ -\sum_k \mu_k Q^{(k)}}}{\tr[e^{ -\sum_k \mu_k Q^{(k)}}]}, 
\end{equation}
where $\{\mu_m^{(j)}\}$ chemical potentials, are determined via the generalised TBA equations~\cite{mossel2012generalized}
\begin{equation} \label{eq:GTBA}
  \log\eta_m(\lambda) = d_m(\lambda)+ 
  \left(T\conL \log\Big[ 1+\frac{1}{\eta}\Big]\right)_m(\lambda),
\end{equation}
where we defined
\begin{equation} \label{eq:Dfun}
\eta_m(\lambda)=\frac{1-\vartheta_m(\lambda)}{\vartheta_m(\lambda)},
  \qquad d_m(\lambda) = \sum_k \mu_k q_m^{(k)}(\lambda)\,.
\end{equation}
If $\rho_{{\rm st},L}$ is the stationary state reached after a quench from an integrable initial state $\ket{\Psi_0}$, an explicit form of $d_m(\lambda)$ can be found by computing the overlaps between the $\ket{\Psi_0}$ and the scattering states $\ket{\bm{\lambda}}$~\cite{caux2013time,caux2016quench}.  The free energy of the GGE in Eq.~\eqref{eq:GGE} is expressed in terms of $\vartheta_m(\lambda)$ as follows
\begin{equation} \label{eq:freeenergy}
  \mkern-6mu \frac{\log \tr[e^{ -\sum_k \mu_k Q^{(k)}}\!]}{L}\mkern-6mu =\mkern-6mu
\sum_{m}\!\int\mkern-6mu\frac{\mathrm{d}\lambda}{2\pi}
p^{\prime}_m(\lambda) \log\!\!\left[1\!+\!\frac{1}{\eta_m(\lambda)}\right]\!.\mkern-4mu
\end{equation}
Here, without loss of generality, we assumed that the conserved charges
annihilate the state without quasiparticles. 

The functions $q_m^{(k)}(\lambda)$ describe the \textit{bare} properties of the
quasiparticles. At finite density it is useful to also introduce their
effective counter parts, $q^{(k)}_{{\rm eff},m}(\lambda)$, which account for the
effects of the interactions. Given a bare function 
$b_m(\lambda)$, its effective version is obtained by solving of the following
integral equations  
\begin{equation} \label{eq:dressing}
  b_{\mathrm{eff},m}
  (\lambda)= 
  b_m(\lambda)- (T\conL b_{\mathrm{eff}}\,\vartheta)_m(\lambda).
\end{equation}
In particular, at finite density one can express the velocity of the quasiparticles in terms of effective quantities as~\footnote{Note that when the kernel is non-symmetric one has to use the convolutions $\conL$ rather than $\conR$ in the dressing equations for $\varepsilon^{\prime}$ and $p^{\prime}$.} 
\begin{equation}\label{eq:effectiveVelocity}
  v_m(\lambda)=
  \frac{(\varepsilon_m^{\prime}(\lambda))_{\rm eff}}{(p_m^{\prime}(\lambda))_{\rm eff}}. 
\end{equation}

We conclude this brief survey by recalling that the TBA formalism can be used
to characterise the FCS of $Q^{(0)}=Q$ in: (i) stationary states like the GGE
in Eq.~\eqref{eq:GGE}; (ii) integrable non-equilibrium states like our initial
state $\ket{\Psi_0}$. Indeed, using Eq.~\eqref{eq:freeenergy} one finds (see
Appendix~\ref{sec:tbaFCS})
\begin{equation} \label{eq:GGEFCS}
  \lim_{L\to\infty}\frac{\log\tr[\rho_{{\rm st},L} e^{\beta Q}]}{L}=
  \sum_m\int \frac{{\rm d}\lambda}{2\pi} p'_m(\lambda)\mathcal{K}_{m}^{(\beta)}(\lambda),
\end{equation}
where we defined 
\begin{equation} \label{eq:logx}
  \begin{aligned}
    \mathcal{K}_{m}^{(\beta)}(\mu) &=
    \log\left[\frac{{\eta_m(\mu)}+e^{-w^{(\beta)}_m(\mu)}}{1+{\eta_m(\mu)}}\right],\\
   w^{(\beta)}_m(\lambda) &= -\beta q_m +
    (T\conL \mathcal{K}^{(\beta)})_m(\lambda),
  \end{aligned}
\end{equation}
and the eta function $\eta_m(\lambda)$ is the one describing the state
$\rho_{{\rm st},L}$. Analogously, using the TBA treatment of the diagonal
ensemble (see, e.g., Ref.~\cite{alba2017quench}) we find
\begin{equation} \label{eq:FCSpsi}
  \lim_{L\to\infty}\frac{\log\expval{e^{\beta Q}}{\Psi_0}}{L} =
  \sum_m\int \!\frac{{\rm d}\lambda}{4\pi} p'_m(\lambda) \mathcal{K}_{m}^{(2\beta)}(\lambda), 
\end{equation}
where now $\eta_m$ in $\mathcal{K}_{m}^{(2\beta)}$ are the eta functions of the
stationary state reached after a quench from $\ket{\Psi_0}$. Note that, to ease
the notation, in the following we suppress the dependence on rapidity
and species index whenever is not ambiguous to do so.

In fact, as we discuss in Appendix~\ref{sec:eqFCStemporal}, the expression in
Eqs.~(\ref{eq:GGEFCS}, \ref{eq:logx}) is not suitable for our space-time swap
and one has to consider the following rewriting  
\begin{align}
  \begin{split} \label{eq:deq2}
    \lim_{L\to\infty}&
    \frac{\log\tr[\rho_{{\rm st},L} e^{\beta Q}]}{L}\\
    &=
    \!\!\int\limits_0^\beta \!\! {\rm d}u \!\sum_{m}
    \!\!\int\!\frac{\mathrm{d}\lambda}{2\pi}
    p^{\prime}_m(\lambda) \vartheta^{(u)}_m(\lambda) q_{\mathrm{eff},m}[\vartheta^{(u)}](\lambda),
  \end{split} \\ 
  \label{eq:thetabeta}
  \vartheta^{(u)}_m &= \frac{1}{1+\eta_{m} e^{w_m^{(u)}}}\,, \\
  \label{eq:logx2}
  \partial_\beta w^{(u)}_m   &=  - {\rm{sgn}}[\rho^t_m[\vartheta^{(u)}]] q_{\mathrm{eff},m}[\vartheta^{(u)}],\qquad w^{(0)}_m=0\,,
\end{align}
where $\rho^t_m[\vartheta]$ and $q_{\mathrm{eff},m}[\vartheta]$ are the total
number of momentum slots and the effective charges in the state described by
the set of filling functions $\vartheta:=\{\vartheta_m\}_m$
(cf.~\eqref{eq:dressing}). Indeed, although one can directly show that
Eqs.~(\ref{eq:GGEFCS}, \ref{eq:logx}) and \eqref{eq:deq2}--\eqref{eq:logx2} are
identical (see Appendix~\ref{sec:tbaFCS}) their space-time swapped counterparts
differ. In Appendix~\ref{sec:eqFCStemporal} we show that the one in
Eqs.~(\ref{eq:deq2})--(\ref{eq:logx2}) is the correct expression to use for the
swap. 

\subsection{Thermodynamics on the Temporal Lattice}
\label{sec:TBArhotilde}

Our basic observation is that in an integrable quantum circuit also the space
evolution is written in terms of integrable local
gates~\cite{pozsgay2013dynamical,piroli2017quantum,piroli2018non,rylands2019loschmidt,klumper2004integrability,suzuki1999spinons,klumper1993thermodynamics}.
This observation can be extended to Hamiltonian
systems via a suitable discretisation of the time evolution --- in fact, it is
the main premise of the well established Quantum Transfer Matrix
approach~\cite{klumper2004integrability,suzuki1999spinons,klumper1993thermodynamics}.
As a consequence, as long as one chooses
appropriate boundary conditions for the temporal lattice (i.e., appropriate
initial states for the space evolution), the space transfer matrix remains
Bethe Ansatz solvable~\cite{klumper2004integrability,faddeev1996how}. 

Here we implement integrable boundary conditions on the time lattice by
considering integrable initial states~\cite{piroli2017what}. Therefore, we
assume ${\tilde{ \rho}}_{{\rm st},t}(\beta_1,\beta_2)$ to be diagonal in a
scattering basis analogous to that in Eq.~\eqref{eq:basis} but defined on the
temporal lattice. Let us denote it by $\{\ket*{\tilde{\boldsymbol \lambda}}\}$
and call $\tilde q_m^{(k)}(\lambda)$ the associated charges. The scattering
matrix for this system coincides with the one for the system on the spatial
lattice. 

Following the TBA treatment described above, to fully specify the
thermodynamics of the system we then need two ingredients 
\begin{itemize}
  \item[(i)] Filling functions $\tilde\vartheta_{m}(\beta_1, \beta_2)$ for
    ${\tilde{\rho}}_{{\rm st},t}(\beta_1,\beta_2)$; 
  \item[(ii)] Dispersion relation $(\tilde{p}_m(\lambda), \tilde{\varepsilon}_m(\lambda))$
    for all quasiparticles on the temporal lattice. 
\end{itemize}
Both these ingredients can be found by using Eq.~\eqref{eq:rhostlink}. We begin
by considering the aforementioned equation in the special case 
\begin{equation}
  a^{(2t)} = a, \qquad a^{(j\neq2t)}=\1,
\end{equation}
which gives  
\begin{equation} \label{eq:rhostlink1}
  \lim_{t\to\infty}\tr[{\tilde \rho}_{{\rm st},t}(\beta_1,\beta_2) \tilde a_t ] 
  = \tr[{\rho}_{{\rm{st}}}(\beta_1,\beta_2) {a}_0]\,,
\end{equation}
for all local operators $a$. Expanding in the scattering basis this gives
\begin{equation}
  \begin{aligned}
    &\lim_{t\to\infty}\sum_{\ket*{\tilde{\boldsymbol \lambda}}} 
    \!\!\expval*{\tilde \rho_{{\rm st},t}(\beta_1,\beta_2)}{\tilde{\boldsymbol \lambda}} 
    \!\!\expval*{\tilde a_0}{\tilde{\boldsymbol \lambda}}\\
    =&\lim_{L\to\infty}\sum_{{\ket{\boldsymbol \lambda}}} 
    \!\!\expval*{ \rho_{{\rm st},L}(\beta_1,\beta_2)}{{\boldsymbol \lambda}} 
    \!\!\expval*{a_0}{{\boldsymbol \lambda}},
  \end{aligned}
\end{equation}
where, to write a symmetric expression, in the second line we explicitly
reported the $L$ dependence of ${\rho}_{\rm st}$ and considered the
thermodynamic limit. Analogously, choosing 
\begin{equation}
  a^{(2t-2\ell)}=a, \qquad a^{(2t)} = b, \qquad a^{(j\neq2t,2t-2\ell)}=\1, 
\end{equation}
in Eq.~\eqref{eq:rhostlink} we find  
\begin{equation}\label{eq:rhostlink2}
  \begin{aligned}
    \lim_{t\to\infty}\!&\tr[{\tilde \rho}_{{\rm st},t}(\beta_1,\beta_2) 
    \tilde{a}_{t-\ell}{\tilde b}_{t}] \\
    =\!&\tr[{\rho}_{{\rm st}}(\beta_1,\beta_2) {a}_0 {b}_0({\ell})].
  \end{aligned}
\end{equation}
Expanding again in the eigenbasis we obtain 
\begin{equation}
  \begin{aligned}
    \lim_{t\to\infty}
    \smashoperator[r]{\sum_{{\ket*{\tilde{\boldsymbol \lambda}_1}},\ket*{\tilde{\boldsymbol \lambda}_2}}}
    &\mel*{\tilde{\boldsymbol \lambda}_1}{\tilde \rho_{{\rm st},t}(\beta_1,\beta_2)}{\tilde{\boldsymbol \lambda}_1} \mel*{\tilde{\boldsymbol \lambda}_1}{{\tilde a}_0}{\tilde{\boldsymbol \lambda}_2}\\
    \times&\mel*{\tilde{\boldsymbol \lambda}_2}{{\tilde b}_0}{\tilde{\boldsymbol \lambda}_1} e^{ i [\tilde P_{{\boldsymbol \lambda}_1}\!-\!\tilde P_{{\boldsymbol \lambda}_2}]\ell}\\
    = \lim_{L\to\infty} \smashoperator[r]{\sum_{{\ket*{{\boldsymbol \lambda}_1}},\ket*{{\boldsymbol \lambda}_2}}}
    &\mel*{{\boldsymbol \lambda}_1}{ \rho_{{\rm st},L}(\beta_1,\beta_2)}{{\boldsymbol \lambda}_1} \mel*{{\boldsymbol \lambda}_1}{a_0}{{\boldsymbol \lambda}_2}\\
    \times&\mel*{{\boldsymbol \lambda}_2}{b_0}{{\boldsymbol \lambda}_1} e^{ i [E_{{\boldsymbol \lambda}_1}\!-\!E_{{\boldsymbol \lambda}_2}]\ell},
  \end{aligned}
\end{equation}
where 
\begin{equation}
  \mkern-1mu E_{{\bm{\lambda}}} =\smashoperator{\sum_{m=1}^{N_s}}
  \sum_{j=1}^{M_m} \varepsilon_m(\lambda^{(m)}_j),
  \quad \tilde P_{{\bm{\lambda}}} = \smashoperator{\sum_{m=1}^{N_s}}
  \sum_{j=1}^{M_m} \tilde p_m(\lambda^{(m)}_j).\mkern-1mu
\end{equation}
Since the same argument can be repeated for an arbitrary number of operators we argue that the only possible solution is 
\begin{align}
  \mel*{\tilde {\boldsymbol \lambda}}{\tilde\rho_{\mathrm{st},t}(\beta_1,\beta_2)}{\tilde{\boldsymbol \lambda}} 
  & = \mel*{{\boldsymbol \lambda}}{ \rho_{{\rm st},L}(\beta_1,\beta_2)}{{\boldsymbol \lambda}}\!, \label{eq:rhorel}\\
  \tilde p_m(\lambda) & = \varepsilon_m(\lambda), \label{eq:ptilde}\\
  \mel*{\tilde{\bm{\lambda}}}{\tilde{a}_0}{\tilde{\bm{\mu}}} & = \mel*{{\bm{\lambda}}}{{a}_0}{{\bm{\mu}}}.
\end{align}
Writing the analogue of \eqref{eq:coefficients} where the operators $\tilde a$ are evolved in space in the Heisenberg picture and repeating the above argument we also find  
\begin{equation} \label{eq:epsilontilde}
  \tilde \varepsilon_m(\lambda) =  p_m(\lambda). 
\end{equation}
The relations \eqref{eq:ptilde} and \eqref{eq:epsilontilde} are very natural.
Since the basis $\{\ket*{\tilde{\boldsymbol \lambda}}\}$ describes scattering
states for the system on the temporal lattice they have a swapped dispersion
relation with respect of the states in $\{\ket*{{\boldsymbol \lambda}}\}$. This
represents a swap of space and time in momentum space. The relation
\eqref{eq:rhorel} on the other hand, implies that ${\tilde \rho}_{{\rm
st},t}(\beta_1,\beta_2)$  and ${\rho}_{{\rm st},L}(\beta_1,\beta_2)$ are
described by the same filling functions. Specifically, a direct application of
Generalised Hydrodynamics~\cite{castroalvaredo2016emergent,bertini2016transport} gives   
\begin{equation}
  \label{eq:thetatilde}
  \begin{aligned}
    \tilde \vartheta_{m}(\beta_1, \beta_2) =& 
    \frac{\vartheta_m\Theta(v_m)}{\vartheta_m+(1-\vartheta_m) x_{m}^{(2 \beta_1)}}\\
    &+\!\frac{\vartheta_m\Theta(-v_m)}{\vartheta_m+(1-\vartheta_m) x_{m}^{(2 \beta_2)}},
  \end{aligned}
\end{equation}
where $\vartheta_m$ are the filling functions of $\rho_{{\rm st}}$,
$v_m(\lambda)$ are the velocities of the excitations on the state with filling
$\vartheta_{m}(\beta_1, \beta_2)$, $x_{m}^{(\beta)}$ is the solution of
Eq.~\eqref{eq:logx} with $\eta_m=(1-\vartheta_m)/\vartheta_m$ and $\Theta(x)$
is the Heaviside function. This equation, together with \eqref{eq:ptilde} and
\eqref{eq:epsilontilde}, addresses (i) and (ii).

\subsection{Closed-form predictions}
\label{sec:TBApredictions}

The TBA formalism discussed in the previous subsections can be combined with
the space-time duality to characterise the leading order behaviour of the
charged moments, i.e., Eqs.~\eqref{eq:density1}, \eqref{eq:slope1}, \eqref{eq:densityn}, \eqref{eq:slopen},
and~\eqref{eq:lambdan}. This represents the second main result of this paper. 

We begin considering the simple full counting statistics (FCS). A prediction for $d_\beta$ in Eq.~\eqref{eq:density1} and $\log \Lambda_{\beta}$ (cf. \eqref{eq:Z1nonequilibrium}) is found immediately by using Eqs.~(\ref{eq:GGEFCS}, \ref{eq:logx}) and Eq.~\eqref{eq:FCSpsi} respectively. Instead, the slope in Eq.~\eqref{eq:slope1} can be written as 
\begin{equation}
  s_{{\beta}} = s^{(\rm R)}_{{\beta}} + s^{(\rm L)}_{{\beta}}, 
\end{equation}
where we defined 
\begin{align}
s^{(\rm L)}_{{\beta}}  &=  \lim_{t \to\infty}\frac{1}{t} \tr[e^{\beta {\tilde Q}_t}  {\tilde{ \rho}}_{{\rm st},t}{(\beta,0)} ], \\
s^{(\rm R)}_{{\beta}} &= \lim_{t \to\infty}\frac{1}{t} \tr[e^{- \beta {\tilde Q}_t} {\tilde \rho}_{{\rm st},t}{(0,\beta)} ]\!.
\end{align}
An explicit prediction for these quantities is found by considering the equilibrium FCS in Eqs.~(\ref{eq:deq2}, \ref{eq:logx2}),
swapping space and time, and using Eqs.~(\ref{eq:ptilde}, \ref{eq:epsilontilde}),
and~\eqref{eq:thetatilde}. This gives  
\begin{align}
  \label{eq:swapdeq2}
  &\!\!\!\!\begin{aligned}
    &\lim_{t\to\infty}\frac{1}{t}\log\tr[{\tilde{ \rho}}_{{\rm st},t}(\beta_1,\beta_2) e^{\beta {\tilde Q}_t}] \\
    &=\!\!\! \int\limits _0^\beta \! \mathrm{d}u
    \sum_{m} \int\!\frac{\mathrm{d}\lambda}{2\pi}
    \varepsilon^{\prime}_m(\lambda) 
    \tilde{\vartheta}^{(u)}_m(\lambda)
   q_{\mathrm{eff},m}[\tilde \vartheta^{(u)}](\lambda),
  \end{aligned}\\ 
  &\!\!\tilde{\vartheta}^{(u)}_m = \frac{1}{1+\eta_m{(\beta_1, \beta_2)} e^{\tilde w^{(u)}_m}},\label{eq:swapthetabeta} \\
  \mkern-12mu
  &\!\!\partial_u \tilde w^{(u)}_m   
  \!=\! -\mathrm{sgn}[\rho^t_m\! v_m[\tilde{\vartheta}^{(u)}_m]] q_{\mathrm{eff},m}[\tilde{\vartheta}^{(u)}],\quad \tilde w^{(0)}_m=0,\label{eq:swaplogx2}
\end{align}
where we used that --- since $Q$ has an equally spaced spectrum --- the single
particle eigenvalues of $\tilde Q$ and $\tilde Q^T$ coincide with the one of
$Q$. Namely $\tilde q=q$ $\Rightarrow$ ${\tilde q_{\mathrm{eff},m}}$.
In agreement with Eq.~\eqref{eq:chargecurrentFCS}, this expression coincides
with the full-counting statistics of the integrated current computed in
Ref.~\cite{myers2020transport} when evaluated on the non-equilibrium steady
state ${{ \rho}}_{{\rm st}}(\beta_1,\beta_2)$. 

We stress that, differently from Eqs.~\eqref{eq:deq2}--\eqref{eq:logx2}, one
cannot generically compute analytically the $u$ integral in
Eqs.~\eqref{eq:swapdeq2}--\eqref{eq:swaplogx2}. This can only be done when 
\begin{equation} \label{eq:signcondition}
  \mathrm{sgn}[\rho^t_m\! v_m[\tilde{\vartheta}^{(u)}_m]]
  = \mathrm{sgn}[\rho^t_m\! v_m[\tilde{\vartheta}^{(0)}_m]], \qquad \forall u\,.
\end{equation}
The resulting simplified expressions obtained in this case are reported in
Appendix~\ref{sec:simplification}. Note that, since Eq.~\eqref{eq:signcondition}
holds for the cases studied in Refs.~\cite{bertini2022growth,
bertini2023nonequilibrium}, the expressions reported in the appendix recover
the results of the aforementioned references while Eqs.~\eqref{eq:swapdeq2}--\eqref{eq:swaplogx2} generalise them.

Considering now the general charged moments in Eqs.~\eqref{eq:densityn}, \eqref{eq:slopen},
and~\eqref{eq:lambdan}, we introduce the following generalised FCS 
\begin{equation} \label{eq:generalFCS}
  f_{\boldsymbol \beta} = \lim_{L \to\infty}\frac{1}{L}
  \log\tr[\prod_{j=1}^{n} e^{\beta_j  Q} \rho_{{\rm st},j} ] 
\end{equation}
where $\rho_{{\rm st},j}$ are, a priori different, stationary states of the
form \eqref{eq:GGE}. As shown in Appendix~\ref{sec:tbaCM}, $f_{\boldsymbol \beta}$ can
be computed in TBA and, in particular, it can be brought to the following
space-time-swap amenable form analogous to Eqs.~\eqref{eq:deq2}--\eqref{eq:logx2}
\begin{equation}
  \label{eq:deqn}
  \begin{aligned}
    f_{\bm{\beta}}  & = \int\limits_0^\beta {\rm d}u
    \sum_{m}\!\!\int\!\frac{\mathrm{d}\lambda}{2\pi}
      p^{\prime}_m(\lambda)
      \vartheta^{(u)}_{n,m}(\lambda) 
      q_{\mathrm{eff},m}[\vartheta^{(u)}_{n}](\lambda) \\
      &+ \sum_m\int \frac{{\rm d}\lambda}{2\pi} 
      p'_m(\lambda)
      \mathcal{K}_{n,m}^{(0)}(\lambda).
  \end{aligned}  
\end{equation}
Here we introduced 
\begin{equation}
     \beta := \sum_{j=1}^n \beta_j,
\end{equation}
and      
\begin{align}     
  &\mathcal{K}_{n, m}^{(u)} 
  =  {{\rm sgn}[\rho^t_m [ \vartheta^{(u)}_{n}]]}  \log\!\!\left[\frac{\prod_{j=1}^n {\eta_{j,m}}\!+e^{-w^{(u)}_{n,m}}}
  {\prod_{j=1}^n (1+\eta_{j,m})}\right]\!\!,\label{eq:Kn}\\
  &\vartheta^{(u)}_{n,m} = 
  \frac{1}{1+ \prod_j \eta_{j,m} e^{w^{(u)}_{n,m}}},
  \label{eq:thetan}\\
  &\partial_u w^{(u)}_{n,m} 
  =- {{\rm sgn}[\rho^t_m [ \vartheta^{(u)}_{n}]]} q_{\mathrm{eff},m}[\vartheta^{(u)}_{n}],
  \label{eq:logxn}
\end{align}
while $\eta_{j, m}$ are the eta functions of $\rho_{{\rm st},j}$.  Finally, 
$w^{(0)}_{n}$ fulfils
\begin{equation}
  {{\rm sgn}[\rho^t_m [ \vartheta^{(0)}_{n}]]} w^{(0)}_{n,m}(\lambda)  = 
  (T\conL \mathcal{K}^{(0)})_m(\lambda).
\end{equation}
The density of charged moments in Eq.~\eqref{eq:densityn} can be obtained from
Eq.~\eqref{eq:generalFCS} by specialising it to the case 
${\rho_{{\rm st},j}=\rho_{\rm st}}$. In fact, as shown in Appendix~\ref{sec:tbaCM},
using 
\begin{equation}
  {{\rm sgn}[\rho^t_m [ \vartheta^{(u)}_{n}]]} =1,
\end{equation}
one can simplify it to a form analogous to Eqs.~(\ref{eq:GGEFCS}, \ref{eq:logx})
that does not involve an integral over $u$.

On the other hand, the slope in Eq.~\eqref{eq:slopen} can be written as
\begin{equation} \label{eq:fullslope}
  s_{\bm{\beta}} = s^{(\rm R)}_{\bm{\beta}} + s^{(\rm L)}_{\bm{\beta}}, 
\end{equation}
where we defined 
\begin{align}
s^{(\rm L)}_{\bm{\beta}}  &=  \lim_{t \to\infty}\frac{1}{t} \tr[\prod_{j=1}^n e^{\beta_j {\tilde Q}_t}  {\tilde{ \rho}}_{{\rm st},t}{(\beta_j,0)} ], \\
s^{(\rm R)}_{\bm{\beta}} &= \lim_{t \to\infty}\frac{1}{t} \tr[\prod_{j=1}^n {\tilde \rho}_{{\rm st},t}{(0,\beta_j)} e^{- \beta_j {\tilde Q}_t}]\!.
\end{align}
These two contributions can be obtained from
Eqs.~\eqref{eq:deqn}--\eqref{eq:logxn} by swapping space and time with the help
of Eqs.~\eqref{eq:ptilde} and~\eqref{eq:epsilontilde}, and using the filling
functions defined in Eq.~\eqref{eq:thetatilde}. The final result reads as
follows 
\begin{equation}
  \label{eq:swappeddeqn}
  \!\!\!\!\begin{aligned}
    s^{(\rm r)}_{\bm{\beta}}&=
    \int\limits_0^{\pm\beta}
      {\rm d} u
    \sum_{m}\!\!\int\!\frac{\mathrm{d}\lambda}{2\pi}
    \varepsilon^{\prime}_m(\lambda) 
    \tilde{\vartheta}^{(\mathrm{r},u)}_{n,m}(\lambda) 
   q_{\mathrm{eff},m}[
    \tilde{\vartheta}^{(\mathrm{r},u)}_{n}](\lambda)\\
    &+ \sum_m\int \frac{{\rm d}\lambda}{2\pi} \varepsilon'_m(\lambda)
    \mathcal{L}_{n,m}^{(\mathrm{r},0)}(\lambda),
  \end{aligned}
\end{equation}
with the choice $+\beta$, $-\beta$ in the integration limit corresponding to
$\mathrm{r}=\mathrm{L}$, $\mathrm{r}=\mathrm{R}$ respectively. We also
introduced
\begin{align}     
  &\mkern-8mu \mathcal{L}_{n, m}^{(\mathrm{r},u)} 
  \!=\!{\rm sgn}[\rho^t_m v_m [\tilde{\vartheta}^{(\mathrm{r},u)}_{n}]] \log\!\!\left[\!\frac{\prod_{j=1}^n 
  \eta^{(\mathrm{r})}_{j,m}
\!+\! e^{-\tilde w^{(u)}_{n,m}}}
{\prod_{j=1}^n (1+\eta^{(\mathrm{r})}_{j,m})}\!\right]\!\!,\label{eq:swappedKn}\\
  &  \mkern-8mu \tilde \vartheta^{(\mathrm{r},u)}_{n,m} = 
\frac{1}{\left[\prod_{j=1}^n \eta^{(\mathrm{r})}_{j,m}\right]
  e^{\tilde w^{(u)}_{n,m}}  +1},
    \label{eq:swappedthetan}\\
  & \mkern-8mu \partial_u \tilde w^{(u)}_{n,m}  =
    - {\rm sgn}[\rho^t_m v_m [\tilde{\vartheta}^{(\mathrm{r},u)}_{n}]] q_{\mathrm{eff},m}[\tilde{\vartheta}^{(\mathrm{r},u)}_{n}],
    \label{eq:swappedlogxn}
\end{align}
where $\tilde w^{(0)}_{n}$ fulfils 
\begin{equation} \label{eq:logy0}
 {\rm sgn}[\rho^t_m v_m [\tilde{\vartheta}^{(\mathrm{r},u)}_{n}]]
  \tilde w^{(0)}_{n,m}(\lambda) =
  (T\conL \mathcal{L}^{(\mathrm{r},0)})_m(\lambda),
\end{equation}
and  $\eta_{j}^{(\rm r)}$ are the eta functions of 
${\tilde{ \rho}}_{{\rm st},t}{(\beta_j,0)}$ and
${\tilde{ \rho}}_{{\rm st},t}{(0,\beta_j)}$
respectively obtained from~\eqref{eq:thetatilde},
\begin{equation} \label{eq:etajmr}
  \eta_{j,m}^{(\rm r)} = 
  \begin{cases}
    \eta_{m}(0,\beta_j), & {\rm r}= {\rm L},\\
    \eta_{m}(\beta_j,0), & {\rm r}= {\rm R}.
  \end{cases}
\end{equation}
As in the case of Eqs.~\eqref{eq:swapdeq2}--\eqref{eq:swaplogx2}, the $u$
integral in these equations can be analytically performed only when the
condition \eqref{eq:signcondition} holds. See Appendix~\ref{sec:simplification}
for the simplified expressions applying in the latter case. 

Finally, using Eq.~\eqref{eq:FCSpsi} we find the following TBA prediction for
the prefactor in Eq.~\eqref{eq:lambdan}
\begin{equation} \label{eq:LambdaTBA}
\log \Lambda_{\bm{\beta}} = \sum_{j=1}^n \sum_m \int \!{\rm d}\lambda 
  \frac{p'_m(\lambda)}{4\pi}\mathcal{K}_{m}^{(2\beta_j)}(\lambda)\,,
\end{equation}
where $\mathcal{K}^{(\beta)}$ is defined in Eq.~\eqref{eq:logx}. 

\section{Numerical and Analytical Tests}
\label{sec:tests}
In this section we perform explicit checks on our predictions for charged moments in TBA integrable models.  Specifically, in Sec.~\ref{sec:FF} we compare them against exact analytical results in free theories and in Sec.~\ref{sec:54} against exact results in an interacting, yet analytically tractable, integrable model: the quantum cellular automaton Rule 54. Lastly, in Sec.~\ref{sec:XXZ} we compare them against numerical simulations in the XXZ model. 

\subsection{Free Theories}
\label{sec:FF}
To begin we test our prediction in a free fermionic model by comparing with an
explicit calculation of the FCS, $Z_{\beta}(A,t)$.
Specifically we consider the system described by the Hamiltonian 
\begin{eqnarray}
  H=\sum_p \epsilon(p)c^\dag_p c_p,
\end{eqnarray}
where $c^\dag_p$ and $c_p$ are canonical fermionic creation and annihilation
operators 
and $\epsilon(p)$ is the  single particle energy.  The model has a $U(1)$
charge, the fermion number  
\begin{equation}
  N=\sum_p c^\dag_p c_p,
\end{equation}
whose FCS we calculate. In this system there is only a single quasiparticle
species and, for simplicity, we take its rapidity to coincide with the
momentum, $\lambda=p$, and denote the velocity by $v=\epsilon'$.
Moreover since there are no interactions the scattering kernel vanishes:
$T_{ml}=0$.  As a result, there is no dressing of quasiparticle properties, which
in particular implies that the sign of the velocity does not change, and  the $u$-integral
in Eq.~\eqref{eq:swappeddeqn} can be explicitly performed 
(cf.\ Appendix~\ref{sec:simplification}).

We shall quench the system from an initial state which is not an eigenstate of
$N$, namely the squeezed state
\begin{eqnarray}
  \ket{\Psi_0}=e^{\sum_{p>0} K(p) c^\dag_p c^\dag_{-p}}\ket{0}
\end{eqnarray}
where $\ket{0}$ is the vacuum state $c_p\ket{0}=0$.  $K(p)$ is some arbitrary, odd, real-valued function whose explicit form is unimportant for what follows.  We note that this type of initial state is the free fermionic version of an integrable initial state in interacting integrable field theories~\cite{ghoshal1994boundary}, and, therefore, it is a natural choice here. This setup is the minimal one in which to check the predictions of the space-time duality approach (higher order charged moments and multiple $U(1)$ symmetries 
can be included through generalisations of the calculations below). We proceed by first formulating the prediction explicitly and then comparing it to an alternative calculation. 

 The occupation function is a conserved quantity and it can be straightforwardly calculated in the initial state 
\begin{eqnarray}
  \vartheta(p)=\frac{\bra{\Psi_0}c^\dag_p c_p\ket{\Psi_0}}{\braket{\Psi_0}{\Psi_0}}=\frac{K^2(p)}{1+K^2(p)}. 
\end{eqnarray}
For the equilibrium regime this is the only information required and upon inserting this into~(\ref{eq:GGEFCS}, \ref{eq:logx}) we have that $\log x^{(\beta)}=-\beta$ and accordingly 
\begin{eqnarray}\label{eq:FFdensity}
d_\beta=\int \frac{{\rm d}p}{2\pi}\log{\left[1-\vartheta(p)+\vartheta(p)e^{\beta}\right]}.
\end{eqnarray}
In the non-equilibrium regime we require the spacetime-swapped occupation functions, which we obtain from~\eqref{eq:swappedthetan}
\begin{align}
\tilde{\vartheta}^{(\rm{L}, \beta)}(p) &=
  \vartheta(p)\Theta(v)+\frac{\vartheta(p)\Theta(-v)}{\vartheta(p)+(1-\vartheta(p))e^{-2\beta}},\\
\tilde{\vartheta}^{(\rm{R}, \beta)}(p)&=
  \vartheta(p)\Theta(-v)+\frac{\vartheta(p)\Theta(v)}{\vartheta(p)+(1-\vartheta(p))e^{-2\beta}}.
\end{align}
Additionally, Eqs.~(\ref{eq:swappedlogxn}, \ref{eq:logy0}) give
${w^{(\beta)}_m=-\beta {\rm sgn}[\epsilon']}$. Combining these together
(cf.\ Eqs.~\eqref{eq:fullslope} and \eqref{eq:swappeddeqn}) we have 
\begin{equation}\label{eq:FFslope}
  s_{\beta}=\int \frac{{\rm d}p}{2\pi}|\epsilon'| \log\left[
    \frac{(1-\vartheta(p)+\vartheta(p)e^\beta)^2}{1-\vartheta(p)+\vartheta(p)e^{2\beta}}
    \right].
\end{equation}
Lastly we need the initial value which using~\eqref{eq:LambdaTBA} is
\begin{equation}\label{eq:FFintial}
  \Lambda_\beta=\int \frac{{\rm d}p}{4\pi}
  \log{\left[1-\vartheta(p)+\vartheta(p)e^{2\beta}\right]}. 
\end{equation}
Equations \eqref{eq:FFdensity} and (\ref{eq:FFslope}, \ref{eq:FFintial}) form
the space-time duality prediction for the FCS in the equilibrium and
non-equilibrium regimes respectively. 

To test these we should calculate the FCS using an alternative method. The
noninteracting nature of the problem and the particular initial state we have
chosen facilitate this.  We use the fact that both the operator $e^{\beta N_A}$
and the initial state are Gaussian, which allows us to calculate charged
moments using the two point correlation function along with the algebra of
Gaussian matrices and the multidimensional stationary phase
approximation~\cite{fagotti2008evolution, fagotti2010entanglement}.  Leaving
the details to Appendix~\ref{sec:FFfcs} the final result for the full time
dynamics  is 
\begin{equation} \label{eq:FFfcs}
  \begin{split}
    \log \frac{Z_{ \beta}(A,t)}{|A|}=
    \int\!\frac{{\rm d}p}{4\pi}\log{\left[1-\vartheta(p)+\vartheta(p) e^{2\beta }\right]}\phantom{.}
    \\
    +\!\int\!\frac{{\rm d}p}{4\pi}\text{min}(1,2|\epsilon'|\zeta)
    \log{\left[\frac{(1-\vartheta(p)+\vartheta(p) e^{\beta })^2}{1-\vartheta(p)+\vartheta(p) e^{2\beta }}\right]},  
  \end{split}
\end{equation}
where $\zeta=t/|A|$.  Taking the two limits $|A|, t\to \infty$ in different
orders gives us the expressions in Eqs.~\eqref{eq:FFdensity} and
(\ref{eq:FFslope}, \ref{eq:FFintial}) confirming our prediction in the case of
free models.


\subsection{Rule 54}
\label{sec:54}
The second non-trivial check of our formula is the comparison with an exact
result obtained for a deterministic cellular automaton ``Rule
54''~\cite{bobenko1993two}.  This is an interacting TBA integrable
model~\cite{friedman2019integrable,gombor2024integrable}, which is simple
enough to allow for a number of exact results on non-equilibrium
quantities~\cite{prosen2016integrability,prosen2017exact,gopalakrishnan2018operator,gopalakrishnan2018hydrodynamics,inoue2018two,friedman2019integrable,buca2019exact,klobas2019time,alba2019entanglement,klobas2020matrix,klobas2020space,klobas2021exact,klobas2021exactrelaxation,klobas2021entanglement}
(see also a recent review~\cite{buca2021rule}). In particular,
Refs.~\cite{klobas2021exact,klobas2021exactrelaxation} introduced a family of
solvable initial states, from which the dynamics of local
observables~\cite{klobas2021exactrelaxation}, and
entanglement~\cite{klobas2021entanglement}, can be exactly described.

As shown in Ref.~\cite{klobas2023inprep}, also the dynamics of charged
moments can be calculated exactly for two simple charges: the total number of
particles, and the particle current. Here we will focus on the former, since
solvable initial states, parametrised by $\vartheta\in(0,1)$, are not symmetric
under it. For this case Ref.~\cite{klobas2023inprep} gives the slope
$s_{\bm{\beta}}=s_{\bm{\beta}}^{(\rm L)}+s_{\bm{\beta}}^{(\rm R)}$ as
\begin{equation} \label{eq:SolR541}
  s_{\bm{\beta}}^{(\rm L)}=
  s_{\bm{\beta}}^{(\rm R)}=\log \lambda_{\bm{\beta}} - \log \Lambda_{\bm{\beta}},
\end{equation}
where $\lambda_{\bm{\beta}}$ is the largest solution to a cubic equation,
\begin{equation} \label{eq:SolR542}
  \lambda_{\bm{\beta}}^3=(\lambda_{\bm{\beta}}(1-\vartheta)^n+\vartheta^n
  e^{\beta} \Lambda_{\bm{\beta}})^2,
\end{equation}
and $\Lambda_{\bm{\beta}}$ is
\begin{equation} \label{eq:SolR543}
  \Lambda_{\bm{\beta}}=\prod_{j=1}^n (1-\vartheta+\mathrm{e}^{2\beta_j}\vartheta).
\end{equation}
Note that we again use the shorthand notation
\begin{equation}
  \beta=\sum_{j=1}^{n}\beta_j.
\end{equation}

To compare this result with the prediction, we first note that the stationary
state reached by this quench obeys the TBA
description introduced in~\cite{friedman2019integrable}: the model exhibits two
species of quasiparticles, denoted by a subscript $\mu\in\{+,-\}$, and there is no
rapidity-dependence, 
\begin{equation}
  \varepsilon^{\prime}_\mu=\mu,\qquad p^{\prime}_\mu=1,\qquad
  T_{\mu\nu}=\mu\nu.
\end{equation}
The dressed velocities in a state with filling functions $\vartheta_{\pm}$ are
\begin{equation}
  v_{\pm}=\pm\frac{1}{1+2\vartheta_{\mp}},
\end{equation}
which in particular means that the sign of the dressed velocity is equal to $\nu$ regardless
of the underlying state. As a result, $\mathrm{sgn}[\rho^t_m\! v_m[\tilde{\vartheta}^{(\beta)}_m]]$ appearing in the conjectured formula has 
a trivial $\beta$-dependence,
\begin{equation}
  \mathrm{sgn}[\rho_{\nu}^{t}v_{\nu}[\tilde{\vartheta}^{(\mathrm{r},\beta)}]]=\nu.
\end{equation}
This implies  that in Eqs.~(\ref{eq:swappeddeqn},\ref{eq:swappedlogxn}) the
integral over $u$ can be performed explicitly (cf.\
Appendix~\ref{sec:simplification}), and we can start with
expressions~\eqref{eq:deqswappedSimpl1} --- \eqref{eq:LeqswappedSimpl1}.
Specializing them to the TBA description of Rule 54 we obtain
\begin{align}
  \label{eq:preR541}
  s_{\bm{\beta}}^{(\rm r)}&=
  \mathcal{L}_{n,+}^{(\mathrm{r}, \beta)}-\mathcal{L}_{n,-}^{(\mathrm{r}, \beta)},\\
  \label{eq:preR542}
  \mathcal{L}_{n,\nu}^{(\mathrm{r}, \beta)}&=
  \nu \log\left[
    \frac{\prod_{j=1}^{n}\eta_{j,\nu}^{(\mathrm{r})} 
    + e^{- \tilde{w}_{n,\nu}^{(\mathrm{r},\beta)}}}
    {\prod_{j=1}^n(1+\eta_{j,\nu}^{(\mathrm{r})})}
    \right],\\
  \label{eq:preR543}
  \mkern-8mu \tilde{w}_{n,\nu}^{(\mathrm{r},\beta)}
  &=
  \mp \nu\beta + 
  \mathcal{L}_{n,+}^{(\mathrm{r},\beta)} -\mathcal{L}_{n,-}^{(\mathrm{r},\beta)},
\end{align}
where $\mathrm{r}\in\{\mathrm{L},\mathrm{R}\}$ denotes the contribution of the
left or right edge, and $-\beta$, $+\beta$ correspond to
taking $\mathrm{r}=\mathrm{L}$ and $\mathrm{r}=\mathrm{R}$ respectively. Here
we also used that the charge $Q$ is the total number of particles, which gives
\begin{equation}
  q_{+}=q_{-}=1.
\end{equation}
Combining~\eqref{eq:preR541}--\eqref{eq:preR543} gives us a general nonlinear
equation for the slope with (so far) unspecified filling functions,
\begin{equation} \label{eq:preR54unspec}
  \begin{aligned}
  s_{\bm{\beta}}^{(\mathrm{r})}&=\log\left[
    \frac{
      e^{\mp\beta}\prod_{j=1}^{n}
    \eta_{j,+}^{(\mathrm{r})} +e^{-s_{\bm{\beta}}^{(\mathrm{r})}}}
    {\prod_{j=1}^{n}
    (1+\eta_{j,+}^{(\mathrm{r})})}\right]\\
  &+\log\left[\frac{e^{\pm\beta} \prod_{j=1}^{n}
     \eta_{j,-}^{(\mathrm{r})}
  +e^{-s_{\bm{\beta}}^{(\mathrm{r})}}}
      {\prod_{j=1}^{n}
    (1+\eta_{j,-}^{(\mathrm{r})})}
  \right],
  \end{aligned}
\end{equation}
where (as in~\eqref{eq:preR543}) for $\mathrm{r}=\mathrm{L}$ we select the top sign (i.e., $-\beta$ and $+\beta$ respectively), while for
$\mathrm{r}=\mathrm{R}$ we take the bottom one ($+\beta$
and $-\beta$ respectively).

To finally connect this expression with the exact result, we need to specify
$\eta_{j,\nu}^{(\mathrm{r})}$. For the quench protocol under consideration, the
stationary filling functions read as~\cite{klobas2021exactrelaxation,klobas2021entanglement} 
\begin{equation}
  \vartheta_{+}=\vartheta_{-}=\vartheta,
\end{equation}
which lead to the following non-interacting form for the quantities in
Eq.~\eqref{eq:logx} 
\begin{equation}
  x_{\nu}^{(u)}=e^{-u},\quad
  \mathcal{K}_{\nu}^{(u)}=(1-\vartheta)+\vartheta e^{u},\quad
  \nu\in\{+,-\}.
\end{equation}
Eq.~\eqref{eq:etajmr} then gives
\begin{equation}
  \mkern-8mu
  \eta_{j,+}^{(\mathrm{L})}=\eta_{j,-}^{(\mathrm{R})}=\frac{1-\vartheta}{\vartheta},\quad
  \eta_{j,-}^{(\mathrm{L})}=\eta_{j,+}^{(\mathrm{R})}=
  e^{-2 \beta_j}\frac{1-\vartheta}{\vartheta}.
\end{equation}
Inserting these into Eq.~\eqref{eq:preR54unspec} yields the following
equation for $s_{\bm{\beta}}^{(\mathrm{r})}$,
\begin{equation}
  s_{\bm{\beta}}^{(\mathrm{r})}=
  2\log\left[(1-\vartheta)^n+\vartheta^n e^{\beta}e^{-s_{\bm{\beta}}^{(\mathrm{r})}}
  \right]-\log\Lambda_{\bm{\beta}},
\end{equation}
which precisely reproduces the exact expression in
Eqs.~\eqref{eq:SolR541}--\eqref{eq:SolR543}.
\subsection{XXZ}
\label{sec:XXZ}

\begin{figure*}[t!]
  \includegraphics[width=0.95\textwidth]{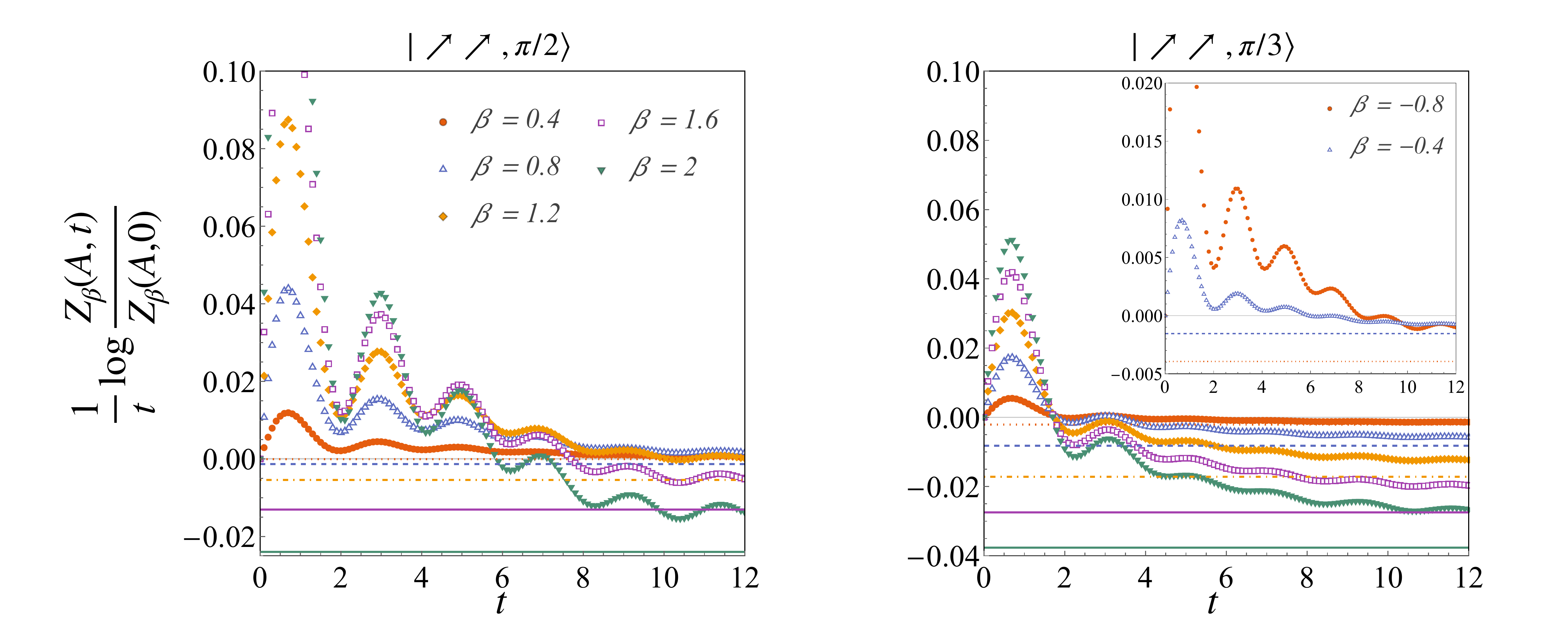}
  \caption{\label{fig:Z1_ferro} 
  Logarithmic slope of the FCS $Z_{\beta}(A,t)$ after a quench in the XXZ model with $\Delta=4$, starting from the tilted ferromagnetic state as in Eq.~\eqref{eq:tilted_ferro}.  Symbols are the iTEBD data computed with $|A|=50$, straight lines are the asymptotic predictions. 
  }
\end{figure*}

In our final round of checks we consider a paradigmatic example of interacting integrable model:  the anisotropic XXZ spin-$1/2$ chain. Having already performed explicit analytic checks in the preceding sections, here we shall instead compare the space-time duality prediction to exact numerics using matrix product state (MPS) based algorithms.  Once again we consider the dynamics of the model when quenched from an integrable initial state.  	

The Hamiltonian is given by 
\begin{eqnarray}
  H=\sum_{j=1}^{2L}\sigma^x_{j}\sigma^x_{j+1}+\sigma^y_{j}\sigma^y_{j+1}+\Delta\sigma^z_{j}\sigma^z_{j+1}
\end{eqnarray}
where $\sigma_j^{x,y,z}$ are spin-$1/2$ operators acting on site $j$,  $\Delta$ is the anisotropy parameter, which we set to be $>1$, and we assume periodic boundary conditions $\sigma^{x,y,z}_{2L+1}={\sigma}^{x,y,z}_1$.  Our $U(1)$ charge will be the $z$-component  of the spin 
\be
Q_A=S^z_A=\sum_{j\in A}\sigma_j^z.
\ee 
A simple family of integrable initial states for this model takes the form of two-site product states.  We focus on states which are not eigenstates of $S^z$ namely the tilted ferromagnetic state $\ket{\nearrow\nearrow, \theta}$ and the tilted N\'eel state $\ket{\nearrow\swarrow, \theta}$,
\begin{eqnarray}
  \ket{\nearrow\nearrow, \theta}&=&e^{i \frac{\theta}{2}\sum_{j=1}^{2L}{\sigma}^x_j}\otimes_{j=L}\ket{\uparrow}_{2j-1}\ket{\uparrow}_{2j},\label{eq:tilted_ferro}\\
  \ket{\nearrow\swarrow, \theta}&=&e^{i \frac{\theta}{2}\sum_{j=1}^{2L}{\sigma}^x_j}\otimes_{j=L}\ket{\uparrow}_{2j-1}\ket{\downarrow}_{2j}.\label{eq:tilted_neel}
\end{eqnarray}
For vanishing tilt the former state becomes stationary while the latter has been considered already in~\cite{bertini2023nonequilibrium}. Being integrable initial states the long time steady state can be determined exactly in terms of its occupation functions $\vartheta_m(\lambda)$.  Moreover in these cases exact analytic expressions are available to describe not only the occupation functions but also the rapidity distributions $\rho_m(\lambda), ~\rho^h_m(\lambda)$. 

The spectrum of the model consists of an infinite number of stable quasiparticle types (also known as strings)  labeled by the index $m\in\mathbb{N}$ and characterized by a rapidity $\lambda\in[-\pi,\pi]$.  Their single particle energy,  momentum and magnetization are expressed through the set of functions
\begin{eqnarray}
  {{a}}_m(\lambda)=\frac{1}{\pi}\frac{\sinh{(m \gamma)}}{\cos{(2\lambda)}-\cosh{(m \eta)}}
\end{eqnarray}
where we have introduced the parameter $\gamma=\text{acosh}{(\Delta)}$.  In terms of these we have that the energy and momentum are
\begin{eqnarray}
  \epsilon_m(\lambda)=-\pi \sinh{(\gamma)}{{a}}_m(\lambda), ~p_m'(\lambda)=2\pi {{a}}_m(\lambda)
\end{eqnarray}
and also the magnetization is $q_m=m$. The scattering kernel is an even function of the rapidity difference $T_{nm}(\lambda,\mu) \to T_{nm}(\lambda-\mu)$, and is  symmetric in the species index $T_{mn}(\lambda)=T_{nm}(\lambda)$.  For $m\geq n$ it is given by 
\begin{equation}
  \!\!\!T_{mn}(\lambda)=2\sum_{l=1}^{n-1}{a}_{|m-n|+2l}(\lambda)+{a}_{m+n}(\lambda)+{a}_{m-n}(\lambda).
\end{equation}
Inserting these expressions  into Eqs.~\eqref{eq:deqn} and~\eqref{eq:swappeddeqn} we obtain the result for the charged moments in the gapped XXZ.  The resulting coupled integral equations can then integrated numerically by truncating the system at a large but finite string number and then proceeding using an iterative Fourier transform scheme. 

\begin{figure*}[t!]
  \includegraphics[width=0.95\textwidth]{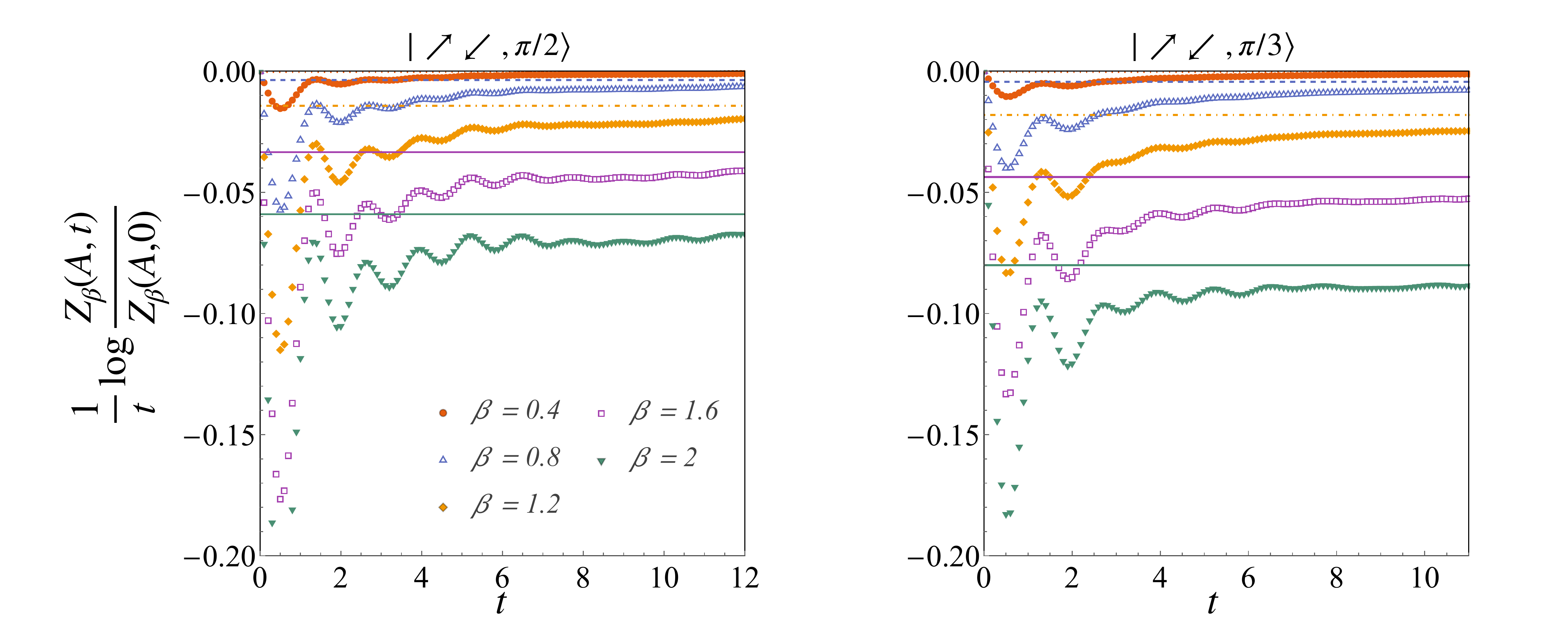}
  \caption{\label{fig:Z1_neel} 
  Same as in Fig.~\eqref{fig:Z1_ferro} after a quench starting from the tilted N\'eel state as in Eq.~\eqref{eq:tilted_neel}.
  }
\end{figure*}
 
To compare with the TBA results we perform numerical simulations of the quench dynamics in the system using Tensor-Network based algorithms. The infinite Matrix Product State (iMPS) representation of the evolved state allows an easy diagonal representation of the semi-infinite reduced density matrix. 
From that representation, it is easy to compute 
\be
F_{\bm{\beta}}(A,t)\equiv\tr[\prod_{j=1}^n 
 \left( e^{\beta_j  S^{z}_A} \rho_{[0,\infty]}(t)\right)], 
\ee   
with reasonable accuracy. The asymptotic behavior of this quantity can be easily related to the correspondent {charged moments} $Z_{\bm{\beta}}(A,t)$ in Eq.~(\ref{eq:chargedmoments}) (see the supplemental material of Ref.~\cite{bertini2023nonequilibrium}).
By increasing the auxiliary dimension of our simulations up to $\chi_{max} = 1024$, we are able to reach a maximum time $t_{max} \simeq 12$. In Figs.~\ref{fig:Z1_ferro} and \ref{fig:Z1_neel} we present the time evolution of the full counting statistics for quenches in the XXZ model toward the gapped phase ($\Delta=4$) starting from tilted ferromagnetic or N\'eel state. 
Note that, in the case of initial tilted antiferromagnetic states (\ref{eq:tilted_neel}) and whenever the subsystem $A$ contains an even number of lattice site, thanks to the symmetry under $\prod_{j}\sigma^{x}_j$, the FCS is an even function of $\beta$, i. e. $Z_{-\beta}(A,t) = Z_{\beta}(A,t)$. 
This is not the case in general for the initial ferromagnetic tilted states (\ref{eq:tilted_ferro}), except for $\theta = \pi/2$ since $\ket{\nearrow\nearrow, \pi/2}$
is eigenstate of $\prod_{j}\sigma^{y}_{j}$.


Therefore, for the specific case $\theta = \pi/3$, to highlight the breaking of the $\beta\to-\beta$ invariance, we also show some representative negative values of $\beta$ in the inset of the right panel of Fig.~\ref{fig:Z1_ferro}. In general, the time-dependent logarithmic slope is approaching the predicted stationary value sooner for $|\beta|$ smaller. Moreover, the quenches from the N\'eel state are relaxing  relatively faster than those from the ferromagnetic states. 

\begin{figure*}[t!]
  \includegraphics[width=0.95\textwidth]{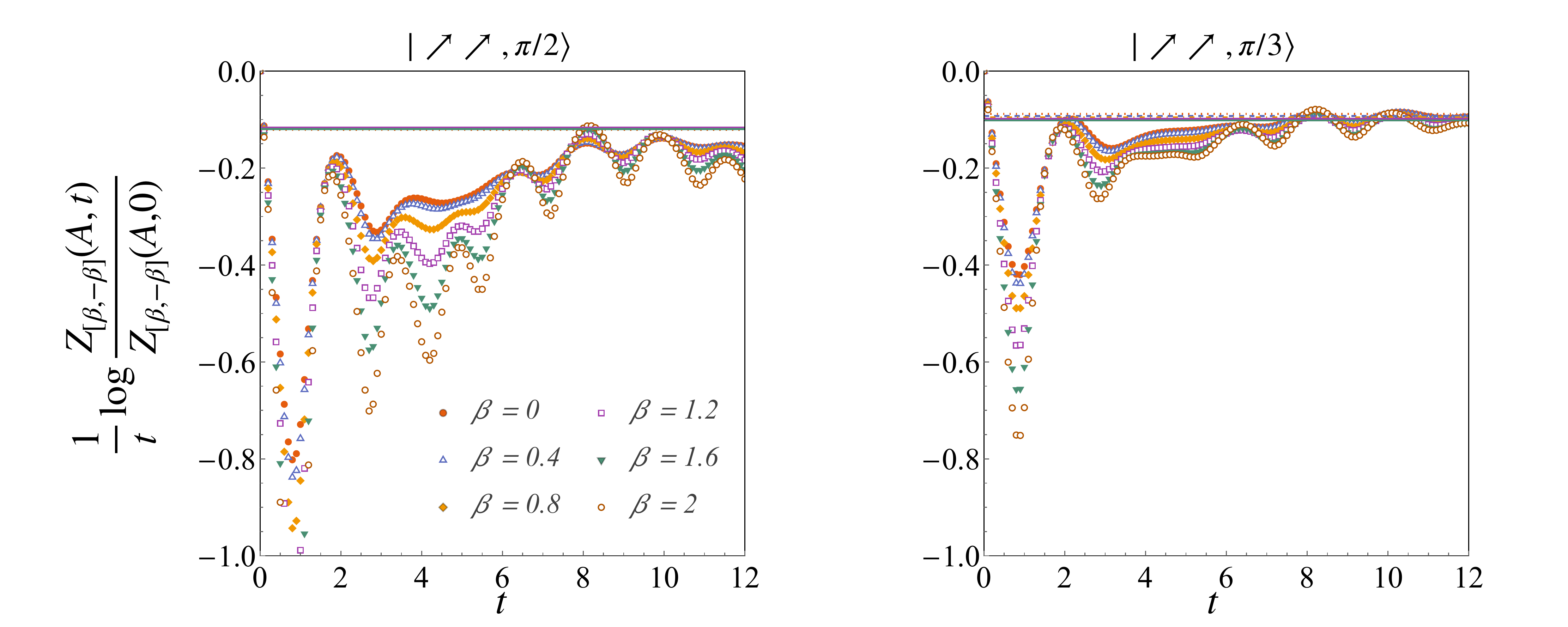}
  \caption{\label{fig:Z2_ferro} 
  Logarithmic slope of the second charged moment $Z_{[\beta,-\beta]}(A,t)$ after a quench in the XXZ model with $\Delta=4$, starting from the tilted ferromagnetic state as in Eq.~\eqref{eq:tilted_ferro}.  Symbols are the iTEBD data computed with $|A|=50$, straight lines are the asymptotic predictions. 
  }
\end{figure*}

\begin{figure*}[t!]
  \includegraphics[width=0.95\textwidth]{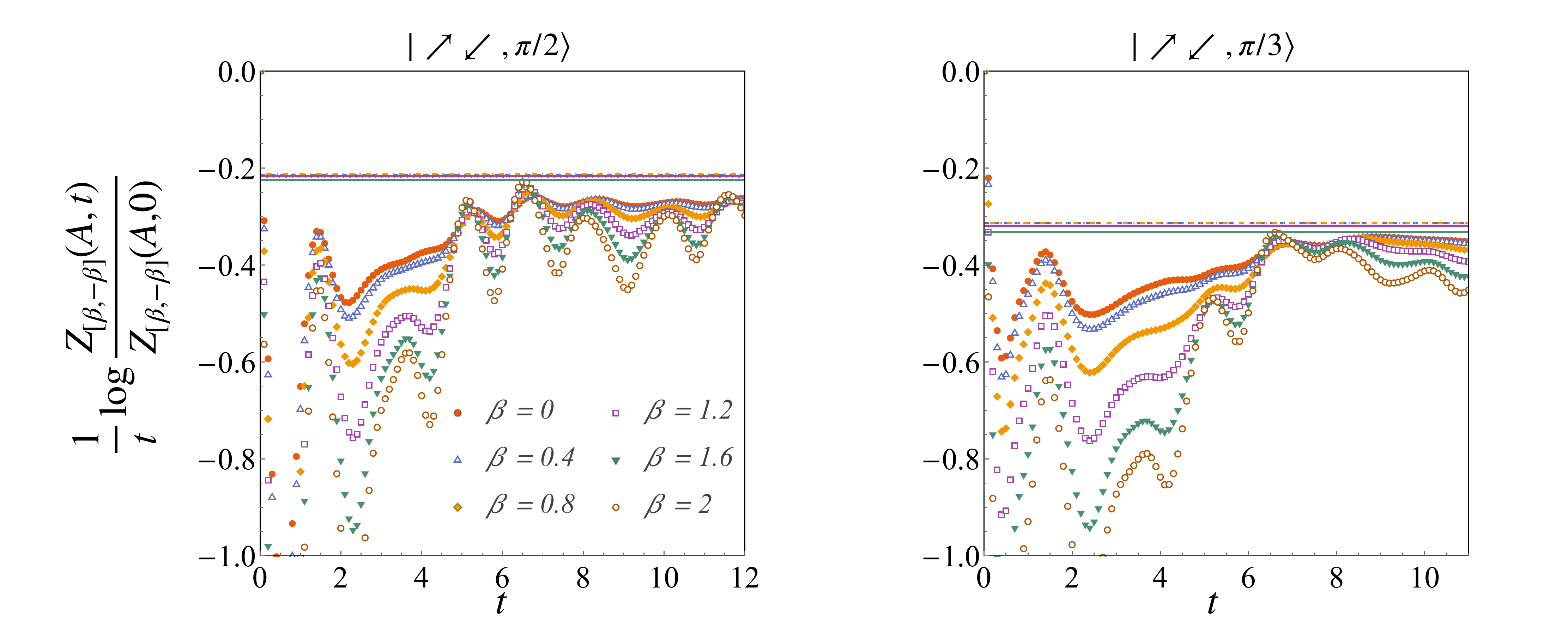}
  \caption{\label{fig:Z2_neel} 
  Same as in Fig.~\eqref{fig:Z2_ferro} after a quench starting from the tilted N\'eel state as in Eq.~\eqref{eq:tilted_neel}.
  }
\end{figure*}

We went beyond the simple full counting statistics by evaluating the second charged moment with
$\bm{\beta} = [\beta,-\beta]$. In Figs.~\ref{fig:Z2_ferro} and \ref{fig:Z2_neel} we show some representative curves for the same quenches in the XXZ model as before. Even though the data are fair agreement with the asymptotic theoretical predictions, the accessible time-window is not sufficient to discriminate predictions for different values $\beta$.



\section{Applications} 
\label{sec:higherrenyi}

So far we have established that the space-time duality approach gives access to the dynamics of charged moments also in the presence of interactions and, furthermore, we presented closed-form expressions for TBA integrable models. In this section we apply these results to the calculation of two physically relevant quantities: the entanglement asymmetry (cf.~ Eq.~\eqref{eq:asymm}) and the charge probability distribution.

\subsection{Entanglement asymmetry}
As discussed in Sec.~\ref{subsec:observables}, the entanglement asymmetry is expressed in terms of the charged moments via a replica trick. In particular, to explicitly determine it, one has to evaluate the integral in Eq.~\eqref{eq:trrhobarn}. Here we compute the leading order behaviour of that integral by means of the saddle point approximation. We begin by recalling Eq.~\eqref{eq:dualityCM} and noting that, since at leading order ${\tilde{ \rho}}_{{\rm st},t}{(\beta_1,\beta_2)}$ commutes with $\tilde Q_t$, the quantity 
\be
\frac{1}{t}\log Z_{i\boldsymbol{\beta}}(A,t)=\frac{|A|}{t}\Lambda_{i\boldsymbol{\beta}}+s_{i\boldsymbol{\beta}},
\ee
is invariant under any permutation of $\beta_j$ and also any shift ${\beta_j\to \beta_j+ m\pi}$ for integer $m$~\footnote{To see this we use that the charges have integer spectrum and that the explicit expressions for $\Lambda_{\boldsymbol{\beta}}$~\eqref{eq:LambdaTBA} and $s^{(\text{r})}_{\boldsymbol{\beta}}$~\eqref{eq:swappeddeqn} in the case of the asymmetry depend only on $2\beta_j$. The last point follows from the form of~\eqref{eq:thetatilde}.  For $n>2$ it is possible that saddle points exist at values of $\beta_j=2\pi m/n$ for integer $m$.  In all cases we have considered these have negligible contribution but also are absent in the $n\to 1$ limit which is the main focus. }. Therefore, the saddle points of the $\beta_j$ integrals are fixed to occur at $\beta_j=0\,{\rm mod}(\pi)$ for all $j$.  After integrating over the delta function this leaves $2^{n-1}$ saddle points which all give equal contribution.  We can  explicitly check this by computing $\partial_{\beta_j}\log Z_{i\boldsymbol{\beta}}(A,t)$ using the fact that, after first fixing $\beta_n=-\sum_j^{n-1}\beta_j$ via the delta function, the space-time swapped occupation functions~\eqref{eq:swappedthetan} obey
\be
\partial_{\beta_j}\tilde{\vartheta}^{(\rm{r},0)}_{n,m}|_{\boldsymbol{\beta}=0}=0. 
\ee
Moreover, we also note that 
\be
\partial_{\beta_j}\partial_{\beta_k}\log Z_{i\boldsymbol{\beta}}(A,t)|_{\boldsymbol{\beta}=0}\!=\!\frac{1}{2}\partial^2_{\beta_j}\log Z_{i\boldsymbol{\beta}}(A,t)|_{\boldsymbol{\beta}=0},
\ee
which again results from the gauge fixing through the delta function. This means that the determinant of the Hessian matrix at the saddle points can be easily computed. Combining these facts along with the expression for the R\'enyi entropy~\eqref{eq:Renyi} and taking the replica limit we arrive at an exact expression for the entanglement asymmetry in the non-equilibrium regime. The result reads as   
\begin{equation}
\Delta S_A(t)=\frac{1}{2}+\frac{1}{2}\log \pi \chi(t),
\end{equation}
where $1 \ll t\ll |A|$, and we introduced 
\begin{align}
\chi(t)&=\lim_{n\to 1}\partial_{\beta_j}^2\log Z_{\boldsymbol{\beta}}(A,t)|_{\boldsymbol{\beta}=0}\\
  &=\sum_m\int {\rm d}\lambda \rho_m(1-\vartheta_m)q_{\mathrm{eff},m}^2\left(|A|-2t|v_m|\right)\!.
\label{eq:chit}
\end{align}
From this compact expression we can gain some intuition about the behaviour of the entanglement asymmetry.  The extensive term in~\eqref{eq:chit} can be recognized as the charge susceptibility in the initial state.  As per the remit of the asymmetry, we expect that this should be greater for states which are further from being symmetric. For example, charge eigenstates exhibit no fluctuations whereas approaching a condensation transition a system exhibits a divergent susceptibility. At finite times we see that~\eqref{eq:chit} decreases following the expectation that under time evolution the state becomes more symmetric.  Moreover, the rate at which this happens coincides with twice the Drude self weight~\cite{doyon2020lecture}. The latter is related to the fluctuations of the time integrated current associated to $Q_A$ at a specific point in space (more precisely it is its second cumulant). Thus the following picture of the entanglement asymmetry in the non-equilibrium regime emerges: The subsystem experiences charge fluctuations throughout its bulk and their strength, quantified through the susceptibility, characterise how far the system is from being symmetric. At finite times these fluctuations are reduced by the transport of charge through the boundaries of the subsystem with the rate at which this happens being determined by the fluctuations of the charge current at the boundaries given by the Drude self weight.  

The condition that $\Delta S_A(t)\geq 0$ sets a limit on the applicability of our calculation as do the requirements of the saddle point approximation.  Therefore, questions regarding symmetry restoration require us to go beyond the non-equilibrium regime and study the full dynamics which is beyond the scope of the current work but will be addressed elsewhere~\cite{rylands2023microscopic}. 

\subsection{Charge Probability Distribution}
Whilst the higher charged moments provide nuanced information on the physics of non-equilibrium systems the most immediately useful quantity remains the simplest of these,  the full counting statistics.  Its utility is evident from the fact that its Fourier transform is the charge probability distribution,
\begin{eqnarray}
P(q,t)=\int_{-\pi}^\pi\frac{{\rm d}\beta}{2 \pi}e^{-i \beta q}Z_{i\beta}(A,t)
\end{eqnarray}
which gives the probability that a measurement of $Q$ inside $A$ at time $t$ returns the value $q$.  We can calculate this integral also by the saddle point approximation, whereupon we find that the saddle point occurs at $\beta=-i\beta^*$, 
\begin{eqnarray}
q&=&\partial_{\beta}\left[\log\Lambda_{\beta}+ts_{\beta}\right]_{\beta=\beta^*}.
\end{eqnarray}
Expanding this for small $\beta^*$ we obtain the saddle point condition for charges close to the initial value,
\begin{eqnarray}
\Delta q=q-\left<Q_A\right>,~~\left<Q_A\right>=\partial_\beta\log\Lambda_\beta|_{\beta=0}. 
\end{eqnarray}
Using this as the saddle point we arrive at,
\begin{eqnarray}\label{eq:chargeprob}
P(\Delta q,t)\simeq\frac{1}{\sqrt{2\pi\mathcal{D}(t) }}e^{-\frac{\Delta q^2}{2\mathcal{D}(t)}}
\end{eqnarray}
where 
\begin{eqnarray}
\!\!\!\mathcal{D}(t)\!=\!2\sum_m\!\int\!\!{\rm d}\lambda\,q^2_{\text{eff},m}\rho_m(1-\vartheta_m) \!\left[|A|-t|v_{m}(\lambda)|\right]\!.
\end{eqnarray}
 Thus the charge probability distribution for small $\Delta q$ is approximately normal distributed with variance $\mathcal{D}(t)$. 

Some comments on this expression are in order.  First, unlike the case for the charge moments of a system quenched from a symmetric state there is no time delay~\cite{parez2021quasiparticle}.  In the symmetric case since there are initially no charge fluctuations, it takes a finite time for systems with a maximal velocity to build up the fluctuations of charge necessary for a nonzero charged moment.  In this case since the initial state is not symmetric, charge fluctuations are present at any time and no time delay occurs.  Second, the variance is similar in form to $\chi(t)$ which governs the entanglement asymmetry.  In the initial state the variance of the probability distribution is the charge susceptibility while at finite time the fluctuations should decrease via the spreading of charge through the boundaries which again is determined by the Drude self weight.  Note however the coefficients of these two terms differ from the asymmetry which results in a different long time behaviour.  Third,  while the expression for the asymmetry is exact in the thermodynamic limit, this is not the case for~\eqref{eq:chargeprob}. Although at small $\Delta q$, $P(\Delta q,t)$ is normal, there are corrections to this and all higher cumulants are non-vanishing leading to breakdown of ~\eqref{eq:chargeprob} for large deviations.  

\section{Conclusions}
\label{sec:conclusions}
Non-equilibrium quantum systems remain relatively poorly understood in comparison to systems which are at,  or close to, equilibrium.  In part this can be attributed to the inherently more complex nature of the former but also due the dearth of widely applicable techniques like those which can be used in the latter.  In this work we have presented, in great detail, the space-time duality approach to non-equilibrium systems.  This method facilitates a far greater understanding of such systems via a mapping of the non-equilibrium system to a dual one which is at \textit{equilibrium}.  Therefore one can apply the extensive toolkit and intuition of equilibrium physics to a non-equilibrium setting.   

We have concentrated in this work on a certain class of physically motivated quantities, the charged moments which allow one to study among other things the spreading of entanglement, the fluctuations of conserved $U(1)$ charges and the interplay between symmetry and relaxation to a steady state.  Using the setting of brickwork quantum circuits we have shown quite generally that the quench dynamics of the charged moments can be understood using the equilibrium properties of a dual system which evolves in the space-like rather than time-like direction.  Directly from this observation one can infer many non-trivial properties of the charge moments in general but also in certain systems derive explicit predictions for the dynamics. 

In Eqs.~\eqref{eq:deqn} and ~\eqref{eq:fullslope} we have presented explicit formulae for the dynamics of the charge moments in TBA integrable systems which are quenched from arbitrary integrable initial states. This extends previous work to encompass initial states which are not symmetric with respect to the charge and which has necessitated a bridging of the space-time duality approach with the theory of generalised hydrodynamics. These expressions represent the first complete and exact analytic analysis of the finite-time dynamics of charge fluctuations in the presence of interactions. We tested them against exact analytic results in the case of free models and the Rule 54 cellular automaton and against numerical simulations of the XXZ spin chain.  They were then used to study symmetry restoration in interacting models via the entanglement asymmetry for the first time.

While in this work we have concentrated on the charged moments for a $U(1)$ conserved charge in one dimension, the space-time duality approach is much more widely applicable.  Indeed, one could also consider the dynamics of non-Abelian charges using the same approach as an example. Moreover,  it is not too difficult to see that the same underlying reasoning could be applied to the calculation of quantities other than the charged moments.  In particular the study of the correlation functions after a quantum quench is an immediate prospect. In this respect, our work complements recent progress in ballistic macroscopic fluctuation theory~\cite{doyon2023ballistic} by providing a general theory to understand many-body systems out of equilibrium. 

\begin{acknowledgments}
We thank {\v Z}iga Krajnik, Enej Ilievski, and Toma{\v z} Prosen for insightful discussions. B.B.\ and K.K.\ thank SISSA for hospitality in the initial stage of the project. This work has been supported by the Royal Society through the University Research Fellowship No.\ 201101 (BB), by the Leverhulme Trust through the Early Career Fellowship No.\ ECF-2022-324 (KK), and by the ERC under Consolidator grant number 771536 NEMO (CR and PC).
\end{acknowledgments}

\appendix

\section{Proof of Eq.~\eqref{eq:chargecons}}\label{sec:proofOfCC}
We start by rewriting the conservation of $Q$, $\mathbb{U} Q \mathbb{U}^{\dagger} = Q$, as 
\begin{equation}\label{eq:appConservationExp}
  \mathbb{U}_{\mathrm{o}} e^{\beta Q} \mathbb{U}_{\mathrm{o}}^{\dagger}=
  \mathbb{U}_{\mathrm{e}}^{\dagger} e^{\beta Q} \mathbb{U}_{\mathrm{e}},
\end{equation}
which holds for any $\beta\in\mathbb{C}$. By performing partial trace of the above relation over all but two consecutive sites {$x$ and $x+1/2$}, we obtain
\begin{equation}\label{eq:applocalRelation}
    U e^{\beta q}\otimes e^{\beta q} U^{\dagger}=
  a\otimes b,
\end{equation}
where $a,b\in\mathrm{End}(\mathbb{C}^{d})$ are one-site operators
\begin{equation}
\begin{aligned}
    a&=\frac{1}{\tr[e^{\beta q}]}
  \tr_1(U^{\dagger} e^{\beta q} \otimes e^{\beta q} U),\\
    b&=\frac{1}{\tr[e^{\beta q}]}
    \tr_2(U^{\dagger} e^{\beta q} \otimes e^{\beta q} U),
  \end{aligned}
\end{equation}
{where $\tr_1$ and $\tr_2$ denote respectively traces over the first and second qudit. Note that that \eqref{eq:applocalRelation} follows because $\mathbb{U}_{\mathrm{o}}$ acts as a product over the bipartition $\{x,x+1/2\}\cup \overline{\{x,x+1/2\}}$ while $\mathbb{U}_{\mathrm{e}}$ does not.} 

Using now again~\eqref{eq:appConservationExp} with~\eqref{eq:applocalRelation}
we obtain 
\begin{equation}\label{eq:applocalRelation2}
    U^{\dagger} e^{\beta q}\otimes e^{\beta q} U= b\otimes a.
\end{equation}
The Hermiticity of $q$ implies that $e^{\beta q}$ is normal, and therefore -- using~\eqref{eq:applocalRelation},~\eqref{eq:applocalRelation2} -- also $a$ and $b$ are normal, which means that all the operators $e^{\beta q}$, $a$, and $b$ are diagonalizable (i.e., \emph{unitarily} similar to appropriate diagonal operators). The spectra of the l.h.s.\ and r.h.s.\ have to be the same,
\begin{equation}
  \mathrm{Spect}(e^{\beta q}\otimes e^{\beta q})=
  \mathrm{Spect}(a\otimes b),
\end{equation}
which --- up to a trivial rescaling by a constant --- can only be fulfilled if
\begin{equation}
  \mathrm{Spect}(e^{\beta q})=\mathrm{Spect}(a)=\mathrm{Spect(b)}.
\end{equation}
Thus, there exist unitary transformations $v,w\in\mathrm{End}(\mathbb{C}^{d})$ such that
\begin{equation}
  a = v e^{\beta q} v^{\dagger},\qquad b = w e^{\beta q} w^{\dagger}. 
\end{equation}
This allows us to introduce
\begin{equation}
  U^{(\mathrm{o})}=v^{\dagger}\otimes w^{\dagger}\, U,\qquad
  U^{(\mathrm{e})}=U\, w\otimes v,\\
\end{equation}
in terms of which~\eqref{eq:applocalRelation}, and~\eqref{eq:applocalRelation2} are rewritten as
\begin{equation}
  \begin{aligned}
    U^{\mathrm{(o)}} e^{\beta q}\otimes e^{\beta q}&=
    e^{\beta q}\otimes e^{\beta q}\, U^{\mathrm{(o)}},\\
    U^{\mathrm{(e)}} e^{\beta q}\otimes e^{\beta q}&=
    e^{\beta q}\otimes e^{\beta q}\, U^{\mathrm{(e)}},
  \end{aligned}
\end{equation}
which is a version of Eq.~\eqref{eq:chargecons} where even and odd time steps are implemented by different unitary gates.
The transformation
\begin{equation}
  \begin{aligned}
  \mathbb{U}_{\mathrm{e}}=U^{\otimes L} &\mapsto
  \left.U^{\mathrm{(e)}}\right.^{\otimes L},\\
  \mathbb{U}_{\mathrm{o}}=\Pi_{2L} U^{\otimes L} \Pi_{2L}^{\dagger} &\mapsto
  \Pi_{2L} \left.U^{\mathrm{(o)}}\right.^{\otimes L}\Pi_{2L}^{\dagger},
  \end{aligned}
\end{equation}
preserves the full time-evolution operator $\mathbb{U}$ and is therefore a gauge transformation. In other words, by redefining the local operator on even and odd time-steps as shown above, \emph{all} the local gates satisfy Eq.~\eqref{eq:chargecons}, and the full time-evolution is unchanged. 

In Sec.~\ref{sec:duality} we assume $U^{(\mathrm{e})}=U^{(\mathrm{o})}=U$ for simplicity. However, since the specific form of the gates is never used, we could repeat the full reasoning and arrive to the same result also taking $U^{(\mathrm{e})}\neq U^{(\mathrm{o})}$ (albeit with additional complications to the notation). Therefore we decided to restrict the discussion to the simplified case.

\section{Current of a $U(1)$ charge in a quantum circuit}
\label{sec:current}

To obtain the form of the current given in~\eqref{eq:appcurrent} we utilize the continuity equation for the charge on the space lattice which can be used as a defining equation for the current operator.  In quantum circuits,  using the definition of Heisenberg evolution~\eqref{eq:Heisenberg}, this is given by 
\begin{equation} \label{eq:continuity}
  \begin{aligned}
    &  q_x(t+1)+  q_{x+\tfrac{1}{2}}(t+1)-  q_x(t)-  q_{x+\tfrac{1}{2}}(t)\\
    &={ j}_{x+1}(t+\tfrac{1}{2})-{ j}_{x}(t+\tfrac{1}{2})+{ j}_{x+1}(t)-{ j}_x(t).
  \end{aligned}
\end{equation}
{Just like the standard continuity equation this equation is just a restatement of the conservation of the charge, which is obtained by summing \eqref{eq:continuity} over $x$. Note that there is a ``gauge ambiguity" associated to this process: one can write many equivalent continuity equations for a given conservation law~\cite{doyon2020lecture}. 

On the left hand side of \eqref{eq:continuity} we have the total change in charge on the sites $x$ and $x+1/2$ between time steps $t$ and $t+1$. On the right hand side we have the total time integrated current which passes into and out of these sites in one time step.  Note that on the right hand side the current operators are defined on time lattice at $t$ and $t+1/2$ and the boundaries in space are $x$ and $x+1$ while on the left hand side the situation is reversed which highlights the equal footing time and space have in a brickwork quantum circuit. 

Recalling now that Eq.~\eqref{eq:chargecons} (see also Appendix~\ref{sec:proofOfCC}) implies} 
\begin{gather}
  \mkern-8mu
  q_x(t+1)+ q_{x+\tfrac{1}{2}}(t+1) 
  =  q_x(t+\tfrac{1}{2})+ q_{x+\tfrac{1}{2}}(t+\tfrac{1}{2}),\mkern-8mu
  \label{eq:evenodd}\\
  \mkern-8mu q_{x+1}(t)+ q_{x+\tfrac{1}{2}}(t) 
  =  q_{x+1}(t+\tfrac{1}{2})+ q_{x+\tfrac{1}{2}}(t+\tfrac{1}{2}),\mkern-8mu
  \label{eq:oddeven}
\end{gather}
one can verify that the current operator fulfilling \eqref{eq:continuity} can
be written as in Eq.~\eqref{eq:appcurrent}.

{
\section{Proof Eq.~(\ref{eq:FCSasymptFINAL})}
\label{sec:proofEI}

In this appendix we derive explicitly Eq.~\eqref{eq:FCSasymptFINAL} from Eq.~\eqref{eq:defSpaceEvolution}. We begin by invoking the following Lemma  
\begin{lemma}
For all $x\geq 2t+1$ one has 
\be
 \tilde{\mathbb{W}}^x_{t,\beta} = \ketbra{R_{t,\beta}}{L_{t,\beta}}
 \label{eq:Widentity}
\ee
where we introduced the states in $\mathcal H_t \otimes \mathcal H_t$
\begin{align}
   &\bra{L_{t,\beta}}\!= \mathbb A_{0,\beta} \mathbb A_{1,\beta}\cdots \mathbb A_{t-1,\beta}\mathbb A_{t,\beta},\\
   &\ket{R_{t,\beta}}\!= \mathbb B_{t,\beta} \mathbb B_{t-1,\beta}\cdots \mathbb B_{1,\beta}\mathbb B_{0,\beta},
\end{align}
the rectangular matrices 
\begin{align}
&\mathbb A_{t\geq 1,\beta} = \frac{1}{d}\left[\sum_{s,r=0}^{d-1} \bra{s,r} \otimes \1^{\otimes (2t-1)} \otimes \bra{s,r}\right] \tilde{\mathbb{W}}_{t,\beta}\,, \label{eq:Adef}\\
&\mathbb A_{0,\beta} =\frac{1}{d} \left(\sum_{s=0}^{d-1} \bra{s,s}\right) \tilde{\mathbb{W}}_{0,\beta}\,,\\
&\mathbb B_{t\geq 1,\beta} = \frac{1}{d} \tilde{\mathbb{W}}_{t,\beta}  \left[\sum_{s,r=0}^{d-1} \ket{s,r} \otimes \1^{\otimes (2t-1)} \otimes \ket{s,r}\right]\,,\\
&\mathbb B_{0,\beta} = \tilde{\mathbb{W}}_{0,\beta}  \left(\sum_{s=0}^{d-1} \ket{s,s}\right)\!,\label{eq:B0def}
\end{align}
and set 
\begin{align}
 \tilde{\mathbb{W}}_{0,\beta} & = \frac{1}{\Lambda_\beta}\sum_{s,r=0}^{d-1} \!\!\ketbra{s}{r}\!\otimes\!\ketbra{s}{r} ((e^{\beta q^{T}} m e^{\beta q})\! \otimes\!
    m^*)\\
     &= \mathbb B_{0,\beta} \mathbb A_{0,\beta}\,. 
\end{align}
\end{lemma}
As per their definition $\mathbb A_{t,\beta}$ maps from $\mathcal H_{t-1} \otimes \mathcal H_{t-1}$ to $\mathcal H_t \otimes \mathcal H_t$ and $\mathbb B_{t,\beta}$ from $\mathcal H_{t} \otimes \mathcal H_{t}$ to $\mathcal H_{t-1} \otimes \mathcal H_{t-1}$. This lemma is readily proven graphically using the unitarity relations \eqref{eq:unitarity} (see e.g. the Supplemental material of Ref.~\cite{bertini2022entanglement}). In Sec.~\ref{sec:prooflemma} we present an equivalent algebraic proof. 

Substituting now Eq.~\eqref{eq:Widentity} in Eq.~\eqref{eq:defSpaceEvolution} we find
\begin{equation}\label{eq:FCSfactorisedMEs}
  \begin{aligned}
    Z_{\beta}(A,t)= & \Lambda_{\beta}^{|A|}
  \mel{L_{t,0}}{e^{\beta \tilde{Q}_t}\otimes \1}{R_{t,\beta}}\\
    &\times\!\!\!\mel{L_{t,\beta}}{e^{-\beta \tilde{Q}_t}\otimes \1}{R_{t,0}}, \quad  |A|\geq 2t,
  \end{aligned}
\end{equation}
where $\tilde{Q}_t$ is defined in Eq.~\eqref{eq:Qtilde}. Finally, we map the states $\bra{L_{t,\beta}},\ket{R_{t,\beta}}\in\mathcal H_t\otimes \mathcal H_t$ into operators $L_{t,\beta}, R_{t,\beta} \in {\rm End}(\mathcal H_t)$ using the correspondence   
\begin{align}
\mel{s_{2t+1} ... s_1}{R_{t,\beta}}{r_{2t+1} ... r_1} = \braket{s_{2t+1} ... s_1 r_1 \cdots r_{2t+1}}{R_{t,\beta}} \notag \\
\mel{s_{2t+1} ... s_1}{L_{t,\beta}}{r_{2t+1} ... r_1} = \braket{s_{2t+1} ... s_1 r_1 \cdots r_{2t+1}}{L_{t,\beta}} \notag 
\end{align}
to rewrite \eqref{eq:FCSfactorisedMEs} as
\be
\!\!\!Z_{\beta}(A,t)=  \Lambda_{\beta}^{|A|}
  {\rm tr}[e^{\beta \tilde{Q}_t} R^{\phantom{\dag}}_{t,\beta} L^\dag_{t,0} ] {\rm tr}[e^{-\beta \tilde{Q}_t} R^{\phantom{\dag}}_{t,0} L^\dag_{t,\beta}]. 
\ee
Finally, Eq.~\eqref{eq:FCSasymptFINAL} follows by observing
\be
L_{t,\beta} =l^{\dag}_{t, \beta}l^{\phantom{\dag}}_{t, 0}, \qquad R_{t,\beta} =r^{\phantom{\dag}}_{t, \beta}r^{\dag}_{t, 0}, 
\ee
where $l_{t, \beta}$ and $r_{t, \beta}$ are defined in Eq.~\eqref{eq:defRhot2}.

\subsection{Proof of the lemma}
\label{sec:prooflemma}

To prove the lemma we begin by observing that, because $U$ is unitary, $\tilde U$ fulfils  
\begin{align}
\label{eq:unitarityrewritten}
    &\left[\smashoperator[r]{\sum_{s=0}^{d-1}}
    \bra{s}\otimes \1^{\otimes x} \otimes \bra{s}\right] \tilde U \otimes O \otimes \widetilde{U^{\dag}} \left[\smashoperator[r]{\sum_{r=0}^{d-1}}
    \ket{r}\otimes \1^{\otimes x} \otimes \ket{r}\right] \notag\\
    &=  \smashoperator[r]{\sum_{s, r=0}^{d-1}}
    \ketbra{r}{s}\otimes O \otimes \ketbra{r}{s},
\end{align}
for any operator $O$ with support on $x\geq 0$ qudits. Next we note that using \eqref{eq:unitarityrewritten} and the definitions \eqref{eq:WmatrixFormula} and \eqref{eq:Adef}--\eqref{eq:B0def} of $\tilde{\mathbb{W}}_{t,\beta},\mathbb A_{t,\beta}$ and $\mathbb B_{t,\beta}$ one can readily establish the following relations
\begin{align}
  \mathbb A_{t,\beta}\tilde{\mathbb{W}}_{t,\beta} &=\tilde{\mathbb{W}}_{t-1,\beta} \mathbb A_{t,\beta}, \label{eq:AW}\\
  \tilde{\mathbb{W}}_{t,\beta}\mathbb B_{t,\beta} &= \mathbb B_{t,\beta} \tilde{\mathbb{W}}_{t-1,\beta}, \label{eq:WB}\\
  \tilde{\mathbb{W}}_{t,\beta}\tilde{\mathbb{W}}_{t,\beta} &=\mathbb B_{t,\beta}\mathbb A_{t,\beta}\,.     \label{eq:WW}
\end{align}
Using the latter we find 
\begin{align}
 & \tilde{\mathbb{W}}^{x>2}_{t,\beta} =  \mathbb B_{t,\beta}\mathbb A_{t,\beta} \tilde{\mathbb{W}}^{x-2}_{t,\beta}\\
 & = \mathbb B_{t,\beta} \mathbb B_{t-1,\beta}\mathbb A_{t-1,\beta} \mathbb A_{t,\beta} \tilde{\mathbb{W}}^{x-4}_{t,\beta} \notag\\
 &\,\, \vdots \notag\\
 & =  
 \mathbb B_{t,\beta} \cdots  \mathbb B_{t+1- \lfloor x/2\rfloor ,\beta} \tilde{\mathbb{W}}^{{\rm mod}(x,2)}_{t- \lfloor x/2\rfloor,\beta}\mathbb A_{t+1- \lfloor x/2\rfloor,\beta} \cdots  \mathbb A_{t,\beta},\notag
\end{align}
and $\tilde{\mathbb{W}}^{2}_{t,\beta}=\mathbb B_{t,\beta}\mathbb A_{t,\beta}$. Eq.~\eqref{eq:Widentity} follows by observing 
\be
 \tilde{\mathbb{W}}_{0,\beta}^y = \mathbb B_{0,\beta} \mathbb A_{0,\beta}\,, \qquad y\geq 1.  
\ee
}

\section{Equivalence between the FCS of the charge and current}
\label{sec:equiv}

In this appendix we use local relaxation to prove Eq.~\eqref{eq:chargecurrentFCS}. We begin by defining  
\be
f_t(\beta):= \frac{1}{t} \log\tr[{\tilde{ \rho}}_{{\rm st},t}(\beta_1,\beta_2) e^{\beta {\tilde Q}_t}],
\label{eq:ftbeta}
\ee
and computing its first derivative with respect to $\beta$. This gives 
\begin{align}
& f_t'(0) = \frac{1}{t}\sum_{\tau\in\mathbb Z_t/2} {\expval{j_0(\tau)}}\,,\label{eq:fprimefinitet}
\end{align}
where we introduced the short-hand notation
\begin{align}
&\expval{A}=\expval{A e^{\beta_1 Q_{L}+ \beta_2 Q_{R}} }{\Psi_0}.
\end{align}
Splitting the sum in Eq.~\eqref{eq:fprimefinitet} as 
\be
\!\!\!\! f_t'(0) = \frac{1}{t}\sum_{\tau=0}^{\lfloor t^\alpha\rfloor} {\expval{j_0(\tau)}} +\frac{1}{t}\sum_{\tau=\lfloor t^\alpha\rfloor+1}^t \!\!\!{\expval{j_0(\tau)}}, \quad \alpha<1\,,
\ee
we have that 
\begin{align}
\left |f_t'(0) - \frac{1}{t}\sum_{\tau=\lfloor t^\alpha\rfloor+1}^t \!\!\!{\expval{j_0(\tau)}} \right| &=\frac{1}{t} \left|\sum_{\tau=0}^{\lfloor t^\alpha\rfloor} {\expval{j_0(\tau)}}\right | \notag\\
&\leq \frac{1}{t} \sum_{\tau=0}^{\lfloor t^\alpha\rfloor} \left|{\expval{j_0(\tau)}}\right | \notag\\
& \leq O(t^{\alpha-1})\,,
\end{align}
where we used 
\be
|{\expval{j_0(\tau)}}| \leq \| j_0\|_\infty \|e^{\beta_1 Q_{L}+ \beta_2 Q_{R}}\|_\infty = O(1).
\ee
Finally, we invoke local relaxation to claim 
\be
{\expval{j_0(\tau)}}\to \tr[ \rho_{\rm st}(\beta_1,\beta_2) j_0],\quad \tau\geq t^{\alpha}. 
\ee
Putting all together we get 
\be
f'(0) = \lim_{t\to\infty}  f_t'(0) =  \tr[ \rho_{\rm st} j_0]\,. 
\label{eq:fpinf}
\ee
Proceeding analogously we have 
\be
\!\!\!\!f_t''(0) \!\!=\!\! \frac{1}{t}\!\!\!\!\!\!\!\!\sum_{\tau_1\leq \tau_2\in\mathbb Z_t/2} \!\!\!\!\!\!\!\!({\expval{j_0(\tau_1)j_0(\tau_2)}\!-\!\expval{j_0(\tau_1)}\!\expval{j_0(\tau_2)}}),
\ee
which gives 
\begin{align}
f''(0) &= \lim_{t\to\infty} f''_t(0)\notag\\
        &= \sum_{\tau=0}^\infty (\tr[ \rho_{\rm st} j_0 j_0(\tau)]\!-\!\tr[ \rho_{\rm st} j_0]^2)\,,
        \label{eq:fppinf}
\end{align}
where we could again reduce the summation over $\tau_1$ to $\tau_1\geq
t^\alpha$ because each term is a connected correlation and therefore it is
$O(t^0)$. This treatment generalises to all derivatives of finite order.
Therefore, assuming that the expansion of Eq.~\eqref{eq:ftbeta} around
$\beta=0$ converges, we find Eq.~\eqref{eq:chargecurrentFCS}. 

\section{Equilibrium FCS on the Temporal Lattice}
\label{sec:eqFCStemporal}
The expression in Eqs.~(\ref{eq:GGEFCS}, \ref{eq:logx}) for the equilibrium FCS
cannot hold for the temporal lattice. To see this let us assume that it holds
and plug in the defining relations in Eqs.~(\ref{eq:ptilde},
\ref{eq:epsilontilde}), and \eqref{eq:thetatilde}. The result reads as 
\begin{equation}
  \label{eq:deqswapped}
  \begin{aligned}
    f(\beta) &:= \lim_{t\to\infty}\frac{1}{t} \log\tr[{\tilde{ \rho}}_{{\rm st},t}(\beta_1,\beta_2) e^{\beta {\tilde Q}_t}] =\\
    &\phantom{:}= \sum_{m}\int\!\frac{\mathrm{d}\lambda}{2\pi}
    \tilde p^{\prime}_m 
    \log\left[\frac{{\tilde \eta_m^{(\beta_1, \beta_2)}}+e^{-\tilde{w}_m^{(\beta)}}}
    {1+{\tilde \eta_m^{(\beta_1, \beta_2)}}}\right] \\ 
    &\phantom{:}= \sum_{m}\int\!\frac{\mathrm{d}\lambda}{2\pi}
    \varepsilon^{\prime}_m   
    \log\left[\frac{{\eta_m^{(\beta_1, \beta_2)}}+e^{-\tilde{w}_m^{(\beta)}}}
    {1+{\eta_m^{(\beta_1, \beta_2)}}}\right],   
  \end{aligned}
\end{equation}
with
\begin{equation}
  \eta^{(\beta_1, \beta_2)}=
  \frac{1-\vartheta^{(\beta_1, \beta_2)}}{\vartheta^{(\beta_1, \beta_2)}},
\end{equation}
and 
\begin{equation}\label{eq:logyeqswapped}
  \tilde{w}^{(\beta)}
    = - \beta 
    q + 
    T \conL \log\Big[
      \frac{\eta^{(\beta_1,\beta_2)}+e^{-\tilde{w}^{(\beta)}}}
      {1+\eta^{(\beta_1,\beta_2)}}
      \Big].
\end{equation}
Specialising this relation to $\beta_1=\beta_2=0$ and to a reflection symmetric
integrable model we have that $\eta_{m}^{(\beta_1, \beta_2)}(\lambda)$ are even
functions of the rapidity while $\varepsilon^{\prime}_m(\lambda)$ are odd. This
means that Eq.~\eqref{eq:deqswapped} gives identically 0, contradicting known
exact results~\cite{parez2021quasiparticle,ares2022entanglement,bertini2022growth,bertini2023nonequilibrium}. 

This unphysical result is due to an arbitrariness in the starting expression
\eqref{eq:GGEFCS}: since ${p_m'(\lambda)>0}$ one can add terms involving
${\rm{sgn}}[p^{\prime}]$ or ${\rm{sgn}}[\rho^t]$ to Eq.~\eqref{eq:logx} without
changing it. Adding these terms, however, does change the swapped expression.
Using a crossing symmetry argument, Ref.~\cite{bertini2022growth} proposed the
following gauge fixing
\begin{equation}
  \begin{aligned}
    \log\left[\frac{\eta+ e^{-\tilde{w}^{(\beta)}}}{1+{\eta}}\right] &\mapsto  
  \tilde{\kappa}
  \log\left[\frac{\eta + e^{\tilde{w}^{(\beta)}}}{1+\eta }\right],\\
    \tilde{w}^{(\beta)} &\mapsto \tilde{\kappa}\tilde{w}^{(\beta)},
  \end{aligned}
\end{equation}
where $\tilde{\kappa}_m(\lambda)$ is an appropriate sign function [to lighten
the notation we drop the dependence on $\beta_{1,2}$ until the end of
the subsection]. In fact, Ref.~\cite{bertini2022growth} also showed that in the
case of $\beta=0$ and even $\eta_m(\lambda)$ one should choose 
\begin{equation}
  \tilde{\kappa}_m(\lambda)=
  \mathrm{sgn}[\varepsilon^{\prime}_m(\lambda)]= 
  \mathrm{sgn}[v^{(0)}_m(\lambda)]= \mathrm{sgn}[v^{(\beta)}_m(\lambda)].
\end{equation}
Here we denote by $v^{(\beta)}_m(\lambda)$ the dressed velocity in the state
with filling function
\begin{equation} \label{eq:tildethetabeta}
  \tilde{\vartheta}^{(\beta)}_m(\lambda)
  =\frac{1}{\eta_{m}(\lambda) e^{\tilde w_m^{(\beta)}(\lambda)}+1}.
\end{equation}
Whenever $\eta_{m}(\lambda)$ are not even, however, there is an ambiguity in
the choice of $\tilde{\kappa}$, as
$\mathrm{sgn}[\varepsilon^{\prime}]\neq \mathrm{sgn}[v] \neq \mathrm{sgn}[v^{(\beta)}]$.
This is exactly the case that we have to consider here. 

To fix this ambiguity we consider the first two derivatives of $f(\beta)$ with
respect to $\beta$ ($\beta_1$ and $\beta_2$ are kept fixed as we are
considering arbitrary states). A standard TBA calculation gives
\begin{equation}
  \label{eq:fprimebeta}
  \mkern-12mu
  \begin{aligned}
    f'\!(\beta) \!&=\!\! \sum_{m}\!\!\int\!\frac{\mathrm{d}\lambda}{2\pi}
    \varepsilon^{\prime}_m \tilde \vartheta^{(\beta)}_m q_{{\rm eff},m}[\tilde \vartheta^{(\beta)}], \\
    f''\!(\beta) \!&=\!\! \sum_{m}\!\!\int\!\frac{\mathrm{d}\lambda}{2\pi}
    \tilde{\kappa}_m v^{(\beta)}_m \rho_m[\tilde \vartheta^{(\beta)}] (1- \tilde \vartheta^{(\beta)}_m) q^2_{{\rm eff},m}[\tilde \vartheta^{(\beta)}],
  \end{aligned}
  \mkern-12mu
\end{equation}
where $\rho[\vartheta]$, and $q_{{\rm eff}}[\vartheta]$ are root
density and effective charge (cf. Eq.~\eqref{eq:dressing}) in the state
described by the filling functions $\vartheta$.
Note that in computing the derivatives in Eq.~\eqref{eq:fprimebeta} we used 
\begin{equation} \label{eq:logyprime}
  \partial_\beta \left(\tilde{\kappa} \tilde{w}^{(\beta)}\right)
  = \tilde{\kappa} \partial_\beta \tilde{w}^{(\beta)}=
  -q_{{\rm eff}}[\tilde \vartheta^{(\beta)}]\, ,
\end{equation}
and we assumed $\tilde{\kappa}$ to be independent of $\beta$ and, therefore, we
excluded the choice $\tilde{\kappa}={{\rm{sgn}}[v^{(\beta)}]}$: we will come
back to this point after Eq.~\eqref{eq:Leqswappedcorrect1}.   
 
At the same time, recalling Eqs.~\eqref{eq:fpinf}--\eqref{eq:fppinf} and
expressing them in TBA~\cite{doyon2020lecture, doyon2017drude} we have  
\begin{equation} \label{eq:fin0}
  \begin{aligned}
    &f'(0) = \sum_{m}\!\int\!\frac{\mathrm{d}\lambda}{2\pi}
    \varepsilon^{\prime}_m \vartheta_m q_{\mathrm{eff},m}[\vartheta],\\
    &f''(0) = \sum_{m}\!\int\!\frac{\mathrm{d}\lambda}{2\pi} |v_m| \rho_m[\vartheta] (1-\vartheta_m) q^2_{{\rm eff},m}[\vartheta].
  \end{aligned}
\end{equation}
We see that the second derivative agrees with \eqref{eq:fprimebeta} for
$\beta=0$ only if we chose $\tilde{\kappa}={{\rm{sgn}}[v^{(0)}]}$. This leads to 
\begin{equation} \label{eq:deqswappedcorrect1}
  f(\beta)= \sum_{m}\!\int\!\frac{\mathrm{d}\lambda}{2\pi}
  \varepsilon^{\prime}_m(\lambda)  \mathcal{L}_{m}^{(\beta)}(\lambda),
\end{equation}
with
\begin{equation} \label{eq:Leqswappedcorrect1}
  \begin{aligned}
    \mathcal{L}^{(\beta)} &= 
    \tilde{\kappa}
    \log \left[\frac{\eta+ e^{-\tilde{w}^{(\beta)}}}{1+\eta}\right],\\
    \tilde{\kappa} \tilde{w}^{(\beta)}&=
    - \beta q + T\conL \mathcal{L}^{(\beta)}.
  \end{aligned}
\end{equation}

In fact, we also have another option. We can assume
$\tilde{\kappa}^{(\beta)}={{\rm{sgn}}[v^{(\beta)}]}$ and reproduce the first
of~\eqref{eq:fin0} by replacing $f(\beta)$ with the integral of the first
of~\eqref{eq:fprimebeta}, i.e.,
\begin{equation} \label{eq:deqswappedcorrect2}
  f(\beta) = \int_0^\beta {\rm d}u \sum_{m}\int\!\frac{\mathrm{d}\lambda}{2\pi}
  \varepsilon^{\prime}_m \tilde \vartheta^{(u)}_m
  q_{{\rm eff},m}[\tilde \vartheta^{(u)}_m].
\end{equation}
This choice also reproduces the second of \eqref{eq:fin0} upon replacing
$\tilde w^{(\beta)}$ in~\eqref{eq:tildethetabeta} by by the integral of its
derivative (obtained from Eq.~\eqref{eq:logyprime} for $\beta$-independent
$\tilde{\kappa}^{(0)}$). Namely 
\begin{equation} \label{eq:Leqswappedcorrect2}
  \tilde{w}^{(\beta)} = 
  -\int^\beta_0\! {\rm d}u\,\, 
  \tilde{\kappa}^{(u)}
  q_{{\rm eff}}[\tilde{\vartheta}^{(u)}].
\end{equation}
Note that in writing~\eqref{eq:deqswappedcorrect2} 
and~\eqref{eq:Leqswappedcorrect2} we used 
\begin{equation}
  f(0)=0,\qquad \tilde w^{(0)}  = 0. 
\end{equation}
The form (\ref{eq:tildethetabeta}, \ref{eq:deqswappedcorrect2}, \ref{eq:Leqswappedcorrect2}) 
ensures that Eqs.~\eqref{eq:fprimebeta} give the expectation value of current
and Drude self-weight for any value of $\beta$, rather than only for $\beta=0$. 

Importantly, we stress that performing the inverse replacement 
\begin{equation}
 \varepsilon^{\prime}_m \mapsto p'_m,
\end{equation}
both (\ref{eq:deqswappedcorrect1}, \ref{eq:Leqswappedcorrect1}) and 
(\ref{eq:tildethetabeta}, \ref{eq:deqswappedcorrect2}, \ref{eq:Leqswappedcorrect2})
reduce to (\ref{eq:GGEFCS}, \ref{eq:logx}). 

To choose among the two expressions we a further consistency check. We use
Eq.~\eqref{eq:chargecurrentFCS} and impose that $f(\beta)$ computed via
space-time swap should coincide with that computed in
Ref.~\cite{myers2020transport} via the ballistic fluctuation formalism.
A direct comparison shows that the prediction
(\ref{eq:deqswappedcorrect2}, \ref{eq:Leqswappedcorrect2})
agrees exactly with the findings of the aforementioned reference (cf. Eqs.~(15)
and (24) of Ref.~\cite{myers2020transport}). Therefore, we conclude that before
performing the space-time swap in Eqs.~(\ref{eq:GGEFCS}, \ref{eq:logx}) should
be rewritten as in Eqs.~(\ref{eq:deq2})--(\ref{eq:logx2}) which coincides with
(\ref{eq:tildethetabeta}, \ref{eq:deqswappedcorrect2}, \ref{eq:Leqswappedcorrect2})
under the swap. 

\section{Details on the TBA description}
\label{sec:detailsTBA}

\subsection{TBA expressions for the equilibrium FCS}
\label{sec:tbaFCS}
To find Eq.~\eqref{eq:GGEFCS} we write the l.h.s.\ as the ratio of two
partition functions and evaluates their logarithm using \eqref{eq:freeenergy}.
Explicitly we have 
\begin{align}
&\lim_{L\to\infty}\frac{\log\tr[\rho_{{\rm st},L} e^{\beta Q}]}{L}\\
& =\lim_{L\to\infty}\frac{\log\tr[e^{ -\sum_k \mu_k Q^{(k)}+\beta Q}]-\log\tr[e^{ -\sum_k \mu_k Q^{(k)}}]}{L}.\notag
\end{align}
Upon applying \eqref{eq:freeenergy} we find  
\begin{equation}
  \begin{aligned}
    &\lim_{L\to\infty}\frac{\log\tr[\rho_{{\rm st},L} e^{\beta Q}]}{L} \\
    &=\sum_m\int \frac{{\rm d}\lambda}{2\pi} p'_m(\lambda)
    \log\left[\frac{1\!+\!{\eta^{(\beta)}_m(\lambda)^{-1}}}
    {1\!+\!{\eta_m(\lambda)^{-1}}}\right],
  \end{aligned}
\end{equation}
where $\eta^{(\beta)}_m(\lambda)$ fulfils~\eqref{eq:GTBA} with
$d_m(\lambda)\mapsto d_m(\lambda)-\beta q $. This expression can be brought to
the form~(\ref{eq:GGEFCS}, \ref{eq:logx}) by setting
\begin{equation}
  e^{w^{(\beta)}} = \frac{\eta^{(\beta)}}{\eta}\,.
\end{equation}
Finally, noting that 
\begin{align}
  \kappa^{(\beta)} :\!&=\mathrm{sgn}[\rho^t[\vartheta^{(\beta)}]]=1,\\
  \partial_\beta \left(\kappa^{(\beta)}w^{(\beta)}\right)&=
  \kappa^{(\beta)}\partial_\beta w^{(\beta)}
  =-q_{{\rm eff}}[\vartheta^{(\beta)}],\quad
\end{align}
we have 
\begin{equation}
  \partial_\beta \mathcal{K}^{(\beta)}  
  =  \vartheta^{(\beta)} q_{\mathrm{eff}}[\vartheta^{(\beta)}]\,, 
\end{equation}
where $\vartheta^{(\beta)}$ is defined in Eq.~\eqref{eq:thetabeta}. Therefore, we can write  
\begin{equation} \label{eq:proofKint}
  \mathcal{K}^{(\beta)} = \int_0^\beta\!{\rm d}u\,
  \vartheta^{(u)} q_{\mathrm{eff}}[\vartheta^{(u)}], 
\end{equation}
where we used 
\begin{equation}
  w^{(0)}=0 \quad \Rightarrow \quad \mathcal{K}^{(0)}=0\,.
\end{equation}
Eq.~\eqref{eq:proofKint} proves the equality between
Eqs.~(\ref{eq:GGEFCS}, \ref{eq:logx}) and \eqref{eq:deq2}--\eqref{eq:logx2}.  

\subsection{TBA expressions for the equilibrium charged moments}
\label{sec:tbaCM}

Proceeding as in Appendix~\ref{sec:tbaFCS} we have 
\begin{equation}
  \begin{aligned}
    &\lim_{L \to\infty}\frac{1}{L}
    \log\tr[\prod_{j=1}^{n} e^{\beta_j  Q} \rho_{{\rm st},j} ] \\
    &=\sum_m\int \frac{{\rm d}\lambda}{2\pi} p'_m(\lambda) \log\left[\frac{1\!+\!{\eta^{(\beta)}_m(\mu)^{-1}}}{\prod_j (1\!+\!{\eta_{j,m}(\mu)^{-1}})}\right],
  \end{aligned}
\end{equation}
where $\eta_{j}$ fulfils~\eqref{eq:GTBA} with 
\begin{equation}
  d\mapsto d_{j}
\end{equation}
while $\eta^{(\beta)}$ fulfils the same equation with 
\begin{equation}
  d\mapsto \sum_j d_{j}-\beta q.
\end{equation}

Setting 
\begin{equation}
  e^{w^{(\beta)}_n} = \frac{\eta^{(\beta)}}{\prod_j \eta_{j}},
\end{equation}
we then have 
\begin{equation}
    \mkern-4mu
  \begin{aligned}
    &\lim_{L \to\infty}\frac{1}{L}
    \log\tr[\prod_{j=1}^{n} e^{\beta_j  Q} \rho_{{\rm st},j} ]\\
    &= \sum_m\int \frac{{\rm d}\lambda}{2\pi} p'_m (\lambda)
    \log\!\!\left[\frac{\prod_j \eta_{j,m}(\lambda)+e^{-w_{n,m}^{(\beta)}(\lambda)}}
    {\prod_j (1+{\eta_{j,m}(\lambda)})}\right],
  \end{aligned}
    \mkern-5mu
\end{equation}
with 
\begin{align}     
  \kappa_n^{(\beta)}w^{(\beta)}_{n}+\beta q &= T\conL \log\!\!\left[
    \frac{\prod_j \eta_{j}+e^{-w_{n}^{(\beta)}}}{\prod_j (1+{\eta_{j}})}\right]\, ,\\
  \kappa_n^{(\beta)}&=\mathrm{sgn}[\rho^{t}[\vartheta^{(\beta)}_n]]=1,\\
  \vartheta^{(\beta)}_{n}&=\frac{1}{1+e^{w^{(\beta)}_n}\prod_{j}\eta_j}.
\end{align}     
Using now
\begin{equation}
  \partial_\beta \left(\kappa_n^{(\beta)} w_n^{(\beta)}\right)=
  \kappa_n^{(\beta)}\partial_\beta w_n^{(\beta)}=
  -q_{\mathrm{eff}}[\vartheta^{(\beta)}_{n}],
\end{equation}
we find 
\begin{equation}
  \begin{aligned}
    &\lim_{L \to\infty}\frac{1}{L}
    \log\tr[\prod_{j=1}^{n} e^{\beta_j  Q} \rho_{{\rm st},j} ] \\
    &= \int_0^\beta {\rm d}u \sum_{m}\!\!\int\!\frac{\mathrm{d}\lambda}{2\pi}
    p^{\prime}_m \vartheta^{(u)}_{n,m} q_{\mathrm{eff},m}[\vartheta^{(u)}_{n,m}] \\
    & + \sum_m\int \frac{{\rm d}\lambda}{2\pi} p'_m 
    \kappa_{n,m}^{(0)}
    \log\!\!\left[\frac{\prod_j \eta_{j,m}+e^{- w^{(0)}_{n,m}}}
    {\prod_j (1+{\eta_{j,m}})}\right],\\
  \end{aligned}
\end{equation}
and  
\begin{equation}
  \kappa_n^{(0)}w_{n}^{(0)}=
  T \conL
  \log \left[\frac{\prod_j \eta_j + e^{-w_{n}^{(0)}}}{\prod_j(1+\eta_j)}\right].
\end{equation}

\subsection{Simplified form of the slope under the condition~\eqref{eq:signcondition}}
\label{sec:simplification}
Whenever the condition~\eqref{eq:signcondition} holds we can explicitly
integrate Eqs.~\eqref{eq:swapdeq2} and \eqref{eq:swaplogx2}. To this end we
note that in this case Eq.~\eqref{eq:swaplogx2} can be rewritten as 
\begin{equation}
  \partial_{u} \left(\tilde{\kappa}^{(u)}\tilde{w}^{(u)}\right)=
  \tilde{\kappa}^{(u)}\partial_u \tilde{w}^{(u)}=
  -q_{\mathrm{eff}}[\tilde{\vartheta}^{(u)}],
\end{equation}
where we introduced the short-hand notation
\begin{equation}
  \tilde{\kappa}^{(u)}=\mathrm{sgn}[\rho^t v[\tilde{\vartheta}^{(u)}]]
  =\mathrm{sgn}[\rho^t v[\tilde{\vartheta}^{(0)}]]
  =\tilde{\kappa}^{(0)}.
\end{equation}
%
We now recall that $q_{\mathrm{eff},m}[\vartheta^{(u)}](\lambda)$ fulfils
Eq.~\eqref{eq:dressing} with $b_m(\lambda) = q_m$. Integrating the latter
equation in $u\in[0,\beta]$ we find  
\begin{equation} \label{eq:IntEqtildew}
    \tilde{\kappa}^{(\beta)}\tilde{w}^{(\beta)}
    =  - \beta q +
    \int\limits_0^{\beta}\,\mathrm{d} u \,
    T\conL \left(\tilde{\vartheta}^{(u)} q_{\mathrm{eff}}[\tilde{\vartheta}^{(u)}]\right).
\end{equation}
Defining now $\mathcal{L}^{(u)}$ as
\begin{equation}
  \mathcal{L}^{(u)}=\tilde{\kappa}^{(u)}
  \log\left[\frac{\eta(\beta_1,\beta_2)+e^{-\tilde{w}^{(\beta)}}}
  {\eta(\beta_1,\beta_2)+1}\right],
\end{equation}
we note that 
\begin{equation}
  \partial_u \mathcal{L}^{(u)}=
  \tilde{\vartheta}^{(u)} q_{\mathrm{eff}}[\tilde{\vartheta}^{(u)}],
\end{equation}
and therefore the integral over $\mathrm{d}u$ in~\eqref{eq:IntEqtildew} gives the
integral equation for $\tilde{w}^{(\beta)}$,
\begin{equation}
    \tilde{\kappa}^{(\beta)}\tilde{w}^{(\beta)}
    =  - \beta q +
    T\conL \mathcal{L}^{(\beta)} .
\end{equation}
Moreover, this also allows us to perform the integral in Eq.~\eqref{eq:swapdeq2} and
we finally find 
\begin{equation}
  \smashoperator{\lim_{t\to\infty}}
  \frac{\log\tr[{\tilde{ \rho}}_{{\rm st},t}(\beta_1,\beta_2) e^{\beta {\tilde Q}_t}]}{t}
  \! =\! \sum_{m} \mkern-7mu \int\mkern-7mu\frac{\mathrm{d}\lambda}{2\pi}
  \varepsilon^{\prime}_m \mathcal{L}^{(\beta)}_m(\lambda). 
\end{equation}

Similarly, one can show that the same simplification applies for
Eqs.~\eqref{eq:swappeddeqn}--\eqref{eq:logy0}. The final result reads as 
\begin{align}
  \label{eq:deqswappedSimpl1}
  s^{(\rm r)}_{\bm{\beta}} &= \sum_m\int \frac{{\rm d}\lambda}{2\pi} \varepsilon'_m
    \mathcal{L}_{n,m}^{(\mathrm{r},\beta)},\\    
  \mathcal{L}_{n}^{(\mathrm{r},u)} 
  &= \tilde{\kappa}_n^{(0)} \log\left[\frac{\prod_{j=1}^n 
  \eta^{(\mathrm{r})}_{j} + e^{-\tilde{w}_{n}^{(\mathrm{r},\beta)}}}
{\prod_{j=1}^n (1+\eta^{(\mathrm{r})}_{j})}\right],\\
  \label{eq:LeqswappedSimpl1}
  \tilde{\kappa}_n^{(0)}\tilde{w}_{n}^{(\mathrm{r},\beta)}
  &= \mp \beta q +
  T\conL \mathcal{L}^{(\mathrm{r},\beta)}_n,
\end{align}
where the top (bottom) choice in the $\mp \beta$ corresponds to $\mathrm{r}=\mathrm{L}$
($\mathrm{r}=\mathrm{R}$).

\section{FCS in a free model}
\label{sec:FFfcs}
Here we derive the result~\eqref{eq:FFfcs} through the two point correlation
functions of the model.  We start by introducing  ${\bm{c}}_x=(c_x,c^\dag_x)$
where $c_x=\int {\rm d }p \,e^{-ipx}c_p/2\pi$.   The relevant two point
function is
\begin{equation}
  \begin{aligned}
    &\Gamma_A(t)=2\,\text{tr}\left[\rho_A(t){\bm{c}}_x^\dag{\bm{c}}_y\right]-\delta_{x,y}\\
  &=
   \int \frac{{\rm d} p}{2\pi}\, e^{i p(x-y)}\left[a_p\sigma^z+b_p\!\left(\sigma^+e^{-2i\epsilon(p)t}+\mathrm{h.c.}\right)\right],
  \end{aligned}
\end{equation}
where 
\begin{align}
  a_p&=(K^2-1)/(K^2+1),\\
  b_p&=2K/(1+K^2)\\
  \sigma^\pm&=\frac{1}{2}[\sigma^x\pm i \sigma^y],
\end{align} 
with $\sigma^{x,y,z}$ being Pauli matrices.  It is also necessary to express $e^{\beta N_A}$ through its two point function by treating it as a density matrix $\rho_N=e^{\beta N_A}/\text{tr}\left[e^{\beta N_A}\right]$ so that
\begin{equation}
    \Gamma_N=2\,\mathrm{tr}\left[\rho_N{\bm{c}}_x^\dag{\bm{c}}_y\right]-\delta_{x,y}
    =\tanh{(\beta/2)}\,\sigma^z\delta_{x,y}.
\end{equation}
Using  the algebra of Gaussian matrices we can express the charged moment as~\cite{fagotti2010entanglement}
\be
Z_{\beta}(A,t)=\text{tr}[e^{\beta N_A}]\sqrt{\text{det}\left[\frac{1}{2}(1+\Gamma_A(t)\Gamma_N)\right]}\,.
\ee
Taking the $\log$ and expanding the resulting expression as a power series gives
\begin{equation} \label{eq:TildeGamma}
  \begin{aligned}
    \log Z_{\beta}(A,t)= &- \sum_{n=1}^\infty \frac{(-\tanh{(\beta/2))^n}}{n}\text{tr}[(\bar{\Gamma}_A(t))^n]\\
    &+|A|\log{\left[\frac{1+e^{\beta}}{2}\right]},
  \end{aligned}
\end{equation}
with $\bar{\Gamma}_A(t)=\Gamma_A(t)\sigma_z$.  The trace over the powers of
$\bar{\Gamma}_A(t)$ can then be evaluated using the multidimensional stationary
phase approximation~\cite{fagotti2008evolution} resulting in
\begin{equation}
  \mkern-12mu
  \begin{aligned}
    &\mathrm{tr}[(\bar{\Gamma}_A(t))^n]=
  L\!\int\!\frac{{\rm d}p}{2\pi}\,\text{min}(1,2|\epsilon'|\zeta)2a_p^n\\
    +& L\!\int\!
    \frac{{\rm d}p}{2\pi}\left(1\!-\!\text{min}(1,2|\epsilon'|\zeta)\right)
    \left[(a_p\!+\! i b_p)^n\!+\!(a_p\!-\! ib_p )^n\right],
  \end{aligned}
  \mkern-12mu
\end{equation}
where $\text{min}(1,2|\epsilon'|\zeta)$ is the characteristic function counting the number of quasiparticle pairs shared between $A$ and its compliment and $\zeta=t/|A|$.  Inserting this  into~\eqref{eq:TildeGamma} and performing the sum we arrive at the stated result. 
\bibliography{./bibliography}

\end{document}